\documentclass[fleqn,usenatbib]{mnras}

\usepackage{newtxtext,newtxmath}
\usepackage[T1]{fontenc}

\DeclareRobustCommand{\VAN}[3]{#2}
\let\VANthebibliography\thebibliography
\def\thebibliography{\DeclareRobustCommand{\VAN}[3]{##3}\VANthebibliography}

\usepackage{graphicx}	% Including figure files
\usepackage{amsmath}	% Advanced maths commands
\usepackage{xspace}
\usepackage{multirow}
\usepackage{dcolumn}
\usepackage{bm}
\usepackage{xcolor} %FIX ME

\DeclareMathAlphabet{\mathbfsf}{\encodingdefault}{\sfdefault}{bx}{sl}

\newcommand{\be}{\begin{equation}}
\newcommand{\ee}{\end{equation}}
\newcommand{\bea}{\begin{eqnarray}}
\newcommand{\eea}{\end{eqnarray}}

\newcommand{\ph}{\textsc{IMRPhenom}\xspace}

\newcommand{\ppvtwo}{\textsc{IMRPhenomPv2}\xspace}
\newcommand{\phD}{\textsc{IMRPhenomD}\xspace}
\newcommand{\phX}{\textsc{IMRPhenomXAS}\xspace}
\newcommand{\phenX}{\textsc{IMRPhenomX}\xspace}
\newcommand{\phXF}{\textsc{IMRPhenomX}\xspace}

\newcommand{\phHM}{\textsc{IMRPhenomHM}\xspace}
\newcommand{\phPvtwo}{\textsc{IMRPhenomPv2}\xspace}

\newcommand{\phPvthreehm}{\textsc{IMRPhenomPv3HM}\xspace}
\newcommand{\phXHM}{\textsc{IMRPhenomXHM}\xspace}
\newcommand{\phXP}{\textsc{IMRPhenomXP}\xspace}
\newcommand{\phXPHM}{\textsc{IMRPhenomXPHM}\xspace}
\newcommand{\phT}{\textsc{IMRPhenomT}\xspace}
\newcommand{\phTHM}{\textsc{IMRPhenomTHM}\xspace}
\newcommand{\phTP}{\textsc{IMRPhenomTP}\xspace}
\newcommand{\phTPHM}{\textsc{IMRPhenomTPHM}\xspace}
\newcommand{\NRHybSur}{\textsc{NRHybSur3dq8}\xspace}
\newcommand{\NRSur}{\textsc{NRSur7dq2}\xspace}
\newcommand{\seobnrvthree}{\textsc{SEOBNRv3}\xspace}

\newcommand{\seobnrvforhm}{\textsc{SEOBNRv4HM}\xspace}
\newcommand{\seobnrvforhmrom}{\textsc{SEOBNRv4HM\_ROM}\xspace}
\newcommand{\seobnrvforphm}{\textsc{SEOBNRv4PHM}\xspace}

\newcommand{\seobnrvtwo}{\textsc{SEOBNRv2}\xspace}

\newcommand{\Msun}{M_\odot}
\newcommand{\chieff}{\chi_\mathrm{eff}}
\newcommand{\chip}{\chi_\mathrm{p}}
\newcommand{\chirpMass}{\mathcal{M}}
\newcommand{\massRatio}{\mathrm{q}}

\def\MOneSourceCIPhTPHMfirst{\ensuremath{35.0_{-2.1}^{+3.4}}\xspace}
\def\MTwoSourceCIPhTPHMfirst{\ensuremath{30.7_{-2.9}^{+2.1}}\xspace}
\def\MtotalSourceCIPhTPHMfirst{\ensuremath{65.7_{-2.3}^{+2.7}}\xspace}
\def\ChirpMassSourceCIPhTPHMfirst{\ensuremath{28.4_{-1.0}^{+1.1}}\xspace}
\def\MassRatioCIPhTPHMfirst{\ensuremath{0.88_{-0.15}^{+0.10}}\xspace}
\def\ChiEffCIPhTPHMfirst{\ensuremath{-0.01_{-0.09}^{+0.09}}\xspace}
\def\ChiPCIPhTPHMfirst{\ensuremath{0.43_{-0.30}^{+0.34}}\xspace}
\def\DLCIPhTPHMfirst{\ensuremath{480_{-130}^{+120}}\xspace}

\def\ThetaJNCIPhTPHMfirst{\ensuremath{2.78_{-0.37}^{+0.23}}\xspace}
\def\FinalSpinCIPhTPHMfirst{\ensuremath{0.80_{-0.08}^{+0.10}}\xspace}
\def\FinalMassSourceCIPhTPHMfirst{\ensuremath{61.7_{-2.4}^{+2.4}}\xspace}

\def\MOneSourceCIPhXPHMfirst{\ensuremath{35.0_{-2.3}^{+3.3}}\xspace}
\def\MTwoSourceCIPhXPHMfirst{\ensuremath{30.2_{-3.2}^{+2.3}}\xspace}
\def\MtotalSourceCIPhXPHMfirst{\ensuremath{65.2_{-2.4}^{+2.5}}\xspace}
\def\ChirpMassSourceCIPhXPHMfirst{\ensuremath{28.2_{-1.1}^{+1.1}}\xspace}
\def\MassRatioCIPhXPHMfirst{\ensuremath{0.87_{-0.15}^{+0.11}}\xspace}
\def\ChiEffCIPhXPHMfirst{\ensuremath{-0.02_{-0.10}^{+0.08}}\xspace}
\def\ChiPCIPhXPHMfirst{\ensuremath{0.44_{-0.30}^{+0.34}}\xspace}
\def\DLCIPhXPHMfirst{\ensuremath{490_{-120}^{+120}}\xspace}

\def\ThetaJNCIPhXPHMfirst{\ensuremath{2.76_{-0.33}^{+0.24}}\xspace}
\def\FinalSpinCIPhXPHMfirst{\ensuremath{0.81_{-0.08}^{+0.10}}\xspace}
\def\FinalMassSourceCIPhXPHMfirst{\ensuremath{61.2_{-2.6}^{+2.4}}\xspace}

\def\MOneSourceCIPhXPfirst{\ensuremath{35.6_{-2.6}^{+3.7}}\xspace}
\def\MTwoSourceCIPhXPfirst{\ensuremath{30.0_{-3.4}^{+2.5}}\xspace}
\def\MtotalSourceCIPhXPfirst{\ensuremath{65.7_{-2.6}^{+2.8}}\xspace}
\def\ChirpMassSourceCIPhXPfirst{\ensuremath{28.4_{-1.2}^{+1.2}}\xspace}
\def\MassRatioCIPhXPfirst{\ensuremath{0.85_{-0.16}^{+0.12}}\xspace}
\def\ChiEffCIPhXPfirst{\ensuremath{-0.02_{-0.11}^{+0.08}}\xspace}
\def\ChiPCIPhXPfirst{\ensuremath{0.42_{-0.29}^{+0.36}}\xspace}
\def\DLCIPhXPfirst{\ensuremath{440_{-140}^{+140}}\xspace}

\def\ThetaJNCIPhXPfirst{\ensuremath{2.63_{-0.54}^{+0.32}}\xspace}
\def\FinalSpinCIPhXPfirst{\ensuremath{0.80_{-0.08}^{+0.11}}\xspace}
\def\FinalMassSourceCIPhXPfirst{\ensuremath{61.7_{-2.8}^{+2.7}}\xspace}

\def\MOneSourceCIPhPvtwofirst{\ensuremath{35.3_{-2.4}^{+3.7}}\xspace}
\def\MTwoSourceCIPhPvtwofirst{\ensuremath{30.3_{-3.4}^{+2.3}}\xspace}
\def\MtotalSourceCIPhPvtwofirst{\ensuremath{65.7_{-2.5}^{+2.7}}\xspace}
\def\ChirpMassSourceCIPhPvtwofirst{\ensuremath{28.4_{-1.1}^{+1.2}}\xspace}
\def\MassRatioCIPhPvtwofirst{\ensuremath{0.86_{-0.16}^{+0.11}}\xspace}
\def\ChiEffCIPhPvtwofirst{\ensuremath{-0.03_{-0.09}^{+0.08}}\xspace}
\def\ChiPCIPhPvtwofirst{\ensuremath{0.39_{-0.27}^{+0.38}}\xspace}
\def\DLCIPhPvtwofirst{\ensuremath{450_{-140}^{+120}}\xspace}

\def\ThetaJNCIPhPvtwofirst{\ensuremath{2.72_{-0.52}^{+0.27}}\xspace}
\def\FinalSpinCIPhPvtwofirst{\ensuremath{0.80_{-0.08}^{+0.12}}\xspace}
\def\FinalMassSourceCIPhPvtwofirst{\ensuremath{61.6_{-2.6}^{+2.7}}\xspace}

\def\MOneSourceCIPhXPHMsecond{\ensuremath{24.2_{-5.1}^{+10.5}}\xspace}
\def\MTwoSourceCIPhXPHMsecond{\ensuremath{13.6_{-4.0}^{+3.5}}\xspace}
\def\MtotalSourceCIPhXPHMsecond{\ensuremath{38.3_{-3.7}^{+6.9}}\xspace}
\def\ChirpMassSourceCIPhXPHMsecond{\ensuremath{15.5_{-1.1}^{+1.5}}\xspace}
\def\MassRatioCIPhXPHMsecond{\ensuremath{0.57_{-0.28}^{+0.32}}\xspace}
\def\ChiEffCIPhXPHMsecond{\ensuremath{0.12_{-0.17}^{+0.21}}\xspace}
\def\ChiPCIPhXPHMsecond{\ensuremath{0.36_{-0.23}^{+0.37}}\xspace}
\def\DLCIPhXPHMsecond{\ensuremath{1060_{-400}^{+460}}\xspace}

\def\ThetaJNCIPhXPHMsecond{\ensuremath{1.66_{-1.25}^{+1.08}}\xspace}
\def\FinalSpinCIPhXPHMsecond{\ensuremath{0.79_{-0.12}^{+0.13}}\xspace}
\def\FinalMassSourceCIPhXPHMsecond{\ensuremath{36.2_{-3.7}^{+7.2}}\xspace}

\def\MOneSourceCIPhXPsecond{\ensuremath{22.0_{-3.8}^{+8.9}}\xspace}
\def\MTwoSourceCIPhXPsecond{\ensuremath{14.1_{-3.8}^{+2.9}}\xspace}
\def\MtotalSourceCIPhXPsecond{\ensuremath{36.6_{-2.8}^{+5.4}}\xspace}
\def\ChirpMassSourceCIPhXPsecond{\ensuremath{15.2_{-0.9}^{+1.2}}\xspace}
\def\MassRatioCIPhXPsecond{\ensuremath{0.65_{-0.31}^{+0.27}}\xspace}
\def\ChiEffCIPhXPsecond{\ensuremath{0.05_{-0.14}^{+0.21}}\xspace}
\def\ChiPCIPhXPsecond{\ensuremath{0.38_{-0.25}^{+0.36}}\xspace}
\def\DLCIPhXPsecond{\ensuremath{1120_{-380}^{+400}}\xspace}

\def\ThetaJNCIPhXPsecond{\ensuremath{1.84_{-1.49}^{+0.96}}\xspace}
\def\FinalSpinCIPhXPsecond{\ensuremath{0.79_{-0.11}^{+0.13}}\xspace}
\def\FinalMassSourceCIPhXPsecond{\ensuremath{34.4_{-2.8}^{+5.6}}\xspace}

\def\MOneSourceCIPhPvtwosecond{\ensuremath{22.6_{-4.3}^{+9.6}}\xspace}
\def\MTwoSourceCIPhPvtwosecond{\ensuremath{13.8_{-3.9}^{+3.2}}\xspace}
\def\MtotalSourceCIPhPvtwosecond{\ensuremath{36.9_{-3.0}^{+5.9}}\xspace}
\def\ChirpMassSourceCIPhPvtwosecond{\ensuremath{15.2_{-0.9}^{+1.2}}\xspace}
\def\MassRatioCIPhPvtwosecond{\ensuremath{0.61_{-0.30}^{+0.30}}\xspace}
\def\ChiEffCIPhPvtwosecond{\ensuremath{0.04_{-0.15}^{+0.20}}\xspace}
\def\ChiPCIPhPvtwosecond{\ensuremath{0.33_{-0.22}^{+0.35}}\xspace}
\def\DLCIPhPvtwosecond{\ensuremath{1090_{-400}^{+390}}\xspace}

\def\ThetaJNCIPhPvtwosecond{\ensuremath{1.99_{-1.64}^{+0.84}}\xspace}
\def\FinalSpinCIPhPvtwosecond{\ensuremath{0.77_{-0.12}^{+0.13}}\xspace}
\def\FinalMassSourceCIPhPvtwosecond{\ensuremath{34.9_{-3.0}^{+6.2}}\xspace}

\def\MOneSourceCIPhXPHMthird{\ensuremath{14.2_{-3.2}^{+6.2}}\xspace}
\def\MTwoSourceCIPhXPHMthird{\ensuremath{7.4_{-1.9}^{+2.0}}\xspace}
\def\MtotalSourceCIPhXPHMthird{\ensuremath{21.7_{-1.4}^{+4.3}}\xspace}
\def\ChirpMassSourceCIPhXPHMthird{\ensuremath{8.8_{-0.2}^{+0.2}}\xspace}
\def\MassRatioCIPhXPHMthird{\ensuremath{0.52_{-0.25}^{+0.34}}\xspace}
\def\ChiEffCIPhXPHMthird{\ensuremath{0.20_{-0.07}^{+0.13}}\xspace}
\def\ChiPCIPhXPHMthird{\ensuremath{0.57_{-0.31}^{+0.28}}\xspace}
\def\DLCIPhXPHMthird{\ensuremath{470_{-160}^{+130}}\xspace}

\def\ThetaJNCIPhXPHMthird{\ensuremath{0.79_{-0.44}^{+1.92}}\xspace}
\def\FinalSpinCIPhXPHMthird{\ensuremath{0.88_{-0.11}^{+0.14}}\xspace}
\def\FinalMassSourceCIPhXPHMthird{\ensuremath{20.3_{-1.4}^{+4.3}}\xspace}

\def\MOneSourceCIPhXPthird{\ensuremath{13.0_{-2.3}^{+5.5}}\xspace}
\def\MTwoSourceCIPhXPthird{\ensuremath{8.0_{-2.1}^{+1.7}}\xspace}
\def\MtotalSourceCIPhXPthird{\ensuremath{21.2_{-1.0}^{+3.3}}\xspace}
\def\ChirpMassSourceCIPhXPthird{\ensuremath{8.8_{-0.2}^{+0.3}}\xspace}
\def\MassRatioCIPhXPthird{\ensuremath{0.62_{-0.30}^{+0.30}}\xspace}
\def\ChiEffCIPhXPthird{\ensuremath{0.18_{-0.06}^{+0.13}}\xspace}
\def\ChiPCIPhXPthird{\ensuremath{0.48_{-0.26}^{+0.32}}\xspace}
\def\DLCIPhXPthird{\ensuremath{460_{-160}^{+150}}\xspace}

\def\ThetaJNCIPhXPthird{\ensuremath{1.01_{-0.62}^{+1.71}}\xspace}
\def\FinalSpinCIPhXPthird{\ensuremath{0.84_{-0.09}^{+0.11}}\xspace}
\def\FinalMassSourceCIPhXPthird{\ensuremath{19.8_{-1.1}^{+3.5}}\xspace}

\def\MOneSourceCIPhPvtwothird{\ensuremath{14.0_{-2.9}^{+5.6}}\xspace}
\def\MTwoSourceCIPhPvtwothird{\ensuremath{7.6_{-1.8}^{+1.8}}\xspace}
\def\MtotalSourceCIPhPvtwothird{\ensuremath{21.6_{-1.3}^{+3.7}}\xspace}
\def\ChirpMassSourceCIPhPvtwothird{\ensuremath{8.9_{-0.2}^{+0.3}}\xspace}
\def\MassRatioCIPhPvtwothird{\ensuremath{0.54_{-0.25}^{+0.30}}\xspace}
\def\ChiEffCIPhPvtwothird{\ensuremath{0.20_{-0.06}^{+0.13}}\xspace}
\def\ChiPCIPhPvtwothird{\ensuremath{0.47_{-0.26}^{+0.32}}\xspace}
\def\DLCIPhPvtwothird{\ensuremath{440_{-160}^{+150}}\xspace}

\def\ThetaJNCIPhPvtwothird{\ensuremath{1.05_{-0.66}^{+1.66}}\xspace}
\def\FinalSpinCIPhPvtwothird{\ensuremath{0.84_{-0.09}^{+0.12}}\xspace}
\def\FinalMassSourceCIPhPvtwothird{\ensuremath{20.3_{-1.4}^{+3.9}}\xspace}

\def\MOneSourceCIPhXPHMfourth{\ensuremath{29.3_{-3.7}^{+5.0}}\xspace}
\def\MTwoSourceCIPhXPHMfourth{\ensuremath{20.8_{-4.1}^{+3.6}}\xspace}
\def\MtotalSourceCIPhXPHMfourth{\ensuremath{50.1_{-3.0}^{+3.7}}\xspace}
\def\ChirpMassSourceCIPhXPHMfourth{\ensuremath{21.2_{-1.4}^{+1.7}}\xspace}
\def\MassRatioCIPhXPHMfourth{\ensuremath{0.71_{-0.22}^{+0.22}}\xspace}
\def\ChiEffCIPhXPHMfourth{\ensuremath{-0.02_{-0.15}^{+0.13}}\xspace}
\def\ChiPCIPhXPHMfourth{\ensuremath{0.43_{-0.28}^{+0.33}}\xspace}
\def\DLCIPhXPHMfourth{\ensuremath{1120_{-390}^{+340}}\xspace}

\def\ThetaJNCIPhXPHMfourth{\ensuremath{0.89_{-0.57}^{+1.87}}\xspace}
\def\FinalSpinCIPhXPHMfourth{\ensuremath{0.81_{-0.09}^{+0.11}}\xspace}
\def\FinalMassSourceCIPhXPHMfourth{\ensuremath{47.1_{-2.9}^{+3.5}}\xspace}

\def\MOneSourceCIPhXPfourth{\ensuremath{30.4_{-4.4}^{+5.5}}\xspace}
\def\MTwoSourceCIPhXPfourth{\ensuremath{20.3_{-3.9}^{+4.0}}\xspace}
\def\MtotalSourceCIPhXPfourth{\ensuremath{51.0_{-3.2}^{+3.9}}\xspace}
\def\ChirpMassSourceCIPhXPfourth{\ensuremath{21.4_{-1.4}^{+1.6}}\xspace}
\def\MassRatioCIPhXPfourth{\ensuremath{0.67_{-0.20}^{+0.25}}\xspace}
\def\ChiEffCIPhXPfourth{\ensuremath{-0.02_{-0.15}^{+0.13}}\xspace}
\def\ChiPCIPhXPfourth{\ensuremath{0.43_{-0.27}^{+0.33}}\xspace}
\def\DLCIPhXPfourth{\ensuremath{1040_{-370}^{+350}}\xspace}

\def\ThetaJNCIPhXPfourth{\ensuremath{1.19_{-0.78}^{+1.51}}\xspace}
\def\FinalSpinCIPhXPfourth{\ensuremath{0.81_{-0.10}^{+0.11}}\xspace}
\def\FinalMassSourceCIPhXPfourth{\ensuremath{47.9_{-3.2}^{+3.8}}\xspace}

\def\MOneSourceCIPhPvtwofourth{\ensuremath{31.2_{-4.8}^{+5.7}}\xspace}
\def\MTwoSourceCIPhPvtwofourth{\ensuremath{19.8_{-3.7}^{+4.1}}\xspace}
\def\MtotalSourceCIPhPvtwofourth{\ensuremath{51.1_{-3.3}^{+4.1}}\xspace}
\def\ChirpMassSourceCIPhPvtwofourth{\ensuremath{21.4_{-1.4}^{+1.6}}\xspace}
\def\MassRatioCIPhPvtwofourth{\ensuremath{0.63_{-0.19}^{+0.26}}\xspace}
\def\ChiEffCIPhPvtwofourth{\ensuremath{-0.05_{-0.15}^{+0.13}}\xspace}
\def\ChiPCIPhPvtwofourth{\ensuremath{0.38_{-0.23}^{+0.32}}\xspace}
\def\DLCIPhPvtwofourth{\ensuremath{990_{-350}^{+350}}\xspace}

\def\ThetaJNCIPhPvtwofourth{\ensuremath{1.10_{-0.70}^{+1.59}}\xspace}
\def\FinalSpinCIPhPvtwofourth{\ensuremath{0.79_{-0.09}^{+0.12}}\xspace}
\def\FinalMassSourceCIPhPvtwofourth{\ensuremath{48.2_{-3.3}^{+4.1}}\xspace}

\def\MOneSourceCIPhXPHMfifth{\ensuremath{10.8_{-1.4}^{+3.2}}\xspace}
\def\MTwoSourceCIPhXPHMfifth{\ensuremath{7.8_{-1.6}^{+1.1}}\xspace}
\def\MtotalSourceCIPhXPHMfifth{\ensuremath{18.6_{-0.5}^{+1.5}}\xspace}
\def\ChirpMassSourceCIPhXPHMfifth{\ensuremath{7.9_{-0.1}^{+0.2}}\xspace}
\def\MassRatioCIPhXPHMfifth{\ensuremath{0.72_{-0.28}^{+0.22}}\xspace}
\def\ChiEffCIPhXPHMfifth{\ensuremath{0.05_{-0.04}^{+0.11}}\xspace}
\def\ChiPCIPhXPHMfifth{\ensuremath{0.35_{-0.23}^{+0.36}}\xspace}
\def\DLCIPhXPHMfifth{\ensuremath{350_{-100}^{+100}}\xspace}

\def\ThetaJNCIPhXPHMfifth{\ensuremath{2.41_{-1.99}^{+0.46}}\xspace}
\def\FinalSpinCIPhXPHMfifth{\ensuremath{0.78_{-0.07}^{+0.11}}\xspace}
\def\FinalMassSourceCIPhXPHMfifth{\ensuremath{17.5_{-0.6}^{+1.6}}\xspace}

\def\MOneSourceCIPhXPfifth{\ensuremath{10.8_{-1.4}^{+3.4}}\xspace}
\def\MTwoSourceCIPhXPfifth{\ensuremath{7.7_{-1.6}^{+1.1}}\xspace}
\def\MtotalSourceCIPhXPfifth{\ensuremath{18.6_{-0.5}^{+1.7}}\xspace}
\def\ChirpMassSourceCIPhXPfifth{\ensuremath{7.9_{-0.1}^{+0.1}}\xspace}
\def\MassRatioCIPhXPfifth{\ensuremath{0.72_{-0.29}^{+0.22}}\xspace}
\def\ChiEffCIPhXPfifth{\ensuremath{0.04_{-0.04}^{+0.11}}\xspace}
\def\ChiPCIPhXPfifth{\ensuremath{0.34_{-0.22}^{+0.37}}\xspace}
\def\DLCIPhXPfifth{\ensuremath{330_{-100}^{+100}}\xspace}

\def\ThetaJNCIPhXPfifth{\ensuremath{2.43_{-1.96}^{+0.44}}\xspace}
\def\FinalSpinCIPhXPfifth{\ensuremath{0.77_{-0.07}^{+0.12}}\xspace}
\def\FinalMassSourceCIPhXPfifth{\ensuremath{17.5_{-0.6}^{+1.8}}\xspace}

\def\MOneSourceCIPhPvtwofifth{\ensuremath{11.0_{-1.6}^{+3.9}}\xspace}
\def\MTwoSourceCIPhPvtwofifth{\ensuremath{7.6_{-1.7}^{+1.2}}\xspace}
\def\MtotalSourceCIPhPvtwofifth{\ensuremath{18.7_{-0.6}^{+2.1}}\xspace}
\def\ChirpMassSourceCIPhPvtwofifth{\ensuremath{7.9_{-0.1}^{+0.1}}\xspace}
\def\MassRatioCIPhPvtwofifth{\ensuremath{0.68_{-0.29}^{+0.25}}\xspace}
\def\ChiEffCIPhPvtwofifth{\ensuremath{0.04_{-0.05}^{+0.14}}\xspace}
\def\ChiPCIPhPvtwofifth{\ensuremath{0.34_{-0.22}^{+0.33}}\xspace}
\def\DLCIPhPvtwofifth{\ensuremath{330_{-100}^{+100}}\xspace}

\def\ThetaJNCIPhPvtwofifth{\ensuremath{2.42_{-1.95}^{+0.45}}\xspace}
\def\FinalSpinCIPhPvtwofifth{\ensuremath{0.78_{-0.08}^{+0.11}}\xspace}
\def\FinalMassSourceCIPhPvtwofifth{\ensuremath{17.6_{-0.7}^{+2.3}}\xspace}

\def\MOneSourceCIPhTPsixth{\ensuremath{53.2_{-10.4}^{+14.9}}\xspace}
\def\MTwoSourceCIPhTPsixth{\ensuremath{31.8_{-7.1}^{+7.8}}\xspace}
\def\MtotalSourceCIPhTPsixth{\ensuremath{85.8_{-9.3}^{+12.1}}\xspace}
\def\ChirpMassSourceCIPhTPsixth{\ensuremath{35.4_{-3.4}^{+4.2}}\xspace}
\def\MassRatioCIPhTPsixth{\ensuremath{0.6_{-0.22}^{+0.29}}\xspace}
\def\ChiEffCIPhTPsixth{\ensuremath{0.36_{-0.16}^{+0.13}}\xspace}
\def\ChiPCIPhTPsixth{\ensuremath{0.50_{-0.25}^{+0.25}}\xspace}
\def\DLCIPhTPsixth{\ensuremath{2870_{-990}^{+1020}}\xspace}

\def\ThetaJNCIPhTPsixth{\ensuremath{0.91_{-0.60}^{+1.83}}\xspace}
\def\FinalSpinCIPhTPsixth{\ensuremath{0.89_{-0.10}^{+0.09}}\xspace}
\def\FinalMassSourceCIPhTPsixth{\ensuremath{79.9_{-8.5}^{+11.3}}\xspace}

\def\MOneSourceCIPhTPHMsixth{\ensuremath{57.3_{-10.9}^{+12.0}}\xspace}
\def\MTwoSourceCIPhTPHMsixth{\ensuremath{29.4_{-7.2}^{+8.8}}\xspace}
\def\MtotalSourceCIPhTPHMsixth{\ensuremath{87.1_{-9.3}^{+11.0}}\xspace}
\def\ChirpMassSourceCIPhTPHMsixth{\ensuremath{35.1_{-3.9}^{+4.5}}\xspace}
\def\MassRatioCIPhTPHMsixth{\ensuremath{0.52_{-0.18}^{+0.27}}\xspace}
\def\ChiEffCIPhTPHMsixth{\ensuremath{0.33_{-0.24}^{+0.18}}\xspace}
\def\ChiPCIPhTPHMsixth{\ensuremath{0.43_{-0.25}^{+0.29}}\xspace}
\def\DLCIPhTPHMsixth{\ensuremath{2490_{-880}^{+1160}}\xspace}

\def\ThetaJNCIPhTPHMsixth{\ensuremath{1.08_{-0.71}^{+1.62}}\xspace}
\def\FinalSpinCIPhTPHMsixth{\ensuremath{0.87_{-0.15}^{+0.10}}\xspace}
\def\FinalMassSourceCIPhTPHMsixth{\ensuremath{81.7_{-8.5}^{+9.9}}\xspace}

\def\MOneSourceCIPhXPHMsixth{\ensuremath{52.5_{-9.5}^{+9.2}}\xspace}
\def\MTwoSourceCIPhXPHMsixth{\ensuremath{30.5_{-7.7}^{+8.4}}\xspace}
\def\MtotalSourceCIPhXPHMsixth{\ensuremath{82.6_{-8.0}^{+9.7}}\xspace}
\def\ChirpMassSourceCIPhXPHMsixth{\ensuremath{34.1_{-4.0}^{+4.6}}\xspace}
\def\MassRatioCIPhXPHMsixth{\ensuremath{0.58_{-0.19}^{+0.29}}\xspace}
\def\ChiEffCIPhXPHMsixth{\ensuremath{0.30_{-0.25}^{+0.17}}\xspace}
\def\ChiPCIPhXPHMsixth{\ensuremath{0.42_{-0.26}^{+0.33}}\xspace}
\def\DLCIPhXPHMsixth{\ensuremath{2750_{-1120}^{+1230}}\xspace}

\def\ThetaJNCIPhXPHMsixth{\ensuremath{1.05_{-0.69}^{+1.68}}\xspace}
\def\FinalSpinCIPhXPHMsixth{\ensuremath{0.86_{-0.15}^{+0.09}}\xspace}
\def\FinalMassSourceCIPhXPHMsixth{\ensuremath{77.5_{-7.5}^{+8.8}}\xspace}

\def\MOneSourceCIPhXPsixth{\ensuremath{51.7_{-9.4}^{+11.4}}\xspace}
\def\MTwoSourceCIPhXPsixth{\ensuremath{31.8_{-7.4}^{+8.1}}\xspace}
\def\MtotalSourceCIPhXPsixth{\ensuremath{83.7_{-8.4}^{+10.5}}\xspace}
\def\ChirpMassSourceCIPhXPsixth{\ensuremath{34.7_{-3.6}^{+4.6}}\xspace}
\def\MassRatioCIPhXPsixth{\ensuremath{0.62_{-0.21}^{+0.28}}\xspace}
\def\ChiEffCIPhXPsixth{\ensuremath{0.35_{-0.19}^{+0.15}}\xspace}
\def\ChiPCIPhXPsixth{\ensuremath{0.50_{-0.27}^{+0.28}}\xspace}
\def\DLCIPhXPsixth{\ensuremath{2890_{-1070}^{+1060}}\xspace}

\def\ThetaJNCIPhXPsixth{\ensuremath{0.94_{-0.65}^{+1.82}}\xspace}
\def\FinalSpinCIPhXPsixth{\ensuremath{0.89_{-0.10}^{+0.08}}\xspace}
\def\FinalMassSourceCIPhXPsixth{\ensuremath{77.9_{-7.7}^{+9.8}}\xspace}

\def\MOneSourceCIPhPvtwosixth{\ensuremath{51.2_{-9.1}^{+12.3}}\xspace}
\def\MTwoSourceCIPhPvtwosixth{\ensuremath{32.4_{-7.7}^{+7.7}}\xspace}
\def\MtotalSourceCIPhPvtwosixth{\ensuremath{84.0_{-8.7}^{+11.1}}\xspace}
\def\ChirpMassSourceCIPhPvtwosixth{\ensuremath{35.0_{-3.7}^{+4.6}}\xspace}
\def\MassRatioCIPhPvtwosixth{\ensuremath{0.63_{-0.22}^{+0.28}}\xspace}
\def\ChiEffCIPhPvtwosixth{\ensuremath{0.36_{-0.21}^{+0.15}}\xspace}
\def\ChiPCIPhPvtwosixth{\ensuremath{0.44_{-0.24}^{+0.28}}\xspace}
\def\DLCIPhPvtwosixth{\ensuremath{2840_{-1070}^{+1060}}\xspace}

\def\ThetaJNCIPhPvtwosixth{\ensuremath{0.93_{-0.64}^{+1.82}}\xspace}
\def\FinalSpinCIPhPvtwosixth{\ensuremath{0.87_{-0.11}^{+0.08}}\xspace}
\def\FinalMassSourceCIPhPvtwosixth{\ensuremath{78.3_{-7.9}^{+10.3}}\xspace}

\def\MOneSourceCIPhTPHMseventh{\ensuremath{32.5_{-3.5}^{+5.7}}\xspace}
\def\MTwoSourceCIPhTPHMseventh{\ensuremath{25.3_{-4.2}^{+3.1}}\xspace}
\def\MtotalSourceCIPhTPHMseventh{\ensuremath{58.0_{-2.9}^{+3.7}}\xspace}
\def\ChirpMassSourceCIPhTPHMseventh{\ensuremath{24.8_{-1.2}^{+1.5}}\xspace}
\def\MassRatioCIPhTPHMseventh{\ensuremath{0.78_{-0.22}^{+0.17}}\xspace}
\def\ChiEffCIPhTPHMseventh{\ensuremath{0.09_{-0.13}^{+0.14}}\xspace}
\def\ChiPCIPhTPHMseventh{\ensuremath{0.37_{-0.24}^{+0.32}}\xspace}
\def\DLCIPhTPHMseventh{\ensuremath{1090_{-290}^{+240}}\xspace}

\def\ThetaJNCIPhTPHMseventh{\ensuremath{2.64_{-0.43}^{+0.31}}\xspace}
\def\FinalSpinCIPhTPHMseventh{\ensuremath{0.79_{-0.08}^{+0.10}}\xspace}
\def\FinalMassSourceCIPhTPHMseventh{\ensuremath{54.5_{-2.8}^{+3.5}}\xspace}

\def\MOneSourceCIPhXPHMseventh{\ensuremath{33.8_{-4.4}^{+6.5}}\xspace}
\def\MTwoSourceCIPhXPHMseventh{\ensuremath{24.1_{-4.4}^{+3.9}}\xspace}
\def\MtotalSourceCIPhXPHMseventh{\ensuremath{58.1_{-3.0}^{+3.9}}\xspace}
\def\ChirpMassSourceCIPhXPHMseventh{\ensuremath{24.6_{-1.2}^{+1.5}}\xspace}
\def\MassRatioCIPhXPHMseventh{\ensuremath{0.72_{-0.22}^{+0.22}}\xspace}
\def\ChiEffCIPhXPHMseventh{\ensuremath{0.08_{-0.13}^{+0.14}}\xspace}
\def\ChiPCIPhXPHMseventh{\ensuremath{0.46_{-0.30}^{+0.35}}\xspace}
\def\DLCIPhXPHMseventh{\ensuremath{1130_{-270}^{+240}}\xspace}

\def\ThetaJNCIPhXPHMseventh{\ensuremath{2.64_{-0.41}^{+0.31}}\xspace}
\def\FinalSpinCIPhXPHMseventh{\ensuremath{0.82_{-0.10}^{+0.12}}\xspace}
\def\FinalMassSourceCIPhXPHMseventh{\ensuremath{54.5_{-2.9}^{+3.7}}\xspace}

\def\MOneSourceCIPhXPseventh{\ensuremath{34.8_{-5.0}^{+6.8}}\xspace}
\def\MTwoSourceCIPhXPseventh{\ensuremath{23.9_{-4.3}^{+4.1}}\xspace}
\def\MtotalSourceCIPhXPseventh{\ensuremath{58.9_{-3.3}^{+4.3}}\xspace}
\def\ChirpMassSourceCIPhXPseventh{\ensuremath{24.9_{-1.3}^{+1.7}}\xspace}
\def\MassRatioCIPhXPseventh{\ensuremath{0.69_{-0.21}^{+0.24}}\xspace}
\def\ChiEffCIPhXPseventh{\ensuremath{0.08_{-0.13}^{+0.13}}\xspace}
\def\ChiPCIPhXPseventh{\ensuremath{0.44_{-0.29}^{+0.36}}\xspace}
\def\DLCIPhXPseventh{\ensuremath{1050_{-320}^{+250}}\xspace}

\def\ThetaJNCIPhXPseventh{\ensuremath{2.53_{-0.47}^{+0.37}}\xspace}
\def\FinalSpinCIPhXPseventh{\ensuremath{0.81_{-0.10}^{+0.12}}\xspace}
\def\FinalMassSourceCIPhXPseventh{\ensuremath{55.3_{-3.2}^{+4.2}}\xspace}

\def\MOneSourceCIPhPvtwoseventh{\ensuremath{34.9_{-5.1}^{+6.9}}\xspace}
\def\MTwoSourceCIPhPvtwoseventh{\ensuremath{23.7_{-4.1}^{+4.2}}\xspace}
\def\MtotalSourceCIPhPvtwoseventh{\ensuremath{58.8_{-3.2}^{+4.4}}\xspace}
\def\ChirpMassSourceCIPhPvtwoseventh{\ensuremath{24.8_{-1.3}^{+1.6}}\xspace}
\def\MassRatioCIPhPvtwoseventh{\ensuremath{0.68_{-0.20}^{+0.24}}\xspace}
\def\ChiEffCIPhPvtwoseventh{\ensuremath{0.06_{-0.12}^{+0.14}}\xspace}
\def\ChiPCIPhPvtwoseventh{\ensuremath{0.38_{-0.25}^{+0.35}}\xspace}
\def\DLCIPhPvtwoseventh{\ensuremath{1030_{-300}^{+240}}\xspace}

\def\ThetaJNCIPhPvtwoseventh{\ensuremath{2.57_{-0.48}^{+0.34}}\xspace}
\def\FinalSpinCIPhPvtwoseventh{\ensuremath{0.80_{-0.09}^{+0.12}}\xspace}
\def\FinalMassSourceCIPhPvtwoseventh{\ensuremath{55.3_{-3.2}^{+4.3}}\xspace}

\def\MOneSourceCIPhTPHMeighth{\ensuremath{31.8_{-3.2}^{+5.1}}\xspace}
\def\MTwoSourceCIPhTPHMeighth{\ensuremath{24.6_{-3.4}^{+2.8}}\xspace}
\def\MtotalSourceCIPhTPHMeighth{\ensuremath{56.7_{-2.4}^{+2.8}}\xspace}
\def\ChirpMassSourceCIPhTPHMeighth{\ensuremath{24.3_{-0.9}^{+1.0}}\xspace}
\def\MassRatioCIPhTPHMeighth{\ensuremath{0.77_{-0.19}^{+0.18}}\xspace}
\def\ChiEffCIPhTPHMeighth{\ensuremath{0.12_{-0.10}^{+0.10}}\xspace}
\def\ChiPCIPhTPHMeighth{\ensuremath{0.58_{-0.35}^{+0.28}}\xspace}
\def\DLCIPhTPHMeighth{\ensuremath{630_{-140}^{+100}}\xspace}

\def\ThetaJNCIPhTPHMeighth{\ensuremath{0.60_{-0.32}^{+1.80}}\xspace}
\def\FinalSpinCIPhTPHMeighth{\ensuremath{0.86_{-0.11}^{+0.09}}\xspace}
\def\FinalMassSourceCIPhTPHMeighth{\ensuremath{52.8_{-2.1}^{+2.5}}\xspace}

\def\MOneSourceCIPhXPHMeighth{\ensuremath{30.6_{-2.7}^{+4.0}}\xspace}
\def\MTwoSourceCIPhXPHMeighth{\ensuremath{24.8_{-3.2}^{+2.4}}\xspace}
\def\MtotalSourceCIPhXPHMeighth{\ensuremath{55.5_{-2.2}^{+2.5}}\xspace}
\def\ChirpMassSourceCIPhXPHMeighth{\ensuremath{23.9_{-0.9}^{+1.1}}\xspace}
\def\MassRatioCIPhXPHMeighth{\ensuremath{0.82_{-0.18}^{+0.14}}\xspace}
\def\ChiEffCIPhXPHMeighth{\ensuremath{0.06_{-0.09}^{+0.10}}\xspace}
\def\ChiPCIPhXPHMeighth{\ensuremath{0.38_{-0.26}^{+0.35}}\xspace}
\def\DLCIPhXPHMeighth{\ensuremath{610_{-190}^{+140}}\xspace}

\def\ThetaJNCIPhXPHMeighth{\ensuremath{0.69_{-0.40}^{+1.77}}\xspace}
\def\FinalSpinCIPhXPHMeighth{\ensuremath{0.79_{-0.08}^{+0.10}}\xspace}
\def\FinalMassSourceCIPhXPHMeighth{\ensuremath{52.1_{-2.1}^{+2.4}}\xspace}

\def\MOneSourceCIPhXPHMeighthNNLO{\ensuremath{30.5_{-2.5}^{+3.8}}\xspace}
\def\MTwoSourceCIPhXPHMeighthNNLO{\ensuremath{25.3_{-3.1}^{+2.3}}\xspace}
\def\MtotalSourceCIPhXPHMeighthNNLO{\ensuremath{55.8_{-2.2}^{+2.6}}\xspace}
\def\ChirpMassSourceCIPhXPHMeighthNNLO{\ensuremath{24.0_{-0.9}^{+1.1}}\xspace}
\def\MassRatioCIPhXPHMeighthNNLO{\ensuremath{0.83_{-0.18}^{+0.13}}\xspace}
\def\ChiEffCIPhXPHMeighthNNLO{\ensuremath{0.08_{-0.09}^{+0.10}}\xspace}
\def\ChiPCIPhXPHMeighthNNLO{\ensuremath{0.44_{-0.3}^{+0.33}}\xspace}
\def\DLCIPhXPHMeighthNNLO{\ensuremath{610_{-170}^{+130}}\xspace}

\def\ThetaJNCIPhXPHMeighthNNLO{\ensuremath{0.67_{-0.36}^{+1.68}}\xspace}
\def\FinalSpinCIPhXPHMeighthNNLO{\ensuremath{0.81_{-0.09}^{+0.10}}\xspace}
\def\FinalMassSourceCIPhXPHMeighthNNLO{\ensuremath{52.3_{-2.1}^{+2.4}}\xspace}

\def\MOneSourceCIPhXPeighth{\ensuremath{30.3_{-2.4}^{+3.9}}\xspace}
\def\MTwoSourceCIPhXPeighth{\ensuremath{25.2_{-3.1}^{+2.2}}\xspace}
\def\MtotalSourceCIPhXPeighth{\ensuremath{55.5_{-2.1}^{+2.6}}\xspace}
\def\ChirpMassSourceCIPhXPeighth{\ensuremath{23.9_{-0.9}^{+1.1}}\xspace}
\def\MassRatioCIPhXPeighth{\ensuremath{0.83_{-0.18}^{+0.13}}\xspace}
\def\ChiEffCIPhXPeighth{\ensuremath{0.06_{-0.09}^{+0.10}}\xspace}
\def\ChiPCIPhXPeighth{\ensuremath{0.40_{-0.27}^{+0.35}}\xspace}
\def\DLCIPhXPeighth{\ensuremath{610_{-200}^{+130}}\xspace}

\def\ThetaJNCIPhXPeighth{\ensuremath{0.68_{-0.42}^{+1.75}}\xspace}
\def\FinalSpinCIPhXPeighth{\ensuremath{0.80_{-0.08}^{+0.10}}\xspace}
\def\FinalMassSourceCIPhXPeighth{\ensuremath{52.1_{-2.1}^{+2.5}}\xspace}

\def\MOneSourceCIPhPvtwoeighth{\ensuremath{30.5_{-2.4}^{+4.1}}\xspace}
\def\MTwoSourceCIPhPvtwoeighth{\ensuremath{25.4_{-3.1}^{+2.3}}\xspace}
\def\MtotalSourceCIPhPvtwoeighth{\ensuremath{56.0_{-2.2}^{+2.7}}\xspace}
\def\ChirpMassSourceCIPhPvtwoeighth{\ensuremath{24.2_{-0.9}^{+1.1}}\xspace}
\def\MassRatioCIPhPvtwoeighth{\ensuremath{0.84_{-0.19}^{+0.13}}\xspace}
\def\ChiEffCIPhPvtwoeighth{\ensuremath{0.07_{-0.09}^{+0.10}}\xspace}
\def\ChiPCIPhPvtwoeighth{\ensuremath{0.57_{-0.36}^{+0.30}}\xspace}
\def\DLCIPhPvtwoeighth{\ensuremath{590_{-160}^{+120}}\xspace}

\def\ThetaJNCIPhPvtwoeighth{\ensuremath{0.69_{-0.39}^{+1.70}}\xspace}
\def\FinalSpinCIPhPvtwoeighth{\ensuremath{0.85_{-0.11}^{+0.09}}\xspace}
\def\FinalMassSourceCIPhPvtwoeighth{\ensuremath{52.3_{-2.2}^{+2.5}}\xspace}

\def\MOneSourceCIPhTPHMnineth{\ensuremath{35.1_{-3.8}^{+5.8}}\xspace}
\def\MTwoSourceCIPhTPHMnineth{\ensuremath{26.9_{-4.0}^{+3.4}}\xspace}
\def\MtotalSourceCIPhTPHMnineth{\ensuremath{62.3_{-3.1}^{+3.6}}\xspace}
\def\ChirpMassSourceCIPhTPHMnineth{\ensuremath{26.6_{-1.3}^{+1.5}}\xspace}
\def\MassRatioCIPhTPHMnineth{\ensuremath{0.77_{-0.20}^{+0.18}}\xspace}
\def\ChiEffCIPhTPHMnineth{\ensuremath{-0.06_{-0.15}^{+0.14}}\xspace}
\def\ChiPCIPhTPHMnineth{\ensuremath{0.52_{-0.31}^{+0.30}}\xspace}
\def\DLCIPhTPHMnineth{\ensuremath{1090_{-320}^{+340}}\xspace}

\def\ThetaJNCIPhTPHMnineth{\ensuremath{2.47_{-0.36}^{+0.36}}\xspace}
\def\FinalSpinCIPhTPHMnineth{\ensuremath{0.84_{-0.10}^{+0.10}}\xspace}
\def\FinalMassSourceCIPhTPHMnineth{\ensuremath{58.3_{-3.1}^{+3.5}}\xspace}

\def\MOneSourceCIPhXPHMnineth{\ensuremath{34.8_{-3.7}^{+5.3}}\xspace}
\def\MTwoSourceCIPhXPHMnineth{\ensuremath{26.9_{-4.0}^{+3.3}}\xspace}
\def\MtotalSourceCIPhXPHMnineth{\ensuremath{61.8_{-3.2}^{+3.7}}\xspace}
\def\ChirpMassSourceCIPhXPHMnineth{\ensuremath{26.5_{-1.3}^{+1.6}}\xspace}
\def\MassRatioCIPhXPHMnineth{\ensuremath{0.78_{-0.19}^{+0.17}}\xspace}
\def\ChiEffCIPhXPHMnineth{\ensuremath{-0.05_{-0.16}^{+0.13}}\xspace}
\def\ChiPCIPhXPHMnineth{\ensuremath{0.55_{-0.34}^{+0.31}}\xspace}
\def\DLCIPhXPHMnineth{\ensuremath{1190_{-340}^{+320}}\xspace}

\def\ThetaJNCIPhXPHMnineth{\ensuremath{2.51_{-0.37}^{+0.35}}\xspace}
\def\FinalSpinCIPhXPHMnineth{\ensuremath{0.85_{-0.11}^{+0.10}}\xspace}
\def\FinalMassSourceCIPhXPHMnineth{\ensuremath{57.7_{-3.1}^{+3.6}}\xspace}

\def\MOneSourceCIPhXPHMninethNNLO{\ensuremath{35.1_{-3.9}^{+5.4}}\xspace}
\def\MTwoSourceCIPhXPHMninethNNLO{\ensuremath{26.4_{-4.1}^{+3.6}}\xspace}
\def\MtotalSourceCIPhXPHMninethNNLO{\ensuremath{61.6_{-3.1}^{+3.6}}\xspace}
\def\ChirpMassSourceCIPhXPHMninethNNLO{\ensuremath{26.3_{-1.3}^{+1.6}}\xspace}
\def\MassRatioCIPhXPHMninethNNLO{\ensuremath{0.76_{-0.19}^{+0.19}}\xspace}
\def\ChiEffCIPhXPHMninethNNLO{\ensuremath{-0.09_{-0.17}^{+0.15}}\xspace}
\def\ChiPCIPhXPHMninethNNLO{\ensuremath{0.57_{-0.32}^{+0.28}}\xspace}
\def\DLCIPhXPHMninethNNLO{\ensuremath{1130_{-310}^{+330}}\xspace}

\def\ThetaJNCIPhXPHMninethNNLO{\ensuremath{2.48_{-0.34}^{+0.35}}\xspace}
\def\FinalSpinCIPhXPHMninethNNLO{\ensuremath{0.85_{-0.10}^{+0.09}}\xspace}
\def\FinalMassSourceCIPhXPHMninethNNLO{\ensuremath{57.5_{-3.2}^{+3.6}}\xspace}

\def\MOneSourceCIPhXPnineth{\ensuremath{34.9_{-3.6}^{+5.8}}\xspace}
\def\MTwoSourceCIPhXPnineth{\ensuremath{27.2_{-4.2}^{+3.4}}\xspace}
\def\MtotalSourceCIPhXPnineth{\ensuremath{62.3_{-3.3}^{+3.9}}\xspace}
\def\ChirpMassSourceCIPhXPnineth{\ensuremath{26.7_{-1.4}^{+1.6}}\xspace}
\def\MassRatioCIPhXPnineth{\ensuremath{0.78_{-0.21}^{+0.17}}\xspace}
\def\ChiEffCIPhXPnineth{\ensuremath{-0.05_{-0.16}^{+0.14}}\xspace}
\def\ChiPCIPhXPnineth{\ensuremath{0.59_{-0.36}^{+0.28}}\xspace}
\def\DLCIPhXPnineth{\ensuremath{1130_{-330}^{+320}}\xspace}

\def\ThetaJNCIPhXPnineth{\ensuremath{2.46_{-0.37}^{+0.37}}\xspace}
\def\FinalSpinCIPhXPnineth{\ensuremath{0.86_{-0.11}^{+0.10}}\xspace}
\def\FinalMassSourceCIPhXPnineth{\ensuremath{58.1_{-3.3}^{+3.8}}\xspace}

\def\MOneSourceCIPhPvtwonineth{\ensuremath{35.4_{-3.9}^{+5.5}}\xspace}
\def\MTwoSourceCIPhPvtwonineth{\ensuremath{26.5_{-4.2}^{+3.6}}\xspace}
\def\MtotalSourceCIPhPvtwonineth{\ensuremath{62.0_{-3.1}^{+3.7}}\xspace}
\def\ChirpMassSourceCIPhPvtwonineth{\ensuremath{26.5_{-1.3}^{+1.5}}\xspace}
\def\MassRatioCIPhPvtwonineth{\ensuremath{0.75_{-0.20}^{+0.19}}\xspace}
\def\ChiEffCIPhPvtwonineth{\ensuremath{-0.11_{-0.16}^{+0.14}}\xspace}
\def\ChiPCIPhPvtwonineth{\ensuremath{0.56_{-0.32}^{+0.27}}\xspace}
\def\DLCIPhPvtwonineth{\ensuremath{1050_{-280}^{+310}}\xspace}

\def\ThetaJNCIPhPvtwonineth{\ensuremath{2.44_{-0.35}^{+0.35}}\xspace}
\def\FinalSpinCIPhPvtwonineth{\ensuremath{0.85_{-0.11}^{+0.09}}\xspace}
\def\FinalMassSourceCIPhPvtwonineth{\ensuremath{57.8_{-3.1}^{+3.8}}\xspace}

\def\MOneSourceCIPhTPHMtenth{\ensuremath{37.8_{-4.7}^{+7.2}}\xspace}
\def\MTwoSourceCIPhTPHMtenth{\ensuremath{29.2_{-5.2}^{+4.4}}\xspace}
\def\MtotalSourceCIPhTPHMtenth{\ensuremath{67.0_{-5.3}^{+7.0}}\xspace}
\def\ChirpMassSourceCIPhTPHMtenth{\ensuremath{28.6_{-2.3}^{+3.0}}\xspace}
\def\MassRatioCIPhTPHMtenth{\ensuremath{0.79_{-0.23}^{+0.17}}\xspace}
\def\ChiEffCIPhTPHMtenth{\ensuremath{0.07_{-0.16}^{+0.17}}\xspace}
\def\ChiPCIPhTPHMtenth{\ensuremath{0.43_{-0.27}^{+0.34}}\xspace}
\def\DLCIPhTPHMtenth{\ensuremath{2050_{-700}^{+640}}\xspace}

\def\ThetaJNCIPhTPHMtenth{\ensuremath{1.80_{-1.49}^{+1.03}}\xspace}
\def\FinalSpinCIPhTPHMtenth{\ensuremath{0.81_{-0.09}^{+0.10}}\xspace}
\def\FinalMassSourceCIPhTPHMtenth{\ensuremath{62.8_{-5.0}^{+6.5}}\xspace}

\def\MOneSourceCIPhXPHMtenth{\ensuremath{37.7_{-4.9}^{+7.2}}\xspace}
\def\MTwoSourceCIPhXPHMtenth{\ensuremath{28.5_{-5.7}^{+4.7}}\xspace}
\def\MtotalSourceCIPhXPHMtenth{\ensuremath{66.1_{-5.2}^{+6.9}}\xspace}
\def\ChirpMassSourceCIPhXPHMtenth{\ensuremath{28.2_{-2.3}^{+3.0}}\xspace}
\def\MassRatioCIPhXPHMtenth{\ensuremath{0.77_{-0.24}^{+0.18}}\xspace}
\def\ChiEffCIPhXPHMtenth{\ensuremath{0.07_{-0.17}^{+0.16}}\xspace}
\def\ChiPCIPhXPHMtenth{\ensuremath{0.52_{-0.32}^{+0.32}}\xspace}
\def\DLCIPhXPHMtenth{\ensuremath{2170_{-720}^{+650}}\xspace}

\def\ThetaJNCIPhXPHMtenth{\ensuremath{1.28_{-0.96}^{+1.53}}\xspace}
\def\FinalSpinCIPhXPHMtenth{\ensuremath{0.84_{-0.10}^{+0.10}}\xspace}
\def\FinalMassSourceCIPhXPHMtenth{\ensuremath{61.8_{-4.9}^{+6.4}}\xspace}

\def\MOneSourceCIPhXPtenth{\ensuremath{39.2_{-5.6}^{+8.2}}\xspace}
\def\MTwoSourceCIPhXPtenth{\ensuremath{28.4_{-6.4}^{+5.1}}\xspace}
\def\MtotalSourceCIPhXPtenth{\ensuremath{67.5_{-5.8}^{+7.4}}\xspace}
\def\ChirpMassSourceCIPhXPtenth{\ensuremath{28.6_{-2.6}^{+3.2}}\xspace}
\def\MassRatioCIPhXPtenth{\ensuremath{0.73_{-0.25}^{+0.21}}\xspace}
\def\ChiEffCIPhXPtenth{\ensuremath{0.06_{-0.17}^{+0.16}}\xspace}
\def\ChiPCIPhXPtenth{\ensuremath{0.52_{-0.33}^{+0.33}}\xspace}
\def\DLCIPhXPtenth{\ensuremath{1930_{-700}^{+670}}\xspace}

\def\ThetaJNCIPhXPtenth{\ensuremath{1.68_{-1.27}^{+1.06}}\xspace}
\def\FinalSpinCIPhXPtenth{\ensuremath{0.84_{-0.11}^{+0.11}}\xspace}
\def\FinalMassSourceCIPhXPtenth{\ensuremath{63.1_{-5.5}^{+6.9}}\xspace}

\def\MOneSourceCIPhPvtwotenth{\ensuremath{39.3_{-5.5}^{+8.2}}\xspace}
\def\MTwoSourceCIPhPvtwotenth{\ensuremath{28.9_{-5.7}^{+5.0}}\xspace}
\def\MtotalSourceCIPhPvtwotenth{\ensuremath{68.2_{-5.7}^{+7.7}}\xspace}
\def\ChirpMassSourceCIPhPvtwotenth{\ensuremath{29.0_{-2.5}^{+3.2}}\xspace}
\def\MassRatioCIPhPvtwotenth{\ensuremath{0.74_{-0.23}^{+0.20}}\xspace}
\def\ChiEffCIPhPvtwotenth{\ensuremath{0.06_{-0.16}^{+0.16}}\xspace}
\def\ChiPCIPhPvtwotenth{\ensuremath{0.47_{-0.30}^{+0.34}}\xspace}
\def\DLCIPhPvtwotenth{\ensuremath{1940_{-710}^{+650}}\xspace}

\def\ThetaJNCIPhPvtwotenth{\ensuremath{1.53_{-1.16}^{+1.24}}\xspace}
\def\FinalSpinCIPhPvtwotenth{\ensuremath{0.83_{-0.10}^{+0.11}}\xspace}
\def\FinalMassSourceCIPhPvtwotenth{\ensuremath{63.9_{-5.4}^{+7.2}}\xspace}

\title[PE with current waveform models for GWTC-1]{Parameter estimation with the current generation of phenomenological waveform models applied to the black hole mergers of GWTC-1}

\author[M. Mateu-Lucena et al.]{
Maite Mateu-Lucena,$^{1}$\thanks{E-mail: mt.mateu@uib.es}
Sascha Husa,$^{1}$
Marta Colleoni,$^{1}$
H\'{e}ctor Estell\'{e}s,$^{1}$
Cecilio Garc{\'i}a-Quir{\'o}s,$^{1,2}$ %FIX ME
\newauthor{
David Keitel,$^{1}$
Maria de Lluc Planas$^{1}$
and Antoni Ramos-Buades$^{1,3}$}
\\
$^{1}$Departament de F\'isica, Universitat de les Illes Balears, IAC3 -- IEEC Crta. Valldemossa km 7.5, E-07122 Palma, Spain\\
$^{2}$Laboratoire Astroparticule et Cosmologie, 10 Rue Alice Domon, E-75013, Paris, France\\
$^{3}$Max Planck Institut für Gravitationsphysik (Albert Einstein Institut), Am M\"uhlenberg 1, Potsdam, Germany\\
}
\pubyear{2022}

\begin{document}
\label{firstpage}
\pagerange{\pageref{firstpage}--\pageref{lastpage}}
\maketitle

% Abstract of the paper
\begin{abstract}
We consider the ten confidently detected gravitational-wave signals in the GWTC-1 catalog which are consistent with mergers of binary black hole systems, and perform a thorough parameter estimation re-analysis. This is made possible by using computationally efficient waveform models of the current (fourth) generation of the \ph family of phenomenological waveform models, which consists of the \phXF frequency-domain modelsand the \phT time-domain models.The analysis is performed with both precessing and non-precessing waveform models with and without subdominant spherical harmonic modes.
Results for all events are validated with convergence tests, discussing in particular the events GW170729 and GW151226.
For the latter and the other two lowest-mass events, we also compare results between two independent sampling codes, Bilby and LALInference.
We find overall consistent results with the original GWTC-1 results, with all Jensen-Shannon divergences between the previous results using \phPvtwo and our default \phXPHM posteriors below 0.045 bits, but we also
discuss cases where including subdominant harmonics and/or precession influences the posteriors.
\end{abstract}

% Select between one and six entries from the list of approved keywords.
% Don't make up new ones.

%%%%%%%%%%%%%%%%%%%%%%%%%%%%%%%%%%%%%%%%%%%%%%%%%%

%%%%%%%%%%%%%%%%% BODY OF PAPER %%%%%%%%%%%%%%%%%%

\section{Introduction}
\label{sec:Introduction}

There has been rapid progress in gravitational-wave (GW) astronomy
over recent years
in developing both waveform models and parameter estimation techniques.
%Despite this progress, there are still open challenges in achieving
Still it remains challenging to achieve
robust inference results on real data,
due to the complexity of fully modelling the physics of binary black hole (BBH) mergers and 
also the current limitations in sampling techniques for Bayesian inference used in this relatively new field.
This makes it interesting to revisit older detections with the latest toolkit at our disposal,
and important to check if robust results can be obtained.
Revisiting the 10 BBH events from the first Gravitational-Wave Transient Catalog \citep[GWTC-1,][]{LIGOScientific:2018mvr}, our purpose with this paper is threefold: 
First, updating the waveform models used for the analysis to the latest generation of phenomenological models and adding subdominant harmonics will allow us to sharpen parameter estimation results. 
Second, the computational efficiency of these models allows us to perform careful studies of convergence, and to validate key results by comparing different sampling methods and priors.
Third, we will gain insight into effects of waveform systematics on the posterior distributions and provide a ``stress-test'' for  the new waveform model families.

A first comprehensive analysis of detections from the first two observing runs of the advanced GW detector network was given in
\citet{LIGOScientific:2018mvr}.
This analysis was based on waveform models that describe only the dominant quadrupole content of the signals and assume quasi-circularity (i.e.~orbital eccentricity is neglected).
Bayesian parameter estimation was performed with the LALInference \citep{Veitch:2014wba} code
and the posterior distributions were constructed from a combination of results with the \phPvtwo and \seobnrvthree waveform models

These models are members of the two principal families of waveform models which have been used for parameter estimation by the LIGO--Virgo(--KAGRA) collaboration (LVC/LVK), and specifically in catalog papers \citep{LIGOScientific:2018mvr,Abbott:2020niy,LIGOScientific:2021usb,LIGOScientific:2021djp}: IMRPhenom and SEOBNR.
For details, references, and a brief summary of the broader context of waveform models see Sec.~\ref{sec:waveforms} and Table~\ref{tab:models}.

The \phPvtwo posteriors have later been reproduced in \citet{Romero-Shaw:2020owr} with the Bilby code \citep{Ashton:2018jfp} for GW parameter estimation, using the nested sampling \citep{10.1214/06-BA127} algorithm {\tt dynesty} \citep{Speagle:2020spe}.
For a study on the effect of waveform systematics on the first detected event, GW150914, see \citet{LIGOScientific:2016ebw}. This study concluded that 
for signals with higher signal-to-noise ratio (SNR) than GW150914, or in other regions of the binary parameter space (lower masses, larger mass ratios, or higher spins), we expect that systematic errors in current waveform models may impact GW measurements, making more accurate models desirable for future observations.

Some limited analysis with models that contain subdominant harmonics has also been performed: Most notably the event GW170729 has been studied with various waveform models \citep{Chatziioannou:2019dsz}, some of which contain subdominant harmonics. For the GWTC-1 catalog \citep{LIGOScientific:2018mvr} all BBH events have been cross-checked with the {\tt RapidPE} algorithm \citep{Pankow:2015cra,Lange:2018pyp} and a waveform bank composed of numerical relativity (NR) simulations supplemented by waveforms from the \NRSur model \citep{Blackman:2017pcm}. The calibration region of \NRSur is restricted to mass ratios between 1 and 5, and dimensionless spins up to $0.8$.
For {\tt RapidPE} results no detailed discussion of results has been provided, but the posterior samples have been released publicly \citep{GWTC1_NR_samples}. %
A similar study has been performed also with the {\tt RIFT} code \citep{Lange:2018pyp,Healy:2020jjs}, which like {\tt RapidPE} is based on a likelihood interpolation method, and a catalog of NR waveforms.
A study of higher modes, without precession, in GWTC-1 based on the approximate method of likelihood reweighting was presented in \citet{Payne:2019wmy}.

Here we present our analysis of the ten confidently detected BBH signals in GWTC-1, which has been the first to be performed based on state-of-the art methods in the following sense:
We use waveform models where at least for the non-precessing sector all spherical harmonics have been calibrated to NR;
the waveform models are sufficiently computationally efficient to use wide priors and perform systematic studies that include convergence tests and comparisons of different samplers;
and we use direct Bayesian sampling evaluating the likelihood with these waveform models, without extra approximations, such as recurring to a discrete template bank or likelihood reweighting without sampling the posterior for each waveform model. 
Analyses of these GWTC-1 events with precessing higher-mode models have also later been performed
by \citet{Nitz:2021uxj} and by the LVC~\citep{LIGOScientific:2021usb},
finding broad agreement with our results,
but using only one of the models (\phXPHM) we use here and will discuss next.
Two events for which the picture is more complex,
GW151226 \citep[see also][]{Chia:2021mxq,Vajpeyi:2022dvs}
and GW170729 \citep[see also][]{Chatziioannou:2019dsz},
will be discussed further below after first summarizing our methodology.

The waveform models we use correspond to the current (fourth) generation of the \ph family of inspiral--merger--ringdown phenomenological waveform models. These models consist of the \phXF sub-family of frequency-domain models \citep{Pratten:2020fqn,Garcia-Quiros:2020qpx,Garcia-Quiros:2020qlt,Pratten:2020ceb}, 
and the \phT time-domain models \citep{Estelles:2020osj, Estelles:2020twz, Estelles:TPHM}.
\phXF constitutes a thorough upgrade of the previous generation of frequency-domain phenomenological waveform models based on \phD \citep{Husa:2015iqa,Khan:2015jqa,Hannam:2013oca,Bohe:PPv2,London:2017bcn,Khan:2018fmp,Khan:2019kot}, which included
the first models in the \ph family to describe subdominant harmonics (\phHM) and both subdominant harmonics and precession (\phPvthreehm). 
\phT offers advantages in the description of precession, in particular concerning the merger and ringdown, as described in detail in \citet{Estelles:TPHM} and briefly summarized in Sec.~\ref{sec:waveforms}.

Our main results are obtained with the parallel Bilby \citep{Ashton:2018jfp,Smith:2019ucc} sampling code.
In addition, for the three lowest-mass events we also cross-check with the independent LALInference code \citep{Veitch:2014wba}, and demonstrate good agreement.
We have added these cross-checks, which confirm our results, after the initial versions of \citet{Nitz:2021uxj} and \citet{Chia:2021mxq} appeared on the arXiv and showed tension for the event GW151226.
In particular, the first version of \citet{Chia:2021mxq} found a bi-modal posterior with support for very unequal masses, in tension with both our results and with \citet{Nitz:2021uxj}.
However, the results in their final journal version (using an updated method and priors) have significantly reduced this tension, with remaining differences consistent with those in priors.
A later study by \citet{Vajpeyi:2022dvs} also studied possible multiple peaks and prior dependency for GW151226.
We will discuss this case in detail in Sec.~\ref{sec:GW151226} and Appendix~\ref{subsec:convergence_test}.
Another special case is GW170729, the highest-mass event of GWTC-1.
We have also performed additional runs and comparisons with the literature for this event and discuss it in detail in Sec.~\ref{sec:GW170729}.

The paper is organized as follows: We first present our methods in Sec.~\ref{sec:preliminaries}: our conventions, the observational data used, our waveform models, and the setup for running parameter estimation. In Sec.~\ref{sec:results} we present our parameter estimation results, and we conclude in Sec.~\ref{sec:conclusions}. Further details are presented in two appendices: In appendix \ref{subsec:convergence_test} we present checks on the sampler convergence, and compare  results for the three lowest mass events between the Bilby and LALInference samplers. Finally in appendix \ref{subsec:multibanding} we test aggressive settings of our multibanding algorithm to accelerate waveform evaluation.

\begin{table*}
 %\label{tab:models}
 %\begin{ruledtabular}
 \begin{tabular}{llccc}
\hline\hline
  \textbf{Family} & \textbf{Full name} & \textbf{Precession} & \textbf{Multipoles $(\ell,\,|m|)$ included} & \textbf{Ref.}\\
\hline
\multirow{4}{*}{SEOBNR} 
  & SEOBNRv2 & $\times$ &  (2, 2) & \citet{Taracchini:2013rva} \\
  & SEOBNRv3 & $\checkmark$ &  (2, 2) & \citet{Babak:2016tgq} \\
  & SEOBNRv4\_ROM & $\times$ &  (2, 2) & \citet{Bohe:2016gbl} \\
  & SEOBNRv4HM\_ROM &  $\times$ &  (2, 2), (2,1), (3, 3), (4, 4), (5,5) & \citet{Cotesta:2018fcv, Cotesta:2020qhw} \\
  & SEOBNRv4P &  \checkmark &  (2, 2), (2, 1) & \citet{Ossokine:2020kjp}\\
  & SEOBNRv4PHM &  \checkmark & (2, 2), (2, 1), (3, 3), (4, 4), (5,5) & \citet{Ossokine:2020kjp}
\\[6pt]
\multirow{4}{*} {IMRPhenom - $3^\mathrm{rd}$ Generation}
 & IMRPhenomD & $\times$ & (2, 2) & \citet{Husa:2015iqa, Khan:2015jqa} \\
 & IMRPhenomHM    &  $\times$ & (2, 2), (2, 1), (3, 3), (3, 2), (4,4), (4, 3) & \citet{London:2017bcn} \\
 & IMRPhenomPv2   &  \checkmark & (2, 2) & \citet{Hannam:2013oca, Bohe:PPv2} \\
 & IMRPhenomPv3   &  \checkmark & (2, 2) & \citet{Khan:2018fmp} \\
 & IMRPhenomPv3HM &  \checkmark & (2, 2), (2, 1), (3, 3), (3, 2),(4, 4), (4, 3) & \citet{Khan:2019kot}
\\[6pt]
\multirow{4}{*}{IMRPhenomX} & IMRPhenomXAS & $\times$ & (2, 2) &
\citet{Pratten:2020fqn}
\\
  & IMRPhenomXHM &  $\times$ & (2, 2), (2, 1), (3, 3), (3, 2), (4,4)  & \citet{Garcia-Quiros:2020qpx,Garcia-Quiros:2020qlt} \\
  & IMRPhenomXP
  &  \checkmark & (2, 2) & \citet{Pratten:2020ceb} \\
  & IMRPhenomXPHM &  \checkmark & (2, 2), (2, 1), (3, 3), (3, 2),(4, 4) & \citet{Pratten:2020ceb}\\[6pt]
\multirow{4}{*}{IMRPhenomT} & IMRPhenomT & $\times$ & (2, 2) &
\citet{Estelles:2020osj,Estelles:2020twz}
\\
  & IMRPhenomTHM &  $\times$ & (2, 2), (2, 1), (3, 3), (4,4), (5,5)  & \citet{Estelles:2020twz} \\
  & IMRPhenomTP
  &  \checkmark & (2, 2) & \citet{Estelles:2020osj,Estelles:TPHM} \\
  & IMRPhenomTPHM &  \checkmark & (2, 2), (2, 1), (3, 3),(4, 4), (5,5) & \citet{Estelles:TPHM}\\
  \hline\hline
 \end{tabular}
 %\end{ruledtabular}
\caption{The table lists waveform models from the IMRPhenom and SEOB families relevant to this paper. Only the latest generation of IMRPhenom waveforms has been used to obtain the results in this paper, however SEOB and older IMRPhenom waveforms have been used in relevant literature we compare to, and we list them for convenience of the reader.} For precessing models, the multipoles correspond to those in a co-precessing frame.
\label{tab:models}
\end{table*}

\section{Methods}
\label{sec:preliminaries}

\subsection{Notation and conventions}
\label{sec:notation}

Masses refer to the source frame unless we note explicitly that we refer to the detector frame.
Source frame masses are
inferred assuming a standard cosmology \citep{Planck2015} (see Appendix B of \citet{LIGOScientific:2018mvr}).
Individual component masses are denoted by $m_i$,
and the total mass is \mbox{$M_T = m_1 + m_2$}.
The chirp mass is \mbox{$\mathcal M = (m_1 \, m_2)^{3/5} M_T^{-1/5}$},
and we define mass ratios as \mbox{$q = m_2 / m_1 \leq 1$} and \mbox{$Q = m_1 / m_2 \geq 1$}.

We also report two effective spin parameters which are commonly used in waveform modelling and parameter estimation.
The parameter $\chi_{\rm{eff}}$ \citep{PhysRevD.82.064016,PhysRevLett.106.241101} captures those dominant spin effects which do not depend on spin precession, and is defined as
 \begin{equation}\label{def:chieff}
    \chi_{\mathrm{eff}}=\frac{m_1 \, \chi_1 + m_2 \, \chi_2}{m_1 + m_2}, 
 \end{equation}
where the $\chi_i$ are the projections of the dimensionless component spin vectors onto the orbital angular momentum.
The effective spin precession parameter $\chi_\mathrm{p}$ \citep{Schmidt:2014iyl} is designed to capture the dominant effect of precession, and corresponds to an approximate average of the spin component in the precessing orbital plane
over many precession cycles, see \citet{Schmidt:2014iyl} and Eq.~(4.7) in \citet{Pratten:2020ceb}.
Both $\chi_{\mathrm{eff}}$ and $\chi_\mathrm{p}$ are dimensionless, and thus independent of the frame (source or detector).

When referring to specific spherical harmonic modes of the GW signal we will always consider pairs of both positive and negative modes, e.g.~when we refer to the example list of multipoles \mbox{$(l,m)=(2,\pm2), (2,\pm1)$} we will use the notation \mbox{$(l,|m|)=(2,2), (2,1)$} or simply $(2,2),(2,1)$.

\subsection{Waveform models used}\label{sec:waveforms}

Three main approaches have been used to construct waveform models for comparable-mass coalescences of quasi-circular BBHs, which have become standard tools of GW data analysis:
\begin{itemize}
    \item The effective one-body (EOB) approach \citep{Buonanno:1998gg}
    models the Hamiltonian and energy flux, which yield a system of ordinary differential equations.
    This is solved numerically, which is however computationally expensive.
    The Hamiltonian and flux, and any further quantities used in the construction of the waveform, can be calibrated to NR simulations.
    Two families of such models have been developed:
    The SEOBNR family has been regularly used for parameter estimation by the LVC/LVK, and a list of models is given in Table \ref{tab:models}.
    An alternative is the TEOBResumS model \citep{Nagar:2018zoe}.
    \item The IMRPhenom models describe the amplitude and phase of the spherical harmonic modes
    piecewise as closed-form expressions in the inspiral, merger and ringdown. These expressions have been calibrated to NR for the late phase of the coalescence, and for the third and current fourth generation to EOB for the early inspiral. These models are computationally much more efficient than EOB models, and can be used for Bayesian parameter estimation without further approximations.
    \item Another approach has been to construct numerical interpolants of discrete data sets; such types of models are often referred to as reduced-order models (ROMs) or surrogate models. This approach has been used to accelerate the evaluation of EOB models --
    e.g. the SEOBNRv4HM\_ROM model~\citep{Cotesta:2018fcv}, see Table \ref{tab:models} --
    or directly to interpolate data sets of NR waveforms \citep{Blackman:2017pcm}. In the latter case, validity is restricted to the parameter space where NR waveforms are available, with small extensions through extrapolations. Since SNRs for current ground-based detectors are relatively small, posteriors are typically wide, and for GW data analysis it is convenient to use EOB or IMRPhenom models, which can be used for larger regions of parameter space.
\end{itemize}
For a list of IMRPhenom and SEOB waveform models, which are most directly relevant for this paper, see Table \ref{tab:models}.

In this work we present new parameter estimation results which we have obtained with the most recent waveform models of the \phXF and \phT families, and compare with previous results from the GWTC-1 catalog \citep{LIGOScientific:2018mvr}, and also with \citet{Chatziioannou:2019dsz} and \citet{Payne:2019wmy} for GW170729.

Parameter estimation for the GWTC-1 catalog paper \citep{LIGOScientific:2018mvr} has used two precessing waveform models: \phPvtwo \citep{Hannam:2013oca,Bohe:PPv2} is a predecessor of our updated precessing phenomenological waveform models \phXPHM \citep{Pratten:2020ceb} and \phTPHM \citep{Estelles:TPHM}), and is based on the non-precessing \phD \citep{Husa:2015iqa,Khan:2015jqa} model.
\seobnrvthree \citep{Babak:2016tgq} is based on the non-precessing \seobnrvtwo \citep{Taracchini:2013rva} model and is a predecessor of \seobnrvforphm \citep{Ossokine:2020kjp},
constructed within the effective-one-body (EOB) framework \citep{Buonanno:1998gg,Buonanno:2000ef}, which is based on solving a system of ordinary differential equations for the dynamics of the binary.

For the GW170729 event an analysis with higher mode models has already been performed \citep{Chatziioannou:2019dsz}, using the non-precessing \phHM~\citep{London:2017bcn} and \seobnrvforhmrom~\citep{Cotesta:2018fcv, Cotesta:2020qhw} models, with which we will also compare. The latter is an example of the reduced-order models (ROMs) which are often used to accelerate the evaluation of EOB waveforms.
Preliminary analyses of GW170729 with our \phX and \phT models have already been presented in \citet{Garcia-Quiros:2020qpx} and \citet{Estelles:2020twz}.

We now turn to describing the new \phXF and \phT families.
Phenomenological frequency-domain models are constructed as piecewise closed-form expressions that are calibrated to post-Newtonian (PN) or EOB inspiral results and NR waveforms. 
Such explicit descriptions of the waveform can be evaluated very rapidly. 
The corner stone of the \phXF family is \phX \citep{Pratten:2020fqn}, which models the $(\ell, \vert m\vert) = (2,2)$ modes
of non-precessing signals, and has been extended to include the 
$(\ell, \vert m\vert) = (2,1), (3,3), (3,2), (4,4)$ modes by \phXHM~\citep{Garcia-Quiros:2020qpx,Garcia-Quiros:2020qlt}. All the modes
have been calibrated to around 500 numerical waveforms as reported in \citet{Garcia-Quiros:2020qlt},
whereas in the previous generation of \ph waveforms
only the dominant $(\ell, \vert m\vert) = (2,2)$ modes have 
been calibrated to 19 NR waveforms \citep{Husa:2015iqa,Khan:2015jqa}.

\phXPHM \citep{Pratten:2020ceb} augments \phXHM to account for spin precession following previous phenomenological models \citep{Schmidt:2010it,Schmidt:2012rh,Hannam:2013oca} in employing an approximate map between
non-precessing and precessing waveforms, where the signal of the precessing system in a non-inertial co-precessing frame
is identified with an appropriate non-precessing signal, and where the final spin is adjusted to account for the effect of spin components orthogonal to the instantaneous axis of the orbital plane. The non-precessing signal with adjusted final spin is then mapped to a precessing signal by employing a time-dependent rotation from the non-inertial co-precessing frame to the inertial frame where the signal is observed.
This procedure is often referred to as ``twisting up'' in the literature.  \phXPHM implements two approximate descriptions for the Euler angles that define the time-dependent rotation: an orbit-averaged effective single-spin description at the next-to-next-to-leading PN order \citep[NNLO,][]{Blanchet:2011zv,Marsat:2013wwa}, or an approximation based on a multiple scale analysis of the PN equations \citep{Chatziioannou:2013dza}, which is the default choice.
For a detailed discussion of the procedure and the conventions employed to describe precessing waveforms in the frequency domain see \citet{Pratten:2020ceb}. 

\phXPHM also allows to choose among different approximations for the final spin of the remnant black hole. Version 0 is equivalent to the estimate implemented in \phPvtwo, where the in-plane spin contribution is captured by the effective precession spin $\chi_\mathrm{p}$.
Version 1 is equivalent to version 0, with the replacement of $\chi_\mathrm{p}$ by the x-component of the spin, $\chi_x$.
Versions 2 and 3 attempt in different ways to incorporate the full double-spin effects, with the latter relying on precession-averaged quantities that arise in the context of the MSA formalism \citep{Chatziioannou:2013dza}.
Version 3 is the default in the LALSuite \citep{lalsuite} implementation.
We refer the reader to Sec. IV D of \citet{Pratten:2020ceb} for a thorough discussion of these options. 
The \phT family of models \citep{Estelles:2020osj,Estelles:2020twz,Estelles:TPHM} extends to the time domain the methods that have been used to construct the frequency-domain model \phXHM and calibrate it to a catalog of numerical waveforms. The main motivation for developing a time domain model is that the morphology of the waveform is often simpler to understand in the time domain, in particular for the late inspiral, merger, and ringdown, where rapid changes are not ``smeared out'' by a Fourier transform. This allows a cleaner separation of inspiral, merger and ringdown, which can be valuable for tests of general relativity, this is however not directly relevant for the present paper. 
Precession is again described by the ``twisting up'' approach \citep{Estelles:2020osj,Estelles:TPHM}, however the time domain model indeed allows some key improvements over \phXPHM:
First, in order to obtain explicit expressions for the spherical harmonic modes of the precessing frequency-domain models, the stationary phase approximation (SPA) is used to compute approximate Fourier transforms. This also makes it difficult to incorporate analytical knowledge about the ringdown frequencies in the ringdown portion of a precessing waveform. This is however simple in the time domain, where the ringdown waveform can be simply modelled as a damped sinusoid, with the complex ringdown frequency determined by the final spin and mass. This allows a simple analytical approximation to the Euler angles during ringdown, based on the effective precessional motion observed in NR signals \citep{PhysRevD.87.044038}, which can be derived analytically in the small opening angle limit. (A derivation is presented in \citet{Estelles:2020osj}, see also \citet{Marsat:2018oam}.)
In consequence, the merger and ringdown of precessing waveforms is typically more accurately described in our time-domain model. 

\phTPHM also offers an improvement of the inspiral description by departing from the previous approach of using closed-form expressions for the Euler angles that describe the precession of the orbital plane, and by default a numerical integration of the equations for the spin dynamics is used, which can be carried out without significant increase of computational cost \citep{Estelles:TPHM}. For a more detailed discussion of the differences between \phXPHM and \phTPHM regarding the treatment of precession see \citet{GW190521}.

For lower-mass events, despite the gain in accuracy of describing precession, there is however some loss of accuracy for the phase of the dominant harmonic, which is constructed based on a calibrated extension of the TaylorT3 PN approximant \citep{Blanchet:2006zz}.
The choice of this approximant was motivated by the explicit time dependence of the orbital frequency as a closed-form expression, but it is known \citep{Buonanno:2009zt} to be less accurate than other numerical PN approximants, and in consequence turns out to be more  difficult to calibrate to NR. Resolving this weakness of the underlying ``carrier phase'' in \phT will be the goal of future work,
and will allow to use this model also for low mass events. As discussed above, the primary motivation for the \phT family was to improve the treatment of the late part of the waveform, which is more relevant for higher-mass systems. We thus perform parameter estimation with the \phT family only for higher-mass events.

For \phXPHM we have implemented the ``multibanding'' method \citep{Vinciguerra:2017ngf,Garcia-Quiros:2020qlt} to accelerate waveform evaluation, based on interpolation and the choice of an accuracy parameter.
In appendix \ref{subsec:multibanding} we discuss parameter estimation results with a more aggressive choice of multibanding settings, as would be appropriate for accelerated parameter estimation runs, e.g. for the initial exploration of newly detected events.

\subsection{Methodology for Parameter estimation}\label{sec:PEsetup}

\subsubsection{Data}\label{sec:data}

We use public GW strain data from the Gravitational Wave Open Science Center \citep[GWOSC, ][]{Vallisneri:2014vxa,GWOSC}, and power spectral densities \citep[PSDs, ][]{gwtc1psd} and calibration uncertainties \citep{gwtc1calib} included in the GWOSC release for GWTC-1.
The Virgo detector officially joined the detector network on August 1, 2017.
However, Virgo data for the event GW170729 is also publicly available 
from GWOSC, and following the example of \citet{LIGOScientific:2018mvr} and \citet{Chatziioannou:2019dsz} we have used Virgo data also to analyse this event, in addition to the later events, but with the exception of GW170823, which was only observed by the LIGO detectors.

\begin{figure*} 
    \includegraphics[width=0.91\columnwidth]{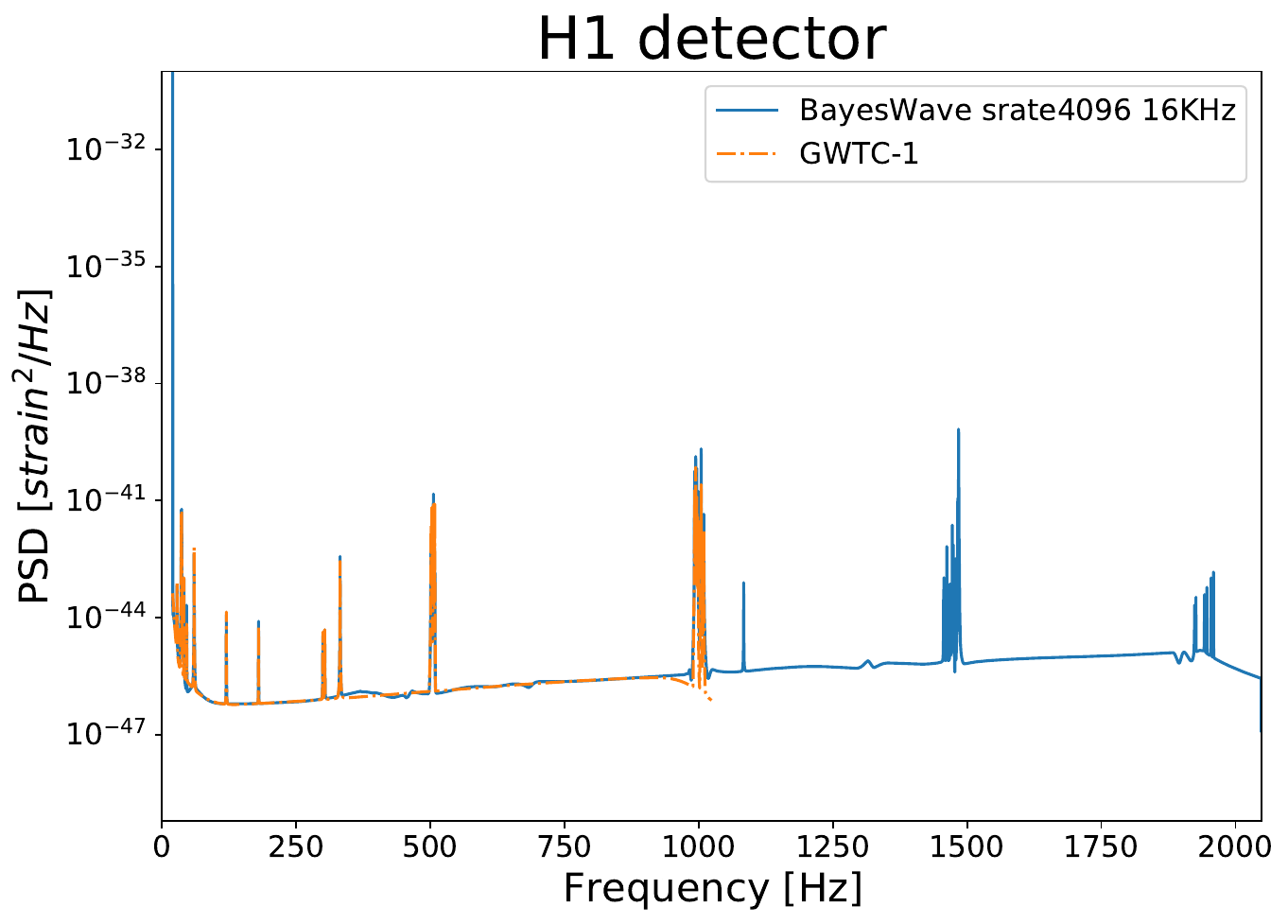}
    \includegraphics[width=0.91\columnwidth]{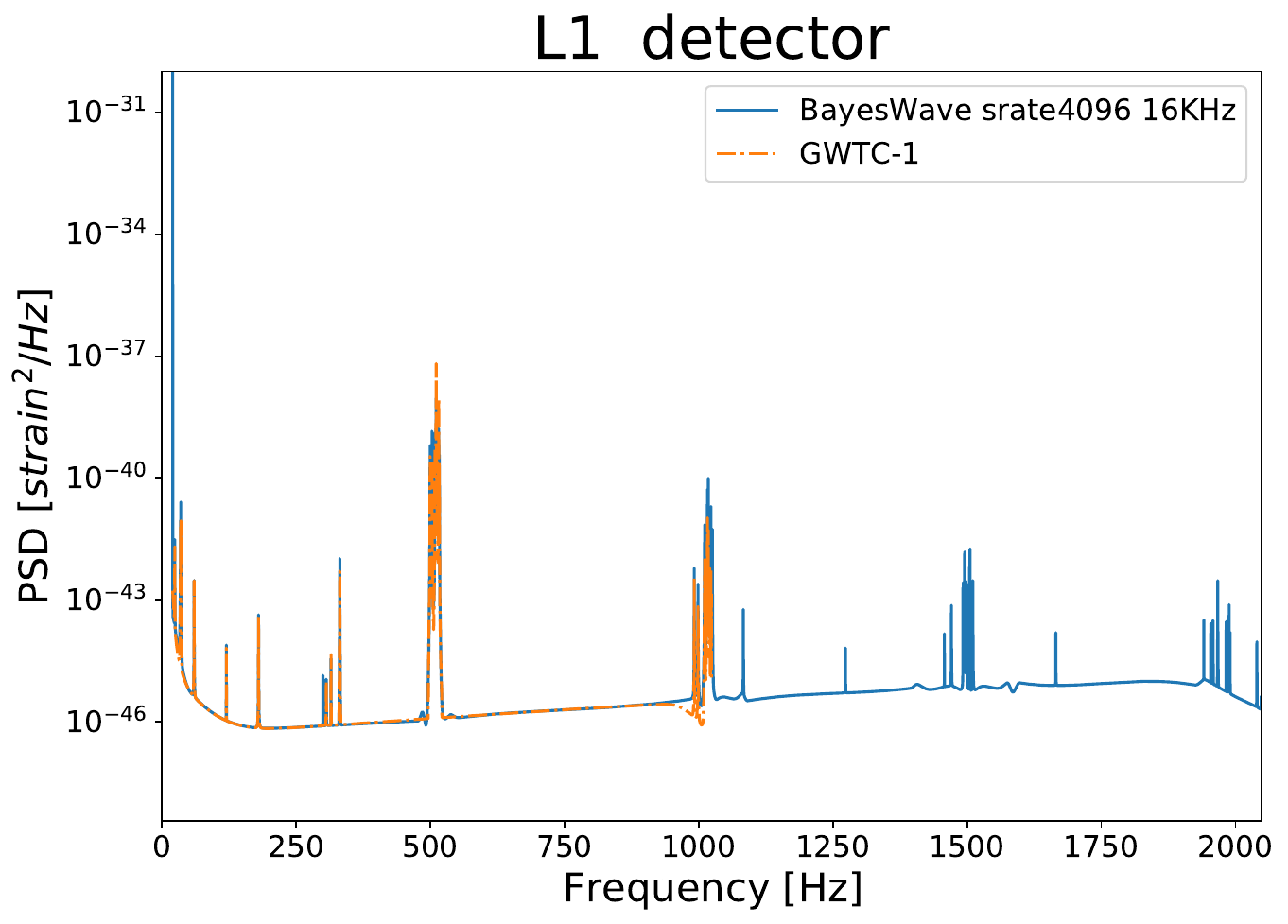}

    \caption{Comparison of PSDs of the two LIGO interferometers (H1 and L1) for GW151226 between the public data files provided in the 16\,kHz GWOSC data release with a sampling rate of 2048\,Hz (labeled with {\tt GWTC-1}), and our PSDs produced with the {\tt BayesWave} code, using a sampling rate of 4096\,Hz.
    One can see that both PSD data sets are very similar up to a frequency of almost 1\,kHz, where artefacts show up in the public PSD.
    These artefacts (decrease of the PSD) are similar to the high-frequency artefacts (near 2\,kHz) that can be seen in the PSDs we have produced with BayesWave.}
    \label{fig:PSDs}
\end{figure*}

Table \ref{tab:dataSettings} lists the trigger times corresponding to the GW events, indicates the detector data sets we use for each event, the duration of data we analyze, the minimal frequency we use to compute the likelihood function (and also as a reference frequency to define the GW phase and, for precessing systems, the component spins), and the sampling rate used. All of these parameters correspond to the choices made in \citet{LIGOScientific:2018mvr}. The sampling rate used in \citet{LIGOScientific:2018mvr} is 2048\,Hz, corresponding to a Nyquist frequency of 1024\,Hz. We have used this as our default setting to facilitate comparisons, together with GWOSC data sets sampled at 4\,KHz.
For high-mass events this is sufficient. For the three low-mass events GW151012, GW151226 and GW170608 however, the $\vert m \vert=3,4$ harmonics can reach higher frequencies than 1024\,Hz, and we have performed additional parameter estimation runs with a sampling rate of 4096\,Hz and the GWOSC 16\,kHz data sets (Nyquist frequency of 2048\,Hz).
For these analyses we have computed PSDs with the BayesWave code \citep{BayesWavepaper,BayesWaveGit}, which has been used by the LVC to produce the PSDs provided in the GWOSC data release.
We have checked that we can reproduce the public PSDs when setting the sampling rate to 2048\,Hz and using the 16\,kHz data sets, and then used the same settings to produce PSDs at twice the sampling rate.
These agree well with the PSDs using a lower sampling rate up to approximately 1\,kHz; see Fig.~\ref{fig:PSDs} for the example of data around the GW151226 event.
We also extend the calibration envelope files to 2048\,Hz. In \citet{CalEnvO2Plots}, we can find the plots for the magnitude and phase quantities of the O2 events. For GW170608, we extracted the data from 20\,Hz to 2048\,Hz from these plots. However, for the O1 events such plots are not publicly available. For that reason, we performed a linear extrapolation in frequency of the calibration uncertainties from 1024\,Hz to 2048\,Hz with the public data available in \citet{gwtc1calib}.

\begin{table}
%\begin{ruledtabular}
\begin{tabular}{lccccc}
\hline\hline
 Event & Trigger time & Duration & Low freq. & Sampling rate\\ \hline
GW150914 & 1126259462.4 & 8  & 20  & 2048 \\
GW151012 & 1128678900.4 & 8  & 20  & 2048/4096 \\
GW151226 & 1135136350.6 & 8  & 20  & 2048/4096 \\
GW170104 & 1167559936.6 & 4  & 20  & 2048\\
GW170608 & 1180922494.5 & 16 & 20  & 2048/4096\\
GW170729$^*$ & 1185389807.3 & 4  & 20  & 2048\\
GW170809$^*$ & 1186302519.7 & 4  & 20  & 2048\\
GW170814$^*$ & 1186741861.5 & 4  & 20  & 2048\\
GW170818$^*$ & 1187058327.1 & 4  & 16  & 2048\\
GW170823 & 1187529256.5 & 4  & 10  & 2048\\
\hline\hline
\end{tabular}
%\end{ruledtabular}
\caption{Data settings of each event from \citet{LIGOScientific:2018mvr} 
where * indicates that the three LIGO and Virgo detectors were taking data and all three data sets were used for our analysis. Note that for GW151012, GW151226 and GW170608 we extended the analysis also for a higher sampling rate.
\label{tab:dataSettings}
}
\end{table}

\subsubsection{Bayesian sampling algorithms}\label{sec:sampling}

For our standard runs we use the parallel Bilby \citep{Ashton:2018jfp,Smith:2019ucc} implementation of the {\tt dynesty} \citep{Speagle:2020spe} algorithm, based on nested sampling \citep{10.1214/06-BA127}.
In Appendix \ref{subsec:convergence_test} we also compare with results obtained with the LALInference code \citep{Veitch:2014wba}.
For both Bilby and LALInference we sample the component masses in terms of the mass ratio and chirp mass.
We also use the default settings of the Bilby code apart from the following choices: we fix the minimal ({\tt walks}) and maximal ({\tt maxmcmc}) number of  Markov Chain Monte Carlo (MCMC) steps to 200 and 15000 respectively. We vary the number of nested sampling live points, setting {\tt nlive} = 512, 1024, 2048, and 4096 for the three lowest mass events, and we vary the number of autocorrelation times to use before accepting a point, {\tt nact} = 10, 30, 50, in order to study the convergence of results. As explained in appendix \ref{subsec:convergence_test}, we choose {\tt nlive} = 2048 and {\tt nact} = 30 as the standard settings to report our main results. In order to speed up calculations we use distance marginalization as described in \citet{thrane_talbot_2019}.

In this paper we only discuss the most relevant results, however the complete posterior data sets are included in our Zenodo data release \citep{Zenodo_release}.

\subsubsection{Priors}\label{sec:priors}
In this paper we analyze the posterior probability densities of the source parameters $\theta$ from compact binary merger signals, computed using Bayes' theorem,
\begin{equation}
    p(\theta|d, \mathcal{H}) = \frac{\mathcal{L}(d|\theta,\mathcal{H})\pi(\theta|\mathcal{H})}{\mathcal{Z}(d|\mathcal{H})}
\end{equation}
where $\mathcal{L}(d|\theta,\mathcal{H})$ is the likelihood, $\pi(\theta|\mathcal{H})$ the prior and $\mathcal{Z}(d|\mathcal{H})$ the evidence for a given model ($\mathcal{H}$) and data ($d$).

The prior incorporates previous knowledge about the source parameters. The intention of our prior choices is to be uninformative, however, ambiguities arise when it is not possible to choose a natural quantity in which to prescribe a flat prior. For several quantities no problems arise: 
We use an isotropic prior for the sky location, the prior for the inclination angle is uniform in its cosine, and the priors of the polarization angle and phase of coalescence are uniform. 
For precessing spins, we use uniform distributions in the dimensionless component spin magnitudes and isotropically distributed spin tilts, as also used in \citet{LIGOScientific:2018mvr} and \citet{Romero-Shaw:2020owr}. We allow dimensionless component spin magnitudes up to $0.99$.

The distance prior used for the GWTC-1 catalog \citep{LIGOScientific:2018mvr} is uniform in luminosity distance volume (described by a second-order power law, \mbox{$\pi(d_L) \propto d_L^2$)}, which distributes mergers uniformly through a stationary Euclidean universe and does not take the expansion of the universe into account. The effect of the expansion of the universe can be neglected for small redshifts, however for larger redshifts it is more appropriate to use a prior that factors in our knowledge about cosmological expansion.
(See the discussions in Appendix C of \citet{Abbott:2020niy} and section 3.2.3 of \citet{Romero-Shaw:2020owr}.)
We will use the simpler prior ``uniform in luminosity distance volume'' for all events, however for the furthest events (GW170729 and GW170823) we also compare with a prior uniform in the comoving volume and source-frame time ({\tt UniformComovingVolume} prior in the Bilby code) using the Planck2015 cosmology \citep{Planck2015}, following the Bilby catalog \citep{Romero-Shaw:2020owr}.

Regarding the masses, the GWCT-1 catalog applies the LALInference \citep{Veitch:2014wba} prior, which is uniform in the component masses $m_1$ and $m_2$ with cuts in $\mathcal{M}_c$ and $q$.
The Bilby re-analysis of GWTC-1 \citep{Romero-Shaw:2020owr} uses a prior that is flat in chirp mass and mass ratio, but then the samples are reweighted to the LALInference choice using the Jacobian given in Eq.~(21) of \citet{Veitch:2014wba}. 
We follow this approach here for our default runs for all events, and present posteriors reweighted for a prior that is flat in component masses, consistent with \citet{LIGOScientific:2018mvr} and \citet{Romero-Shaw:2020owr}.
A very different choice has been used recently in \citet{Nitz:2020mga} to re-analyse the signal GW190521, which is consistent with the creation of an intermediate-mass black hole in a heavy black hole merger. Using a prior that is flat in mass ratio $Q \geq 1$, they found a multi-modal source-frame total mass posterior. We analyse this event in a parallel paper \citep{GW190521}.
with improved waveform models and confirm the multi-modality, although we also find significant differences with respect to \citet{Nitz:2020mga}.
In order to check whether similar multi-modalities appear for the most massive events in GWTC-1, we rerun GW170729 with a prior that is flat in mass ratio $Q \geq 1$ and {\it detector-frame} total mass
(a prior that is flat in source-frame masses adds complications to using distance marginalization, which we use to reduce the computational cost).

We also select GW170729 (see Sec.~\ref{sec:GW170729}) and the three lowest-mass events (see appendix~\ref{subsec:convergence_test}) to compare results with the new prior implemented in Bilby that applies the uniform prior in component masses while the mass ratio and chirp mass are sampled (by using the two classes {\tt UniformInComponentsChirpMass} and {\tt UniformInComponentsMassRatio} in the Bilby code), using the same component mass prior choices as the LALInference code.

The limits of the mass and luminosity distance priors vary between different events. We take these limits from the LALInference configuration files of the runs from \citet{LIGOScientific:2018mvr} and we check that the inferred parameters do not rail against the prior bounds.

\subsubsection{Management of parameter estimation runs}\label{sec:management}

In total we have performed 364 runs with parallel Bilby \citep{Smith:2019ucc}. In order to simplify the management of runs, and to avoid human errors when editing configuration files for each run, we have developed a \texttt{Python} code to generate configuration files from simpler descriptions which we call pseudo-config files. These pseudo-config files allow us to loop over parameters such as the waveform model, {\tt nlive} or {\tt nact}. Furthermore, we use this \texttt{Python} code to submit several runs to the {\tt slurm} \citep{slurm} queuing system with one command, and to track version numbers of the codes we use when the jobs start running.

\section{Results}\label{sec:results}

\subsection{Summary of the black hole mergers in GWTC-1}\label{sec:results_summary}

\begin{table*}

%\begin{ruledtabular}
\begin{tabular}{rccccccccccc} 

\hline\hline
Approx. & $q$ & $m_1 / M_\odot$ & $m_2 / M_\odot$ & $M_T / M_\odot$ & $\mathcal M / M_\odot$ &  $M_\mathrm{f} / M_\odot$ & $\chi_\mathrm{f}$ & $\chi_{\rm eff}$ & $\chi_\mathrm{p}$ &  $D_\mathrm{L} / \textrm{Mpc}$ &  $ \theta_{JN} / rad$  \\ 
 \hline

 \textbf{GW150914} \\
    Pv2 & \MassRatioCIPhPvtwofirst & \MOneSourceCIPhPvtwofirst & \MTwoSourceCIPhPvtwofirst & \MtotalSourceCIPhPvtwofirst & \ChirpMassSourceCIPhPvtwofirst & \FinalMassSourceCIPhPvtwofirst & \FinalSpinCIPhPvtwofirst & \ChiEffCIPhPvtwofirst & \ChiPCIPhPvtwofirst & \DLCIPhPvtwofirst &  \ThetaJNCIPhPvtwofirst \\ 
    XP & \MassRatioCIPhXPfirst & \MOneSourceCIPhXPfirst & \MTwoSourceCIPhXPfirst & \MtotalSourceCIPhXPfirst & \ChirpMassSourceCIPhXPfirst & \FinalMassSourceCIPhXPfirst & \FinalSpinCIPhXPfirst & \ChiEffCIPhXPfirst & \ChiPCIPhXPfirst & \DLCIPhXPfirst & \ThetaJNCIPhXPfirst\\
    XPHM  &  \MassRatioCIPhXPHMfirst & \MOneSourceCIPhXPHMfirst & \MTwoSourceCIPhXPHMfirst & \MtotalSourceCIPhXPHMfirst & \ChirpMassSourceCIPhXPHMfirst & \FinalMassSourceCIPhXPHMfirst & \FinalSpinCIPhXPHMfirst & \ChiEffCIPhXPHMfirst & \ChiPCIPhXPHMfirst & \DLCIPhXPHMfirst  & \ThetaJNCIPhXPHMfirst\\
    TPHM  &  \MassRatioCIPhTPHMfirst & \MOneSourceCIPhTPHMfirst & \MTwoSourceCIPhTPHMfirst & \MtotalSourceCIPhTPHMfirst & \ChirpMassSourceCIPhTPHMfirst & \FinalMassSourceCIPhTPHMfirst & \FinalSpinCIPhTPHMfirst & \ChiEffCIPhTPHMfirst & \ChiPCIPhTPHMfirst & \DLCIPhTPHMfirst  & \ThetaJNCIPhTPHMfirst\\[6pt]

 \textbf{GW151012}\\
 Pv2 & \MassRatioCIPhPvtwosecond & \MOneSourceCIPhPvtwosecond & \MTwoSourceCIPhPvtwosecond & \MtotalSourceCIPhPvtwosecond & \ChirpMassSourceCIPhPvtwosecond & \FinalMassSourceCIPhPvtwosecond & \FinalSpinCIPhPvtwosecond & \ChiEffCIPhPvtwosecond & \ChiPCIPhPvtwosecond & \DLCIPhPvtwosecond  &  \ThetaJNCIPhPvtwosecond \\ 
  XP & \MassRatioCIPhXPsecond & \MOneSourceCIPhXPsecond & \MTwoSourceCIPhXPsecond & \MtotalSourceCIPhXPsecond & \ChirpMassSourceCIPhXPsecond & \FinalMassSourceCIPhXPsecond & \FinalSpinCIPhXPsecond & \ChiEffCIPhXPsecond & \ChiPCIPhXPsecond & \DLCIPhXPsecond & \ThetaJNCIPhXPsecond\\
  XPHM  &  \MassRatioCIPhXPHMsecond & \MOneSourceCIPhXPHMsecond & \MTwoSourceCIPhXPHMsecond & \MtotalSourceCIPhXPHMsecond & \ChirpMassSourceCIPhXPHMsecond & \FinalMassSourceCIPhXPHMsecond & \FinalSpinCIPhXPHMsecond & \ChiEffCIPhXPHMsecond & \ChiPCIPhXPHMsecond & \DLCIPhXPHMsecond & \ThetaJNCIPhXPHMsecond \\[6pt]

 \textbf{GW151226}\\
 Pv2 & \MassRatioCIPhPvtwothird & \MOneSourceCIPhPvtwothird & \MTwoSourceCIPhPvtwothird & \MtotalSourceCIPhPvtwothird & \ChirpMassSourceCIPhPvtwothird & \FinalMassSourceCIPhPvtwothird & \FinalSpinCIPhPvtwothird & \ChiEffCIPhPvtwothird & \ChiPCIPhPvtwothird & \DLCIPhPvtwothird &  \ThetaJNCIPhPvtwothird \\ 
  XP & \MassRatioCIPhXPthird & \MOneSourceCIPhXPthird & \MTwoSourceCIPhXPthird & \MtotalSourceCIPhXPthird & \ChirpMassSourceCIPhXPthird & \FinalMassSourceCIPhXPthird & \FinalSpinCIPhXPthird & \ChiEffCIPhXPthird & \ChiPCIPhXPthird & \DLCIPhXPthird & \ThetaJNCIPhXPthird\\
  XPHM  &  \MassRatioCIPhXPHMthird & \MOneSourceCIPhXPHMthird & \MTwoSourceCIPhXPHMthird & \MtotalSourceCIPhXPHMthird & \ChirpMassSourceCIPhXPHMthird & \FinalMassSourceCIPhXPHMthird & \FinalSpinCIPhXPHMthird & \ChiEffCIPhXPHMthird & \ChiPCIPhXPHMthird & \DLCIPhXPHMthird  & \ThetaJNCIPhXPHMthird \\[6pt]

  \textbf{GW170104}\\
  Pv2 & \MassRatioCIPhPvtwofourth & \MOneSourceCIPhPvtwofourth & \MTwoSourceCIPhPvtwofourth & \MtotalSourceCIPhPvtwofourth & \ChirpMassSourceCIPhPvtwofourth & \FinalMassSourceCIPhPvtwofourth & \FinalSpinCIPhPvtwofourth & \ChiEffCIPhPvtwofourth & \ChiPCIPhPvtwofourth & \DLCIPhPvtwofourth &  \ThetaJNCIPhPvtwofourth \\ 
  XP & \MassRatioCIPhXPfourth & \MOneSourceCIPhXPfourth & \MTwoSourceCIPhXPfourth & \MtotalSourceCIPhXPfourth & \ChirpMassSourceCIPhXPfourth & \FinalMassSourceCIPhXPfourth & \FinalSpinCIPhXPfourth & \ChiEffCIPhXPfourth & \ChiPCIPhXPfourth & \DLCIPhXPfourth & \ThetaJNCIPhXPfourth\\
  XPHM  &  \MassRatioCIPhXPHMfourth & \MOneSourceCIPhXPHMfourth & \MTwoSourceCIPhXPHMfourth & \MtotalSourceCIPhXPHMfourth & \ChirpMassSourceCIPhXPHMfourth & \FinalMassSourceCIPhXPHMfourth & \FinalSpinCIPhXPHMfourth & \ChiEffCIPhXPHMfourth & \ChiPCIPhXPHMfourth & \DLCIPhXPHMfourth  & \ThetaJNCIPhXPHMfourth\\[6pt]

 \textbf{GW170104}\\
 Pv2 & \MassRatioCIPhPvtwofifth & \MOneSourceCIPhPvtwofifth & \MTwoSourceCIPhPvtwofifth & \MtotalSourceCIPhPvtwofifth & \ChirpMassSourceCIPhPvtwofifth & \FinalMassSourceCIPhPvtwofifth & \FinalSpinCIPhPvtwofifth & \ChiEffCIPhPvtwofifth & \ChiPCIPhPvtwofifth & \DLCIPhPvtwofifth &  \ThetaJNCIPhPvtwofifth \\ 
  XP & \MassRatioCIPhXPfifth & \MOneSourceCIPhXPfifth & \MTwoSourceCIPhXPfifth & \MtotalSourceCIPhXPfifth & \ChirpMassSourceCIPhXPfifth & \FinalMassSourceCIPhXPfifth & \FinalSpinCIPhXPfifth & \ChiEffCIPhXPfifth & \ChiPCIPhXPfifth & \DLCIPhXPfifth & \ThetaJNCIPhXPfifth\\
  XPHM  &  \MassRatioCIPhXPHMfifth & \MOneSourceCIPhXPHMfifth & \MTwoSourceCIPhXPHMfifth & \MtotalSourceCIPhXPHMfifth & \ChirpMassSourceCIPhXPHMfifth & \FinalMassSourceCIPhXPHMfifth & \FinalSpinCIPhXPHMfifth & \ChiEffCIPhXPHMfifth & \ChiPCIPhXPHMfifth & \DLCIPhXPHMfifth  & \ThetaJNCIPhXPHMfifth\\[6pt]

 \textbf{GW170729}\\
 Pv2 & \MassRatioCIPhPvtwosixth & \MOneSourceCIPhPvtwosixth & \MTwoSourceCIPhPvtwosixth & \MtotalSourceCIPhPvtwosixth & \ChirpMassSourceCIPhPvtwosixth & \FinalMassSourceCIPhPvtwosixth & \FinalSpinCIPhPvtwosixth & \ChiEffCIPhPvtwosixth & \ChiPCIPhPvtwosixth & \DLCIPhPvtwosixth &  \ThetaJNCIPhPvtwosixth \\ 
  XP & \MassRatioCIPhXPsixth & \MOneSourceCIPhXPsixth & \MTwoSourceCIPhXPsixth & \MtotalSourceCIPhXPsixth & \ChirpMassSourceCIPhXPsixth & \FinalMassSourceCIPhXPsixth & \FinalSpinCIPhXPsixth & \ChiEffCIPhXPsixth & \ChiPCIPhXPsixth & \DLCIPhXPsixth & \ThetaJNCIPhXPsixth\\
  TP &  \MassRatioCIPhTPsixth & \MOneSourceCIPhTPsixth & \MTwoSourceCIPhTPsixth & \MtotalSourceCIPhTPsixth & \ChirpMassSourceCIPhTPsixth & \FinalMassSourceCIPhTPsixth & \FinalSpinCIPhTPsixth & \ChiEffCIPhTPsixth & \ChiPCIPhTPsixth & \DLCIPhTPsixth  & \ThetaJNCIPhTPsixth\\
  XPHM  &  \MassRatioCIPhXPHMsixth & \MOneSourceCIPhXPHMsixth & \MTwoSourceCIPhXPHMsixth & \MtotalSourceCIPhXPHMsixth & \ChirpMassSourceCIPhXPHMsixth & \FinalMassSourceCIPhXPHMsixth & \FinalSpinCIPhXPHMsixth & \ChiEffCIPhXPHMsixth & \ChiPCIPhXPHMsixth & \DLCIPhXPHMsixth  & \ThetaJNCIPhXPHMsixth\\
   TPHM  &  \MassRatioCIPhTPHMsixth & \MOneSourceCIPhTPHMsixth & \MTwoSourceCIPhTPHMsixth & \MtotalSourceCIPhTPHMsixth & \ChirpMassSourceCIPhTPHMsixth & \FinalMassSourceCIPhTPHMsixth & \FinalSpinCIPhTPHMsixth & \ChiEffCIPhTPHMsixth & \ChiPCIPhTPHMsixth & \DLCIPhTPHMsixth  & \ThetaJNCIPhTPHMsixth\\[6pt]

 \textbf{GW170809}\\
 Pv2 & \MassRatioCIPhPvtwoseventh & \MOneSourceCIPhPvtwoseventh & \MTwoSourceCIPhPvtwoseventh & \MtotalSourceCIPhPvtwoseventh & \ChirpMassSourceCIPhPvtwoseventh & \FinalMassSourceCIPhPvtwoseventh & \FinalSpinCIPhPvtwoseventh & \ChiEffCIPhPvtwoseventh & \ChiPCIPhPvtwoseventh & \DLCIPhPvtwoseventh &  \ThetaJNCIPhPvtwoseventh \\ 
  XP & \MassRatioCIPhXPseventh & \MOneSourceCIPhXPseventh & \MTwoSourceCIPhXPseventh & \MtotalSourceCIPhXPseventh & \ChirpMassSourceCIPhXPseventh & \FinalMassSourceCIPhXPseventh & \FinalSpinCIPhXPseventh & \ChiEffCIPhXPseventh & \ChiPCIPhXPseventh & \DLCIPhXPseventh & \ThetaJNCIPhXPseventh\\
  XPHM  &  \MassRatioCIPhXPHMseventh & \MOneSourceCIPhXPHMseventh & \MTwoSourceCIPhXPHMseventh & \MtotalSourceCIPhXPHMseventh & \ChirpMassSourceCIPhXPHMseventh & \FinalMassSourceCIPhXPHMseventh & \FinalSpinCIPhXPHMseventh & \ChiEffCIPhXPHMseventh & \ChiPCIPhXPHMseventh & \DLCIPhXPHMseventh  & \ThetaJNCIPhXPHMseventh\\
   TPHM  &  \MassRatioCIPhTPHMseventh & \MOneSourceCIPhTPHMseventh & \MTwoSourceCIPhTPHMseventh & \MtotalSourceCIPhTPHMseventh & \ChirpMassSourceCIPhTPHMseventh & \FinalMassSourceCIPhTPHMseventh & \FinalSpinCIPhTPHMseventh & \ChiEffCIPhTPHMseventh & \ChiPCIPhTPHMseventh & \DLCIPhTPHMseventh  & \ThetaJNCIPhTPHMseventh\\[6pt]

 \textbf{GW170814}\\
 Pv2 & \MassRatioCIPhPvtwoeighth & \MOneSourceCIPhPvtwoeighth & \MTwoSourceCIPhPvtwoeighth & \MtotalSourceCIPhPvtwoeighth & \ChirpMassSourceCIPhPvtwoeighth & \FinalMassSourceCIPhPvtwoeighth & \FinalSpinCIPhPvtwoeighth & \ChiEffCIPhPvtwoeighth & \ChiPCIPhPvtwoeighth & \DLCIPhPvtwoeighth &  \ThetaJNCIPhPvtwoeighth \\ 
  XP & \MassRatioCIPhXPeighth & \MOneSourceCIPhXPeighth & \MTwoSourceCIPhXPeighth & \MtotalSourceCIPhXPeighth & \ChirpMassSourceCIPhXPeighth & \FinalMassSourceCIPhXPeighth & \FinalSpinCIPhXPeighth & \ChiEffCIPhXPeighth & \ChiPCIPhXPeighth & \DLCIPhXPeighth & \ThetaJNCIPhXPeighth\\
  XPHM  &  \MassRatioCIPhXPHMeighth & \MOneSourceCIPhXPHMeighth & \MTwoSourceCIPhXPHMeighth & \MtotalSourceCIPhXPHMeighth & \ChirpMassSourceCIPhXPHMeighth & \FinalMassSourceCIPhXPHMeighth & \FinalSpinCIPhXPHMeighth & \ChiEffCIPhXPHMeighth & \ChiPCIPhXPHMeighth & \DLCIPhXPHMeighth  & \ThetaJNCIPhXPHMeighth\\
   XPHM* &  \MassRatioCIPhXPHMeighthNNLO & \MOneSourceCIPhXPHMeighthNNLO & \MTwoSourceCIPhXPHMeighthNNLO & \MtotalSourceCIPhXPHMeighthNNLO & \ChirpMassSourceCIPhXPHMeighthNNLO & \FinalMassSourceCIPhXPHMeighthNNLO & \FinalSpinCIPhXPHMeighthNNLO & \ChiEffCIPhXPHMeighthNNLO & \ChiPCIPhXPHMeighthNNLO & \DLCIPhXPHMeighthNNLO  & \ThetaJNCIPhXPHMeighthNNLO\\
  TPHM  &  \MassRatioCIPhTPHMeighth & \MOneSourceCIPhTPHMeighth & \MTwoSourceCIPhTPHMeighth & \MtotalSourceCIPhTPHMeighth & \ChirpMassSourceCIPhTPHMeighth & \FinalMassSourceCIPhTPHMeighth & \FinalSpinCIPhTPHMeighth & \ChiEffCIPhTPHMeighth & \ChiPCIPhTPHMeighth & \DLCIPhTPHMeighth  & \ThetaJNCIPhTPHMeighth\\[6pt]

 \textbf{GW170818}\\
 Pv2 & \MassRatioCIPhPvtwonineth & \MOneSourceCIPhPvtwonineth & \MTwoSourceCIPhPvtwonineth & \MtotalSourceCIPhPvtwonineth & \ChirpMassSourceCIPhPvtwonineth & \FinalMassSourceCIPhPvtwonineth & \FinalSpinCIPhPvtwonineth & \ChiEffCIPhPvtwonineth & \ChiPCIPhPvtwonineth & \DLCIPhPvtwonineth &  \ThetaJNCIPhPvtwonineth \\ 
  XP & \MassRatioCIPhXPnineth & \MOneSourceCIPhXPnineth & \MTwoSourceCIPhXPnineth & \MtotalSourceCIPhXPnineth & \ChirpMassSourceCIPhXPnineth & \FinalMassSourceCIPhXPnineth & \FinalSpinCIPhXPnineth & \ChiEffCIPhXPnineth & \ChiPCIPhXPnineth & \DLCIPhXPnineth & \ThetaJNCIPhXPnineth\\
  XPHM  &  \MassRatioCIPhXPHMnineth & \MOneSourceCIPhXPHMnineth & \MTwoSourceCIPhXPHMnineth & \MtotalSourceCIPhXPHMnineth & \ChirpMassSourceCIPhXPHMnineth & \FinalMassSourceCIPhXPHMnineth & \FinalSpinCIPhXPHMnineth & \ChiEffCIPhXPHMnineth & \ChiPCIPhXPHMnineth & \DLCIPhXPHMnineth  & \ThetaJNCIPhXPHMnineth\\
   XPHM* &  \MassRatioCIPhXPHMninethNNLO & \MOneSourceCIPhXPHMninethNNLO & \MTwoSourceCIPhXPHMninethNNLO & \MtotalSourceCIPhXPHMninethNNLO & \ChirpMassSourceCIPhXPHMninethNNLO & \FinalMassSourceCIPhXPHMninethNNLO & \FinalSpinCIPhXPHMninethNNLO & \ChiEffCIPhXPHMninethNNLO & \ChiPCIPhXPHMninethNNLO & \DLCIPhXPHMninethNNLO  & \ThetaJNCIPhXPHMninethNNLO\\
   TPHM  &  \MassRatioCIPhTPHMnineth & \MOneSourceCIPhTPHMnineth & \MTwoSourceCIPhTPHMnineth & \MtotalSourceCIPhTPHMnineth & \ChirpMassSourceCIPhTPHMnineth & \FinalMassSourceCIPhTPHMnineth & \FinalSpinCIPhTPHMnineth & \ChiEffCIPhTPHMnineth & \ChiPCIPhTPHMnineth & \DLCIPhTPHMnineth  & \ThetaJNCIPhTPHMnineth\\[6pt]

 \textbf{GW170823}\\
 Pv2 & \MassRatioCIPhPvtwotenth & \MOneSourceCIPhPvtwotenth & \MTwoSourceCIPhPvtwotenth & \MtotalSourceCIPhPvtwotenth & \ChirpMassSourceCIPhPvtwotenth & \FinalMassSourceCIPhPvtwotenth & \FinalSpinCIPhPvtwotenth & \ChiEffCIPhPvtwotenth & \ChiPCIPhPvtwotenth & \DLCIPhPvtwotenth &  \ThetaJNCIPhPvtwotenth \\ 
  XP & \MassRatioCIPhXPtenth & \MOneSourceCIPhXPtenth & \MTwoSourceCIPhXPtenth & \MtotalSourceCIPhXPtenth & \ChirpMassSourceCIPhXPtenth & \FinalMassSourceCIPhXPtenth & \FinalSpinCIPhXPtenth & \ChiEffCIPhXPtenth & \ChiPCIPhXPtenth & \DLCIPhXPtenth & \ThetaJNCIPhXPtenth\\
  XPHM  &  \MassRatioCIPhXPHMtenth & \MOneSourceCIPhXPHMtenth & \MTwoSourceCIPhXPHMtenth & \MtotalSourceCIPhXPHMtenth & \ChirpMassSourceCIPhXPHMtenth & \FinalMassSourceCIPhXPHMtenth & \FinalSpinCIPhXPHMtenth & \ChiEffCIPhXPHMtenth & \ChiPCIPhXPHMtenth & \DLCIPhXPHMtenth  & \ThetaJNCIPhXPHMtenth\\
   TPHM  &  \MassRatioCIPhTPHMtenth & \MOneSourceCIPhTPHMtenth & \MTwoSourceCIPhTPHMtenth & \MtotalSourceCIPhTPHMtenth & \ChirpMassSourceCIPhTPHMtenth & \FinalMassSourceCIPhTPHMtenth & \FinalSpinCIPhTPHMtenth & \ChiEffCIPhTPHMtenth & \ChiPCIPhTPHMtenth & \DLCIPhTPHMtenth  & \ThetaJNCIPhTPHMtenth\\

\hline\hline
\end{tabular}
%\end{ruledtabular}
\caption{
\label{tab:PEresults}
Inferred source parameter values of the first 10 BBH
detections, given as posterior median values with 90\%
credible intervals. Masses correspond to the source frame. We
compare the public \phPvtwo (Pv2) results
\citep{LIGOScientific:2018mvr} with those obtained with the new
generation of phenomenological waveform models. \phXP (XP) and the
default version of \phXPHM (XPHM) are listed for all events,
\phTPHM (TPHM) is included for high-mass events, \phTP (TP) for GW170729 and an
alternative version of \phXPHM (NNLO angles and final spin
version 2) is included for GW170814 and GW170818, indicated
by a 
*.}
\end{table*}

Table \ref{tab:PEresults} contains a summary of our results using precessing waveform models, with error estimates obtained from 90\% posterior credible intervals, including for comparison also the results obtained for GWTC-1 \citep{LIGOScientific:2018mvr} with \phPvtwo. Our new results are in general within the error estimates of the \phPvtwo results, as can be expected from the initial analysis with subdominant harmonics performed for the GWTC-1 catalog, see appendix B of \citet{LIGOScientific:2018mvr}: There all BBH events have been cross-checked with the RapidPE algorithm \citep{Pankow:2015cra,Lange:2018pyp} and a waveform catalog of NR simulations supplemented by waveforms from the \NRSur model \citep{Blackman:2017pcm}; samples were released with \citet{GWTC1_NR_samples}. \NRSur is however restricted to mass ratios $q \leq 2$ and dimensionless $\chi_i \leq 0.8$. A ``modest influence on the interpretation of observations'' below the statistical measurement uncertainty is found, and for GW170729 a Bayes factor of ``approximately 1.4 for higher modes versus a pure quadrupole model'', which we compare with our results below.

\begin{figure} 
\begin{center}
\includegraphics[width=0.95\columnwidth]{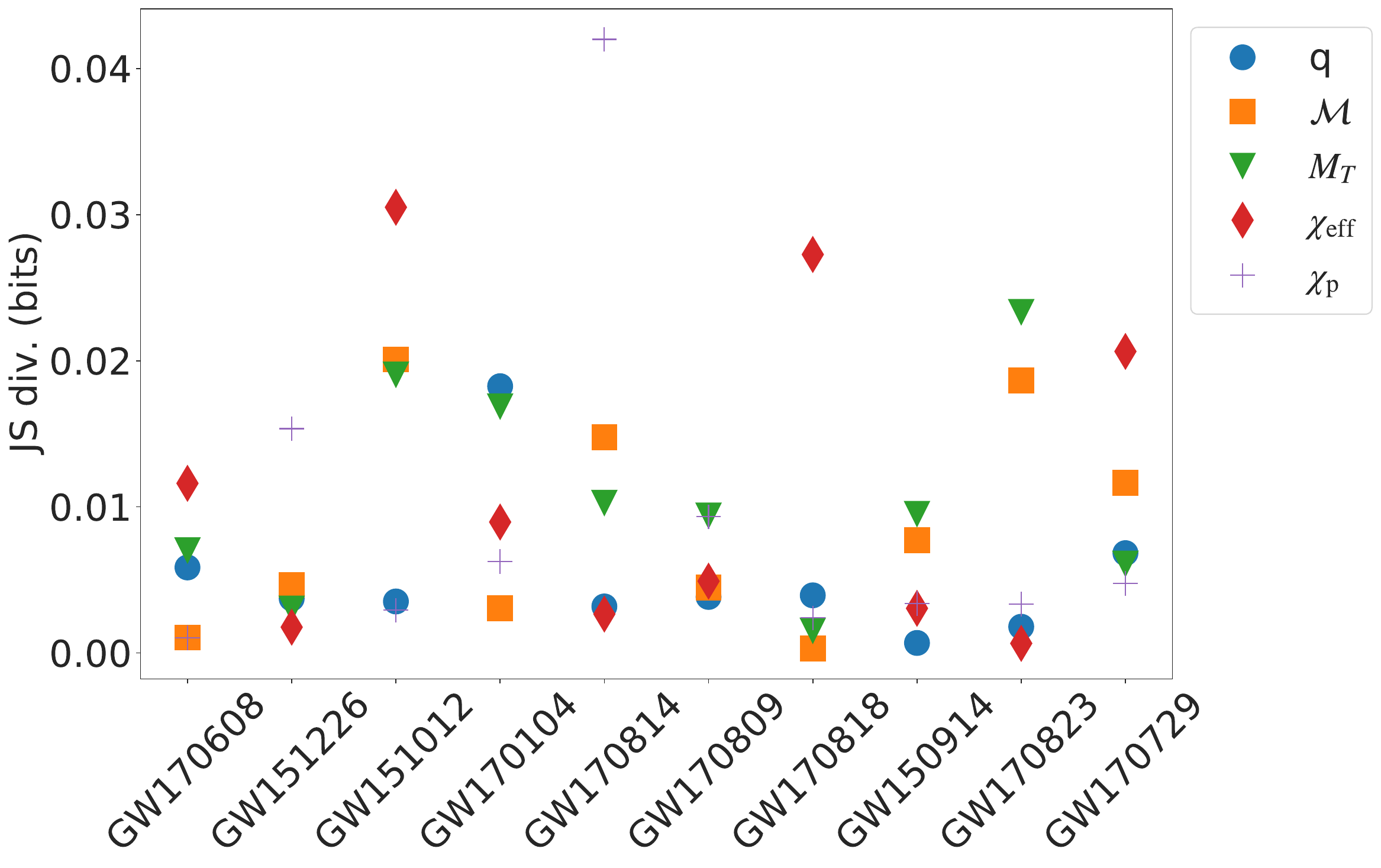}\\
\includegraphics[width=0.95\columnwidth]{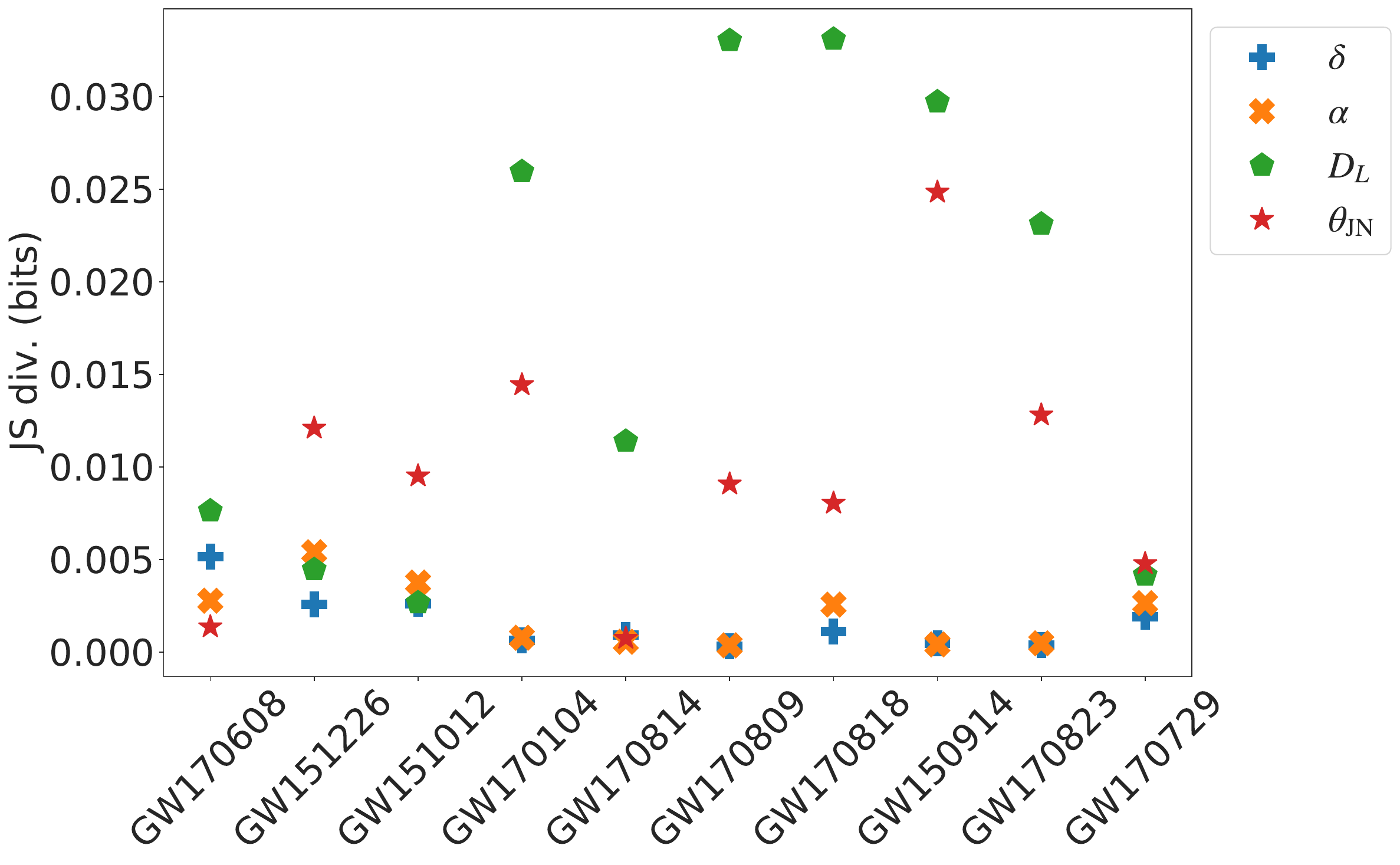}\\
\caption{Jensen-Shannon divergences (JS div) between \phPvtwo \citep{LIGOScientific:2018mvr} and \phXPHM results of each event, which are ordered from left to right with increasing chirp mass $\mathcal{M}$. Top panel: intrinsic parameters. Bottom panel: extrinsic parameters.
\label{fig:js_div_events}
}
\end{center}
\end{figure}

However, the use of multi-mode waveforms in full Bayesian parameter estimation is now standard in GW data analysis, and it
is thus interesting to update the results of GWTC-1 to the methods used for O3 events~\citep{LIGOScientific:2021usb,LIGOScientific:2021djp}, and to compare results.
Differences between our default runs using \phXPHM and the \phPvtwo results from the GWTC-1 release are shown in Fig. \ref{fig:js_div_events} in terms of the Jensen-Shannon (JS) divergence, for a definition see e.g. Eq. (B1) of \citet{LIGOScientific:2018mvr}.
The JS divergence takes values between 0 and 1 bits, where 0 means that there is no difference between two distributions, and 1 means that both distributions have a maximum divergence.
The posteriors for the model with higher modes and precession are generally close to the results obtained with \phPvtwo -- all divergence values between both posterior distributions are smaller than 0.045, which is smaller than the divergence between \phPvtwo and \seobnrvthree runs performed in the GWTC-1 paper, where the largest value is for GW151226 and the effective spin $\chieff$, $JS_{\chieff}$ = 0.14 bits, see Fig.~16 of \citet{LIGOScientific:2018mvr}.

Among the extrinsic parameters, the sky location (declination $\delta$ and right ascension $\alpha$) does not show significant changes, while the 
distance $D_L$ and the angle $\theta_{JN}$ between total angular momentum and the line of sight have the largest divergences.

\begin{figure} 
    \centering

    \includegraphics[width=0.9\columnwidth]{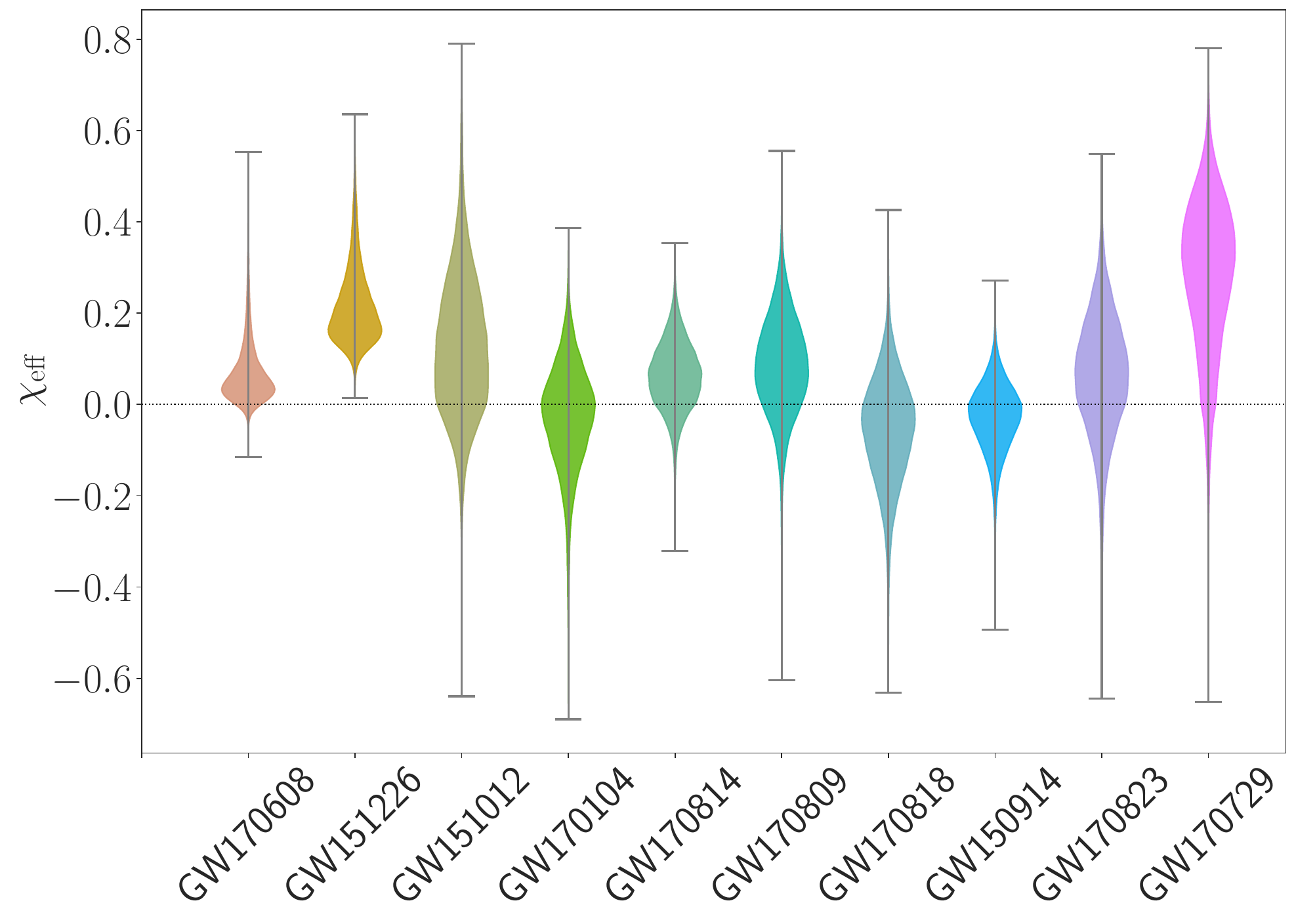}\\
    \includegraphics[width=0.9\columnwidth]{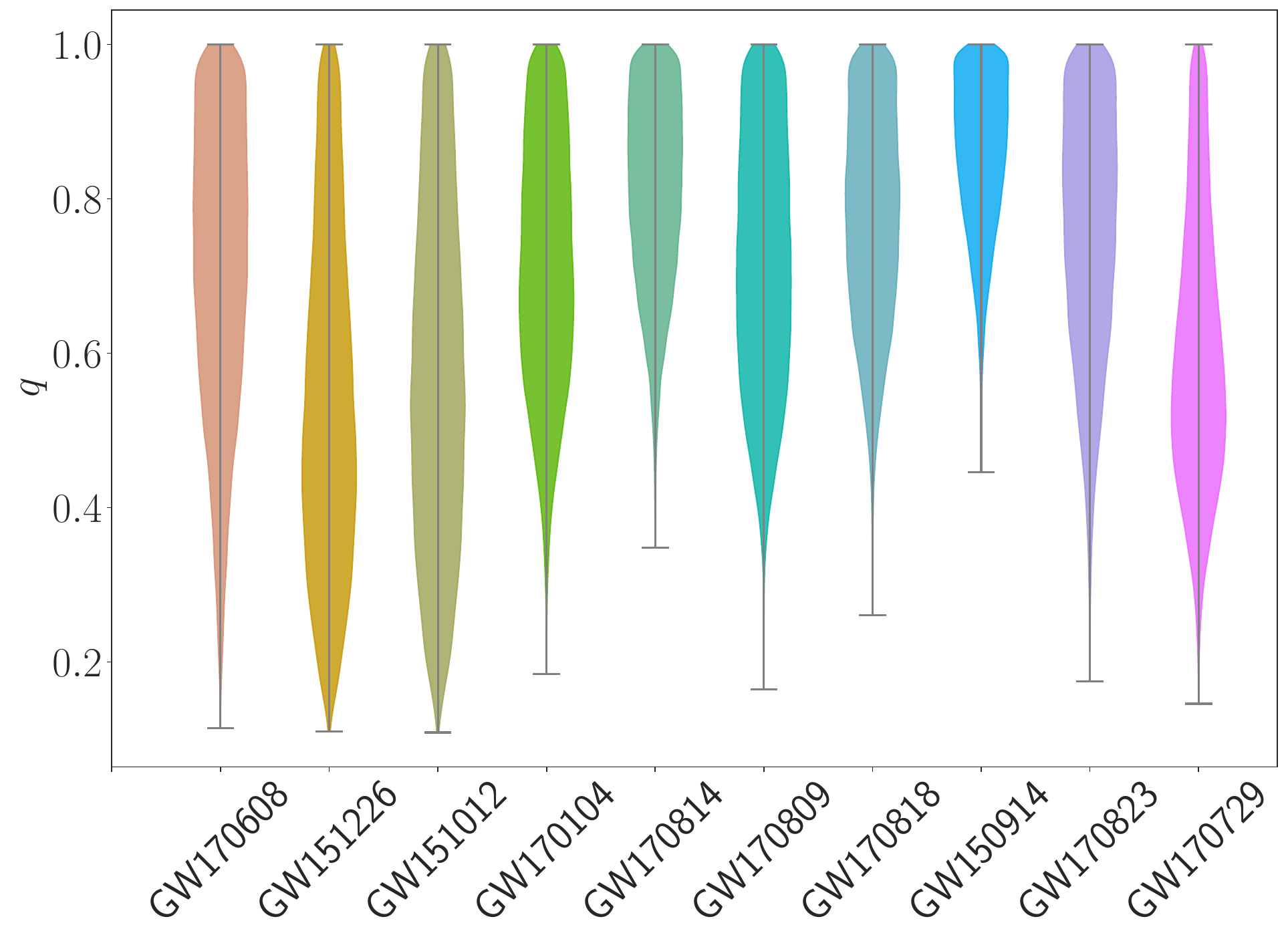}
    \caption{Violin plots of the \phXPHM default results for $\chieff$ and $\massRatio$, where the short horizontal lines indicate the 90\% credible intervals. The events are sorted by the $\chirpMass$ values. Plots created with the PESummary code \citep{Hoy:2020vys}.
    \label{fig:violin_plots}
    }
\end{figure}

An overview of recovered values for the effective spin and mass ratio for the different events is shown in Fig.~\ref{fig:violin_plots}.
One of the trends that can be observed (see also the detailed results in Table \ref{tab:PEresults}) is that the effective spins $\chieff$ inferred with \phXPHM have typically increased over the values inferred with \phPvtwo, although they are still consistent within the error estimates for all events. 
The event least following this trend is the highest-mass event GW170729, which is also one of the two events where $\chieff$ has support only for positive values (GW151226 being the other).

In Table \ref{tab:SNR_BF_table}, we show the matched-filter SNRs obtained for runs with different models, as well as the Bayes factors between the signal and noise hypotheses for each of these runs, and compare these Bayes factors against our default \phXPHM runs.
Bayes factors and SNRs are in general very consistent for different waveform models, confirming the expectation that neither precession, higher mode effects, nor general waveform systematics effects can be clearly identified for any of the GWTC-1 events.

In the rest of this section, we will discuss all ten BBH events from GWTC-1 individually, grouping them into low-mass (GW151012, GW151226, GW170608), medium-mass (GW170104 and GW170814) and high-mass (GW150914, GW170729, GW170809, GW170818, GW170823) events. Note that we just make this distinction for convenience, and the only events that clearly stand out in terms of total mass are the two lightest ones (GW170608 and GW151226) and the heaviest (GW170729).

\begin{table} 
\caption{Network matched-filter SNRs with 90\% credible intervals and $\log_{10}$ signal-to-noise Bayes factors for runs with the \phenX and \phT families, where XPHM$^*$ denotes the version with NNLO Euler angles and final spin option 2.
The last column shows the Bayes factor against our default \phXPHM runs.
\label{tab:SNR_BF_table}
}
\resizebox*{!}{0.9\textheight}{
%\begin{ruledtabular}
\begin{tabular}{llccc} 
\hline\hline
 Event & Approx. & SNR & $\log_{10} \mathcal{BF}$ & $\mathcal{BF}_\mathrm{XPHM}$ \\ \hline

 \multirow{6}{*}{GW150914} & XAS & \ensuremath{25.1_{-0.1}^{+0.1}}\xspace & \ensuremath{124.07\pm 0.07}\xspace & \ensuremath{0.99_{-0.19}^{+0.24}}\xspace \\
 & XHM & \ensuremath{25.1_{-0.1}^{+0.1}}\xspace & \ensuremath{124.08\pm 0.07}\xspace & \ensuremath{1.01_{-0.20}^{+0.24}}\xspace\\
 & THM & \ensuremath{25.1_{-0.1}^{+0.1}}\xspace & \ensuremath{123.83\pm 0.07}\xspace & \ensuremath{0.57_{-0.11}^{+0.14}}\xspace\\
 & XP  & \ensuremath{25.1_{-0.1}^{+0.1}}\xspace & \ensuremath{124.05\pm 0.07}\xspace & \ensuremath{0.95_{-0.19}^{+0.23}}\xspace\\
 & TPHM & \ensuremath{25.1_{-0.1}^{+0.1}}\xspace & \ensuremath{123.86\pm 0.07}\xspace & \ensuremath{0.61_{-0.12}^{+0.15}}\xspace\\
 & XPHM & \ensuremath{25.1_{-0.1}^{+0.1}}\xspace & \ensuremath{124.07\pm 0.07}\xspace & \ensuremath{1.00}\xspace\\[6pt]

 \multirow{6}{*}{GW151012} & XAS & \ensuremath{9.2_{-0.3}^{+0.2}}\xspace & \ensuremath{10.13\pm 0.06}\xspace & \ensuremath{0.58_{-0.10}^{+0.12}}\xspace \\
 & XHM & \ensuremath{9.4_{-0.4}^{+0.3}}\xspace & \ensuremath{10.27\pm 0.06}\xspace & \ensuremath{0.81_{-0.14}^{+0.16}}\xspace \\
 & THM & \ensuremath{9.4_{-0.4}^{+0.3}}\xspace & \ensuremath{10.37\pm 0.06}\xspace & \ensuremath{1.02_{-0.17}^{+0.20}}\xspace\\
 & XP  & \ensuremath{9.2_{-0.3}^{+0.2}}\xspace & \ensuremath{10.25\pm 0.06}\xspace & \ensuremath{0.77_{-0.13}^{+0.15}}\xspace\\
 & XPHM & \ensuremath{9.3_{-0.4}^{+0.3}}\xspace & \ensuremath{10.36\pm 0.06}\xspace & \ensuremath{1.00}\xspace \\[6pt] 
 
 \multirow{6}{*}{GW151226} & XAS & \ensuremath{12.4_{-0.2}^{+0.2}}\xspace & \ensuremath{22.04\pm 0.06}\xspace & \ensuremath{0.26_{-0.05}^{+0.06}}\xspace \\
 & XHM & \ensuremath{12.4_{-0.3}^{+0.2}}\xspace & \ensuremath{22.03\pm 0.06}\xspace & \ensuremath{0.26_{-0.05}^{+0.06}}\xspace \\
 & THM & \ensuremath{12.4_{-0.3}^{+0.2}}\xspace & \ensuremath{22.22\pm 0.06}\xspace & \ensuremath{0.40_{-0.08}^{+0.10}}\xspace \\
 & XP  & \ensuremath{12.5_{-0.3}^{+0.2}}\xspace & \ensuremath{22.60\pm 0.07}\xspace & \ensuremath{0.95_{-0.18}^{+0.23}}\xspace \\
 & XPHM & \ensuremath{12.6_{-0.3}^{+0.3}}\xspace & \ensuremath{22.62\pm 0.07}\xspace & \ensuremath{1.00}\xspace \\[6pt] 
 
  \multirow{6}{*}{GW170104} & XAS & \ensuremath{13.9_{-0.2}^{+0.1}}\xspace & \ensuremath{32.10\pm 0.06}\xspace & \ensuremath{0.79_{-0.14}^{+0.17}}\xspace\\
 & XHM & \ensuremath{13.9_{-0.2}^{+0.1}}\xspace & \ensuremath{32.11\pm 0.06}\xspace & \ensuremath{0.81_{-0.14}^{+0.18}}\xspace\\
 & THM & \ensuremath{13.9_{-0.2}^{+0.1}}\xspace & \ensuremath{31.95\pm 0.06}\xspace & \ensuremath{0.57_{-0.10}^{+0.12}}\xspace\\
 & XP  & \ensuremath{13.9_{-0.2}^{+0.1}}\xspace & \ensuremath{32.28\pm 0.06}\xspace & \ensuremath{1.21_{-0.21}^{+0.26}}\xspace\\
 & XPHM & \ensuremath{14.0_{-0.2}^{+0.2}}\xspace & \ensuremath{32.20\pm 0.06}\xspace & \ensuremath{1.00}\xspace\\[6pt] 
 
  \multirow{6}{*}{GW170608} & XAS & \ensuremath{15.5_{-0.2}^{+0.1}}\xspace & \ensuremath{39.27\pm 0.07}\xspace & \ensuremath{0.93_{-0.18}^{+0.23}}\xspace\\
 & XHM & \ensuremath{15.5_{-0.2}^{+0.1}}\xspace & \ensuremath{39.24\pm 0.07}\xspace & \ensuremath{0.86_{-0.17}^{+0.21}}\xspace\\
 & THM & \ensuremath{15.5_{-0.2}^{+0.1}}\xspace & \ensuremath{39.33\pm 0.07}\xspace & \ensuremath{1.06_{-0.21}^{+0.26}}\xspace\\
 & XP  & \ensuremath{15.5_{-0.2}^{+0.1}}\xspace & \ensuremath{39.29\pm 0.07}\xspace & \ensuremath{0.96_{-0.19}^{+0.24}}\xspace\\
 & XPHM & \ensuremath{15.5_{-0.2}^{+0.2}}\xspace & \ensuremath{39.31\pm 0.07}\xspace & \ensuremath{1.00}\xspace\\[6pt] 
 
  \multirow{6}{*}{GW170729} & XAS & \ensuremath{10.7_{-0.3}^{+0.3}}\xspace & \ensuremath{15.97\pm 0.06}\xspace & \ensuremath{0.64_{-0.11}^{+0.14}}\xspace\\
 & T & \ensuremath{10.7_{-0.3}^{+0.3}}\xspace & \ensuremath{16.12\pm 0.06}\xspace & \ensuremath{0.93_{-0.16}^{+0.20}}\xspace\\
 & XHM & \ensuremath{10.9_{-0.4}^{+0.3}}\xspace & \ensuremath{16.32\pm 0.06}\xspace & \ensuremath{1.47_{-0.26}^{+0.31}}\xspace\\
 & THM & \ensuremath{10.9_{-0.4}^{+0.3}}\xspace & \ensuremath{16.35\pm 0.06}\xspace & \ensuremath{1.56_{-0.28}^{+0.34}}\xspace\\
 & XP  & \ensuremath{10.7_{-0.3}^{+0.3}}\xspace & \ensuremath{16.03\pm 0.06}\xspace & \ensuremath{0.74_{-0.13}^{+0.16}}\xspace\\
 & TP & \ensuremath{10.8_{-0.3}^{+0.3}}\xspace	& \ensuremath{16.13\pm 0.06}\xspace & \ensuremath{0.94_{-0.17}^{+0.20}}\xspace\\
  & TPHM & \ensuremath{10.9_{-0.4}^{+0.3}}\xspace & \ensuremath{16.39\pm 0.06}\xspace & \ensuremath{2.16_{-0.39}^{+0.47}}\xspace\\
 & XPHM & \ensuremath{10.8_{-0.3}^{+0.3}}\xspace & \ensuremath{16.16\pm 0.06}\xspace & \ensuremath{1.00}\xspace\\[6pt]
 
  \multirow{6}{*}{GW170809} & XAS & \ensuremath{12.6_{-0.2}^{+0.2}}\xspace & \ensuremath{24.12\pm 0.06}\xspace & \ensuremath{0.97_{-0.17}^{+0.21}}\xspace\\
 & XHM & \ensuremath{12.6_{-0.2}^{+0.2}}\xspace & \ensuremath{24.10\pm 0.06}\xspace & \ensuremath{0.93_{-0.17}^{+0.21}}\xspace\\
 & THM & \ensuremath{12.6_{-0.2}^{+0.2}}\xspace & \ensuremath{23.95\pm 0.06}\xspace & \ensuremath{0.66_{-0.12}^{+0.15}}\xspace\\
 & XP  & \ensuremath{12.6_{-0.2}^{+0.2}}\xspace & \ensuremath{24.13\pm 0.06}\xspace & \ensuremath{1.00_{-0.18}^{+0.22}}\xspace\\
 & TPHM & \ensuremath{12.6_{-0.2}^{+0.2}}\xspace & \ensuremath{23.79\pm 0.06}\xspace & \ensuremath{0.45_{-0.08}^{+0.10}}\xspace\\
 & XPHM & \ensuremath{12.6_{-0.3}^{+0.2}}\xspace & \ensuremath{24.13\pm 0.06}\xspace & \ensuremath{1.00}\xspace\\[6pt] 
 
  \multirow{6}{*}{GW170814} & XAS & \ensuremath{17.4_{-0.2}^{+0.1}}\xspace & \ensuremath{53.73\pm 0.07}\xspace & \ensuremath{1.38_{-0.27}^{+0.33}}\xspace\\
 & XHM & \ensuremath{17.5_{-0.2}^{+0.1}}\xspace & \ensuremath{53.69\pm 0.07}\xspace & \ensuremath{1.25_{-0.24}^{+0.30}}\xspace\\
 & THM & \ensuremath{17.4_{-0.2}^{+0.1}}\xspace & \ensuremath{53.57\pm 0.07}\xspace & \ensuremath{0.95_{-0.18}^{+0.23}}\xspace\\
 & XP  & \ensuremath{17.4_{-0.2}^{+0.1}}\xspace & \ensuremath{53.62\pm 0.07}\xspace & \ensuremath{1.06_{-0.21}^{+0.26}}\xspace\\
 & TPHM & \ensuremath{17.5_{-0.2}^{+0.2}}\xspace & \ensuremath{54.14\pm 0.07}\xspace & \ensuremath{3.48_{-0.68}^{+0.85}}\xspace\\
 & XPHM* & \ensuremath{17.5_{-0.2}^{+0.1}}\xspace & \ensuremath{53.89\pm 0.07}\xspace & \ensuremath{1.97_{-0.39}^{+0.48}}\xspace\\
 & XPHM & \ensuremath{17.5_{-0.2}^{+0.1}}\xspace & \ensuremath{53.60\pm 0.07}\xspace & \ensuremath{1.00}\xspace\\[6pt] 
 
  \multirow{6}{*}{GW170818} & XAS & \ensuremath{11.7_{-0.3}^{+0.2}}\xspace & \ensuremath{19.91\pm 0.06}\xspace & \ensuremath{0.25_{-0.05}^{+0.06}}\xspace\\
 & XHM & \ensuremath{11.8_{-0.3}^{+0.2}}\xspace & \ensuremath{20.11\pm 0.06}\xspace & \ensuremath{0.39_{-0.07}^{+0.09}}\xspace\\
 & THM & \ensuremath{11.8_{-0.3}^{+0.2}}\xspace & \ensuremath{19.97\pm 0.06}\xspace & \ensuremath{0.29_{-0.05}^{+0.07}}\xspace\\
 & XP  & \ensuremath{11.9_{-0.3}^{+0.2}}\xspace & \ensuremath{20.27\pm 0.07}\xspace & \ensuremath{0.57_{-0.11}^{+0.13}}\xspace\\
 & TPHM & \ensuremath{11.9_{-0.3}^{+0.2}}\xspace & \ensuremath{20.32\pm 0.06}\xspace & \ensuremath{0.63_{-0.12}^{+0.15}}\xspace\\
 & XPHM* & \ensuremath{12.0_{-0.3}^{+0.2}}\xspace & \ensuremath{20.73\pm 0.07}\xspace & \ensuremath{1.64_{-0.31}^{+0.39}}\xspace\\
 & XPHM & \ensuremath{11.9_{-0.3}^{+0.2}}\xspace & \ensuremath{20.51\pm 0.06}\xspace &\ensuremath{1.00}\xspace\\[6pt] 
 
 \multirow{6}{*}{GW170823} & XAS & \ensuremath{12.0_{-0.2}^{+0.1}}\xspace & \ensuremath{22.91\pm 0.05}\xspace & \ensuremath{0.78_{-0.13}^{+0.15}}\xspace\\
 & XHM & \ensuremath{12.0_{-0.2}^{+0.1}}\xspace & \ensuremath{22.90\pm 0.06}\xspace & \ensuremath{0.77_{-0.13}^{+0.15}}\xspace\\
 & THM & \ensuremath{12.0_{-0.2}^{+0.1}}\xspace & \ensuremath{22.81\pm 0.06}\xspace & \ensuremath{0.63_{-0.10}^{+0.12}}\xspace\\
 & XP  & \ensuremath{12.0_{-0.2}^{+0.1}}\xspace & \ensuremath{23.07\pm 0.05}\xspace & \ensuremath{1.13_{-0.19}^{+0.22}}\xspace\\
 & TPHM & \ensuremath{12.0_{-0.2}^{+0.2}}\xspace & \ensuremath{22.73\pm 0.06}\xspace & \ensuremath{0.52_{-0.09}^{+0.10}}\xspace\\ 
 & XPHM & \ensuremath{12.0_{-0.2}^{+0.1}}\xspace & \ensuremath{23.01\pm 0.06}\xspace & \ensuremath{1.00}\xspace\\

 \hline\hline
\end{tabular}
%\end{ruledtabular}
}
\end{table}

\subsection{Low-mass events}\label{sec:lowmass}

To leading order in the post-Newtonian expansion the duration $\Delta T$ of the inspiral starting at a frequency $f_0$ is 
$$
\Delta T= \frac{5}{256 f_0 \eta \pi ^{8/3} }
\left(\frac{f_0 G M}{c^3}\right)^{-5/3},
$$
where $M$ is the total mass of the system, $\eta = m_1 m_2/(m_1+m_2)^2$ is the symmetric mass ratio, $G$ is Newton's constant, and $c$ is the speed of light. The observable part of the waveform is thus significantly longer for lower-mass systems.

In general, the chirp mass is the best measured intrinsic parameter \citep{Cutler:1994ys}, since it is the quantity that appears in the PN waveform at leading order \citep{Blanchet:2006zz}. 

In Fig. \ref{fig:low_mass_collage}, we compare several parameter estimation runs for GW151012, GW151226 and GW170608 with the standard settings mentioned in Sec. \ref{sec:sampling}: using distance marginalization, \mbox{{\tt nlive} $=2048$} and \mbox{{\tt nact} $=30$}.
We compare the whole \phenX family (the aligned-spin, precessing, dominant and subdominant mode models), the more recent aligned-spin time domain model \phTHM with higher-modes content, and the previous \phPvtwo results taken from the fist catalog \citep{LIGOScientific:2018mvr}.
As mentioned in Sec.~\ref{sec:data}, for these low-mass events we use a sampling rate of 4096\,Hz and the GWOSC 16\,kHz data sets, and PSDs and calibration envelopes files which go up to 2048\,Hz.
All models show good agreement in total mass and $\chieff$. 
GW151012 and GW151226 are the events with the largest uncertainty in mass ratio.

We note that the initial preprint version of \citet{Chia:2021mxq} and also \citet{Nitz:2021uxj} have reported results for GW151226 that are in tension with ours, including multi-modal posteriors.
In order to further increase the confidence in our results
we have therefore also performed tests with a different sampling method in Bilby and with the independent LALInference code for the three lowest mass events. These checks confirm our results as discussed in Appendix \ref{subsec:convergence_test}. We also show how lowering the sampling rate can result in bi-modal posteriors.
As discussed in the appendix, the remaining differences between our results and the published results of  \citet{Chia:2021mxq}, as well as the results of 
\citet{Nitz:2021uxj}, seem to be largely consistent with differences in prior choices.

\subsubsection{GW151012}\label{sec:GW151012}

This is the low-mass event with the smallest SNR ($\ensuremath{9.3_{-0.4}^{+0.3}}\xspace$ for \phXPHM) and Bayes factor, see Table \ref{tab:SNR_BF_table}.
Apart from GW151226, this is the only event where we find some visual differences in the mass posteriors between our results and those of \citet{Nitz:2021uxj}.
\citet{Nitz:2021uxj} recover slightly more unequal masses posteriors in their results but consistent with our own results.

From our analysis, Bayes factor differences for different waveform models are within the error estimates, except for a small suppression of \phX and \phXP in comparison to \phXPHM of $\ensuremath{0.58_{-0.10}^{+0.12}}\xspace$ and $\ensuremath{0.77_{-0.13}^{+0.15}}\xspace$ respectively.
As can be seen in Fig.~\ref{fig:low_mass_collage}, the models with higher modes shift the mass ratio posterior toward more unequal masses. Runs with our two different precession versions (NNLO and MSA angles) and with different final spin descriptions show very consistent posterior distributions and Bayes factors that are consistent within the error estimates.

\subsubsection{GW151226}\label{sec:GW151226}

\begin{figure*} 
 \includegraphics[width=2\columnwidth]{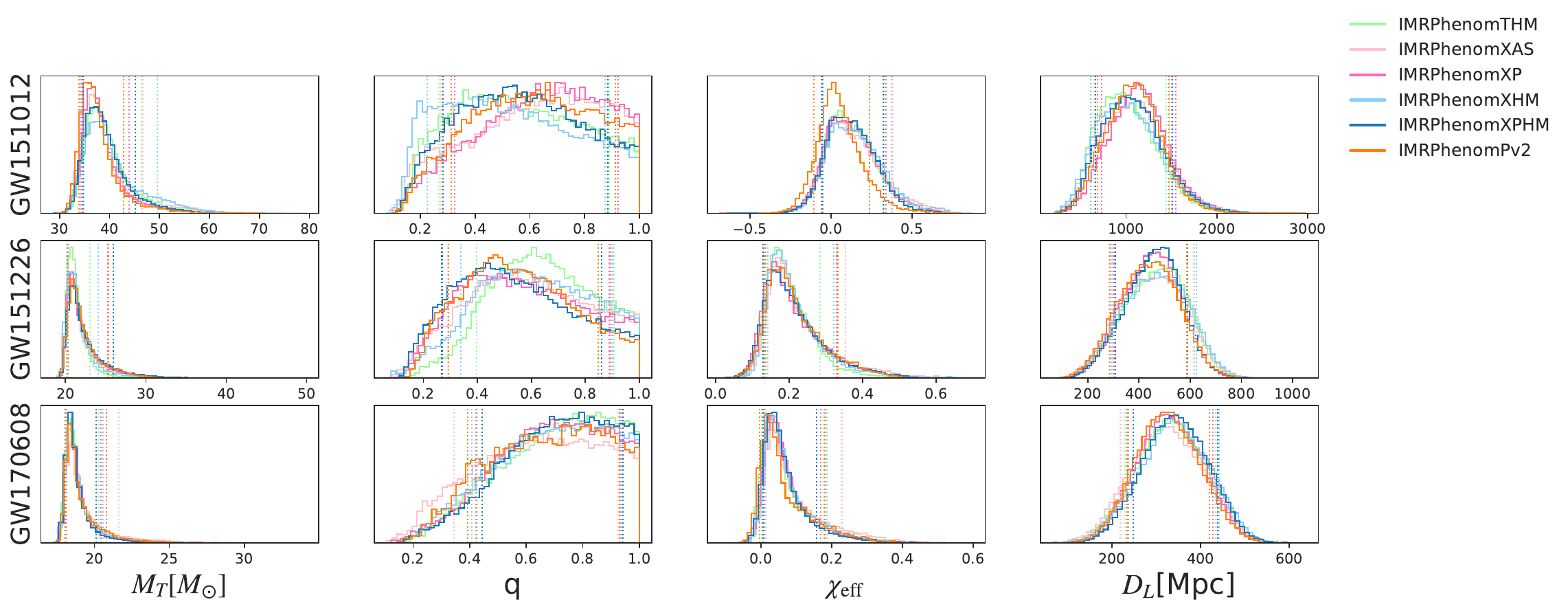}\\
 \caption{Posterior distributions for total mass, mass ratio, effective spin and luminosity distance for all three low-mass events and different waveform models. \label{fig:low_mass_collage}
 }
\end{figure*}

\begin{figure} 
    \centering
    \includegraphics[width=0.9\columnwidth]{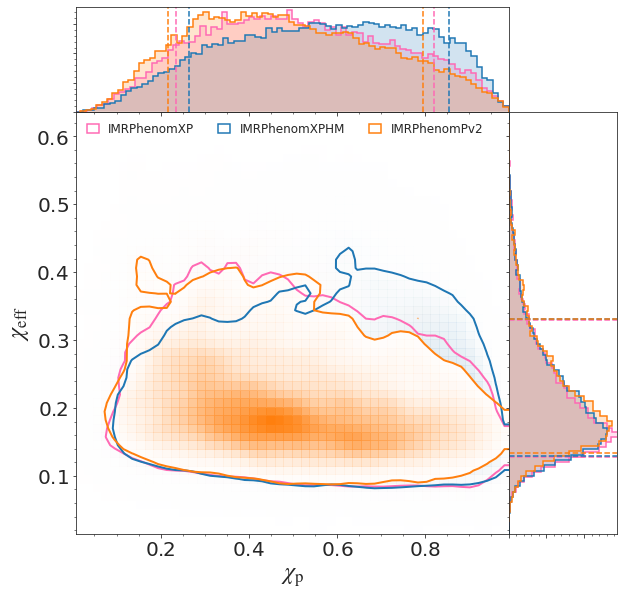}
    \caption{Joint posterior distributions for $\chi_\mathrm{p}$ and $\chieff$ of GW151226, estimated with \phXP (pink), \phXPHM (blue) and \phPvtwo (orange) \citep{LIGOScientific:2018mvr}.
    Here and in similar figures throughout the paper, the central panel shows the 2D joint posterior with contours marking 90\% credible intervals, while the smaller panels on top and to the right show the corresponding 1D distributions for the individual parameters, with the 90\% credible interval indicated by the dashed lines.
    \label{fig:gw151226_chip_chieff}
    }
\end{figure}

This event shows support only for positive values of $\chieff$, and very mild support for precession, with  a Bayes factor of $3.85_{-0.74}^{+0.92}$ between \phXHM and \phXPHM in favor of precession, see Table \ref{tab:tabBF_prec}. Some differences are seen in Fig.~\ref{fig:low_mass_collage} regarding the mass ratio posterior, and in particular \phTHM shifts the mass ratio support toward more equal masses, compared with the other models.
Fig. \ref{fig:gw151226_chip_chieff} also shows that adding higher mode content shifts the precession parameter $\chip$ toward larger values, but again all results are consistent within error estimates.
See also Appendix \ref{subsec:convergence_test} for further consistency tests and comparisons with other results on this event.

\begin{table}

\begin{tabular}{cccc}
\hline\hline
 & GW151226 & GW170814 & GW170818   \\
\hline

 XP vs XAS & \ensuremath{3.63_{-0.69}^{+0.86}}\xspace & \ensuremath{0.77_{-0.15}^{+0.18}}\xspace  & \ensuremath{2.27_{-0.43}^{+0.53}}\xspace  \\
 XPHM vs XHM & \ensuremath{3.85_{-0.74}^{+0.92}}\xspace & \ensuremath{0.80_{-0.16}^{+0.19}}\xspace  & \ensuremath{2.54_{-0.48}^{+0.60}}\xspace  \\
  TPHM vs THM & - & \ensuremath{3.66_{-0.71}^{+0.89}}\xspace  & \ensuremath{2.22_{-0.42}^{+0.52}}\xspace \\
  TPHM vs XPHM & - & \ensuremath{3.48_{-0.68}^{+0.85}}\xspace  & \ensuremath{0.63_{-0.12}^{+0.15}}\xspace \\
  TPHM vs XPHM* & - & \ensuremath{1.76_{-0.35}^{+0.43}}\xspace & \ensuremath{0.38_{-0.07}^{+0.09}}\xspace\\
  XPHM* vs XHM & - & \ensuremath{1.58_{-0.31}^{+0.38}}\xspace & \ensuremath{4.18_{-0.80}^{+0.98}}\xspace \\
  XPHM* vs XPHM & - & \ensuremath{1.97_{-0.39}^{+0.48}}\xspace & \ensuremath{1.64_{-0.31}^{+0.39}}\xspace \\
 
\hline\hline
\end{tabular}
%\end{ruledtabular}
\caption{
\label{tab:tabBF_prec}
Comparison of Bayes factors between precessing and aligned-spin \phenX and \phT waveform models for GW151226, GW170814 and GW170818.}
\end{table}

\subsubsection{GW170608}\label{sec:GW170608}

GW170608 is the event with the lowest mass, the closest luminosity distance, and with the highest SNR of the low-mass events. Bayes factors for all waveform models are consistent within the error estimates, and posterior distributions are visually very similar, including for all versions of \phXPHM which we have run (MLA and MSA Euler angles and the four final spin versions for each angle description).
JS divergences between the various posteriors back up this finding,
with the divergences for all pairs of waveforms and most parameters below 0.02.
Only the divergences for spin magnitudes between precessing and aligned-spin waveforms,
which are expected to differ more,
reach higher values
(up to 0.16 between \phXPHM and \phTHM).

\subsection{Medium-mass events}\label{sec:intermass}

\begin{figure*} 
 \includegraphics[width=2\columnwidth]{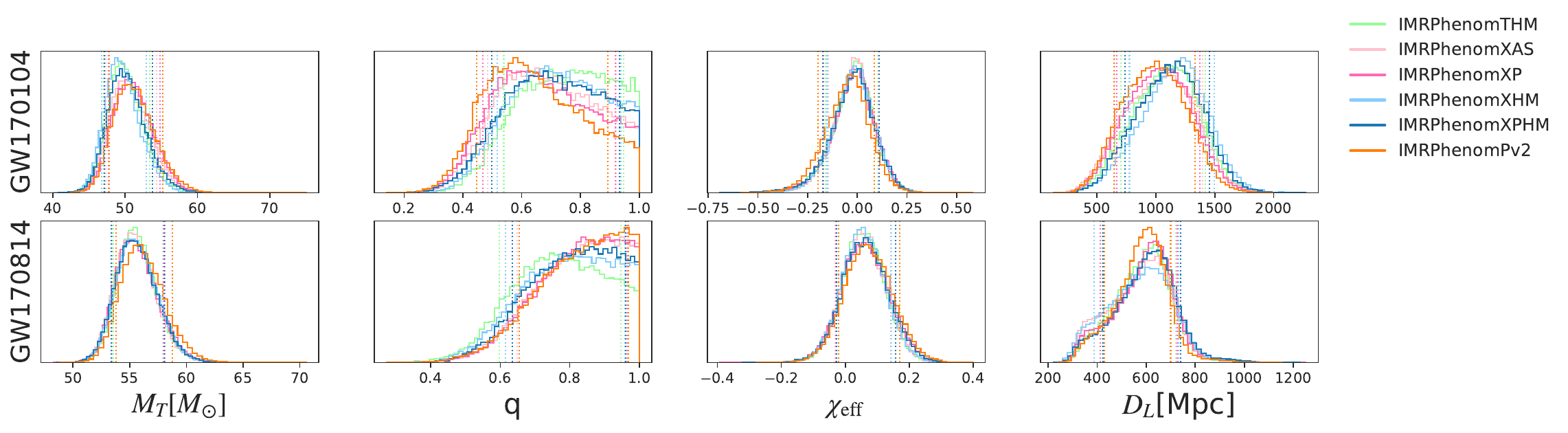}\\
 \caption{Posterior distributions for total mass, mass ratio, effective spin and luminosity distance are compared for the medium-mass events and different waveform models.
  \label{fig:inter_mass_collage}
 }
\end{figure*}

Fig.~\ref{fig:inter_mass_collage} provides a comparison of posteriors for several key quantities for the two medium-mass events GW170104 and GW170814. For these, all different models recover essentially the same total mass and $\chieff$ posterior distributions. We discuss further details below.

\subsubsection{GW170104}\label{sec:GW170104}

This is the first event of the O2 observing run, and the fourth loudest BBH event at an SNR of $\ensuremath{14.0_{-0.2}^{+0.2}}\xspace$ for all the waveform models that we have considered. Bayes factors for different models are consistent within error estimates, apart from the Bayes factor of \phTHM, which is suppressed relative to \phXPHM by $0.57^{+0.12}_{-0.10}$, see Table \ref{tab:SNR_BF_table}. (\phTPHM was only run for higher-mass events.) The higher-mode models shift the posterior marginally toward more equal masses and larger distance.
We also find that \phXPHM versions give consistent results regarding posterior distributions and Bayes factors, and we do not list these results separately but rather refer to our data release for full details.

\subsubsection{GW170814}\label{sec:GW170814}

GW170814 is the second loudest BBH event with an SNR of $\ensuremath{17.5_{-0.2}^{+0.1}}\xspace$, and a well recovered sky position due to being observed by both LIGO (Hanford and Livingston) detectors and the Virgo detector.
Indeed GW170814 was the first BBH published as a coincident observation between the three LIGO--Virgo detectors \citep{Abbott:2017oio}.

Apart from the effective precession parameter $\chi_\mathrm{p}$, the posterior distributions for different waveform models are rather similar for other quantities, with only marginal differences for masses and mass ratio. Just as GW151226, and to a lesser degree GW170818, GW170814 does however show marginal support for precession, as expressed by the Bayes factors listed in Table \ref{tab:tabBF_prec}, and it is an interesting case from the point of view of waveform systematics.
GW170814 shows the largest value of the JS divergence when comparing our \phXPHM results to the GWTC-1 \phPvtwo results, corresponding to $JS_{\chi_\mathrm{p}}=0.042$ for the effective in-plane spin $\chi_\mathrm{p}$. The default version of \phXPHM in fact recovers a lower value of $\chi_\mathrm{p}$ than \phPvtwo, see Table \ref{tab:PEresults}. However it turns out that when using the NNLO Euler angles rather than the MSA angles for \phXPHM, a larger value of $\chi_\mathrm{p} = 0.44_{-0.30}^{+0.33}$ is recovered, and also a larger Bayes factor of $1.97_{-0.39}^{+0.48}$ with respect to the default \phXPHM version (see Tables \ref{tab:PEresults} and \ref{tab:SNR_BF_table}).
Indeed both \phX and \phXHM recover marginally better Bayes factors than the default version of \phXPHM.
Also \phTPHM has a higher Bayes factor, $3.48_{-0.68}^{+0.85}$ against the default version of \phXPHM and $1.76_{-0.35}^{+0.43}$ against the NNLO angles version of \phXPHM. \phTPHM also recovers a higher value of $\chi_\mathrm{p} = 0.58_{-0.35}^{+0.28}$, very close to the value for \phPvtwo, even though the non-precessing time-domain model \phTHM is marginally disfavored when comparing to \phXHM, with a Bayes factor ratio of about $1.3$ in Table \ref{tab:SNR_BF_table}.
These results suggest that this event is a case where the default choice for modelling precession in \phXPHM, the MSA description of the precession Euler angles, is less accurate than the NNLO description, and thus an improved precession treatment for the frequency domain is needed to obtain robust results for this event. Furthermore
\phTPHM might recover an even higher Bayes factor, were it not for the shortcomings of the model in the inspiral, already in the absence of precession, and it will thus be interesting to re-analyze this event again in the future with improved waveform models.

\begin{figure*} 
    \includegraphics[width=0.91\columnwidth]{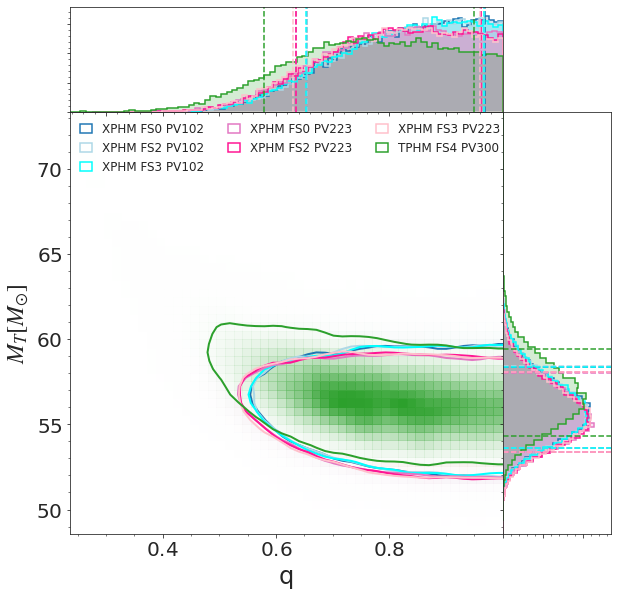}
    \includegraphics[width=0.91\columnwidth]{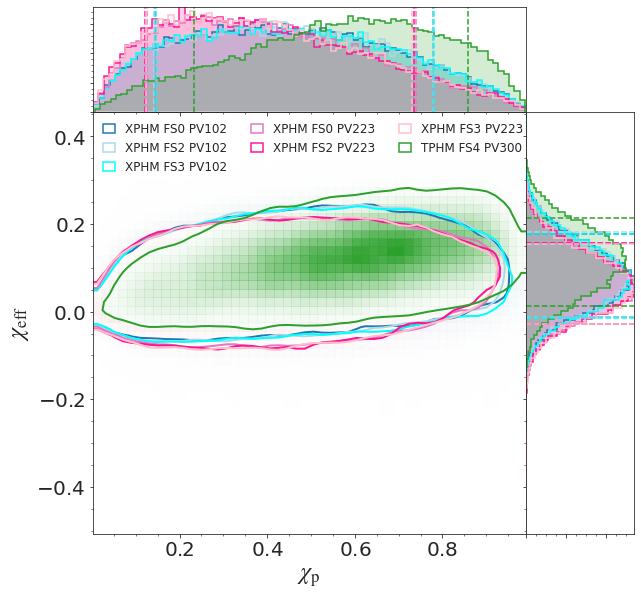}
        \includegraphics[width=0.91\columnwidth]{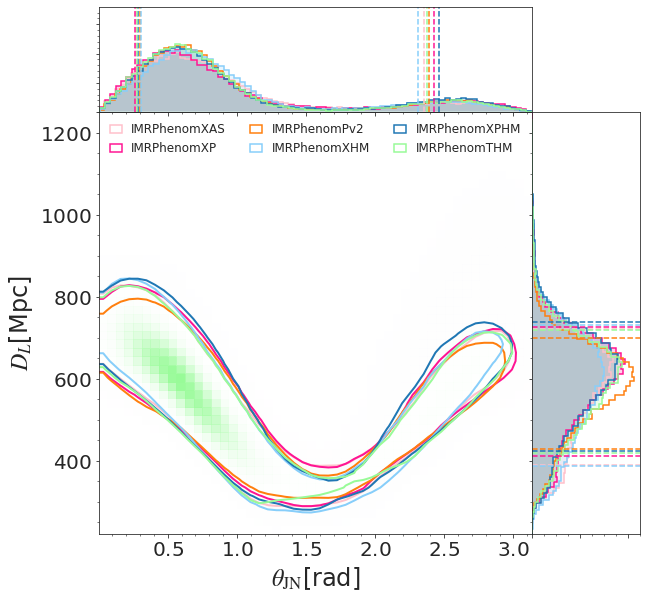}
        \includegraphics[width=0.91\columnwidth]{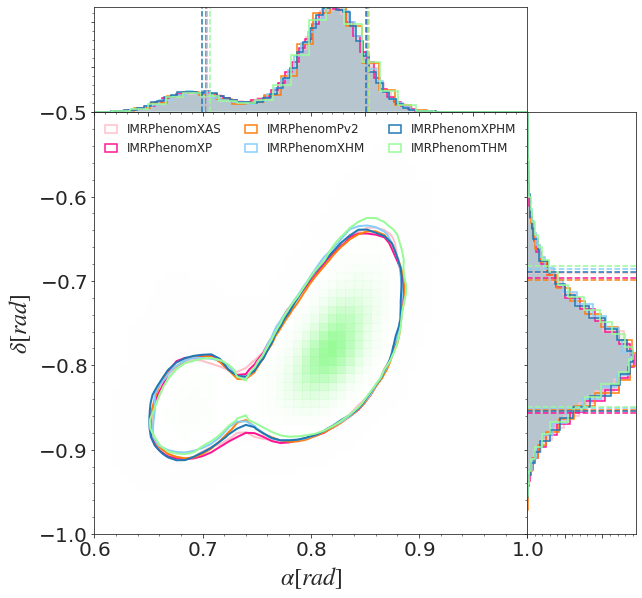}
    \caption{Comparison of posteriors for GW170814, indicating 90\% credible intervals. Upper panel: total mass, mass ratio, and spin parameters for the MSA (blue) and NNLO (pink) versions of \phXPHM and for \phTPHM (green).
Lower panel: distance and inclination (left) and sky position (right) are shown for the dominant-mode models \phX, \phXP (light and dark pink) and \phPvtwo (orange), and the multi-mode models \phXHM, \phXPHM (light and dark blue) and \phTHM (green).
    \label{fig:gw170814_XPHM_versions}
    }
\end{figure*}

Comparisons of posterior distributions for the mass ratio, total mass, $\chieff$ and $\chi_\mathrm{p}$ for this event using our different precessing waveform models are shown in Fig. \ref{fig:gw170814_XPHM_versions}.

\subsection{High-mass events}\label{sec:highmass}

\begin{figure*} 
 \includegraphics[width=2\columnwidth]{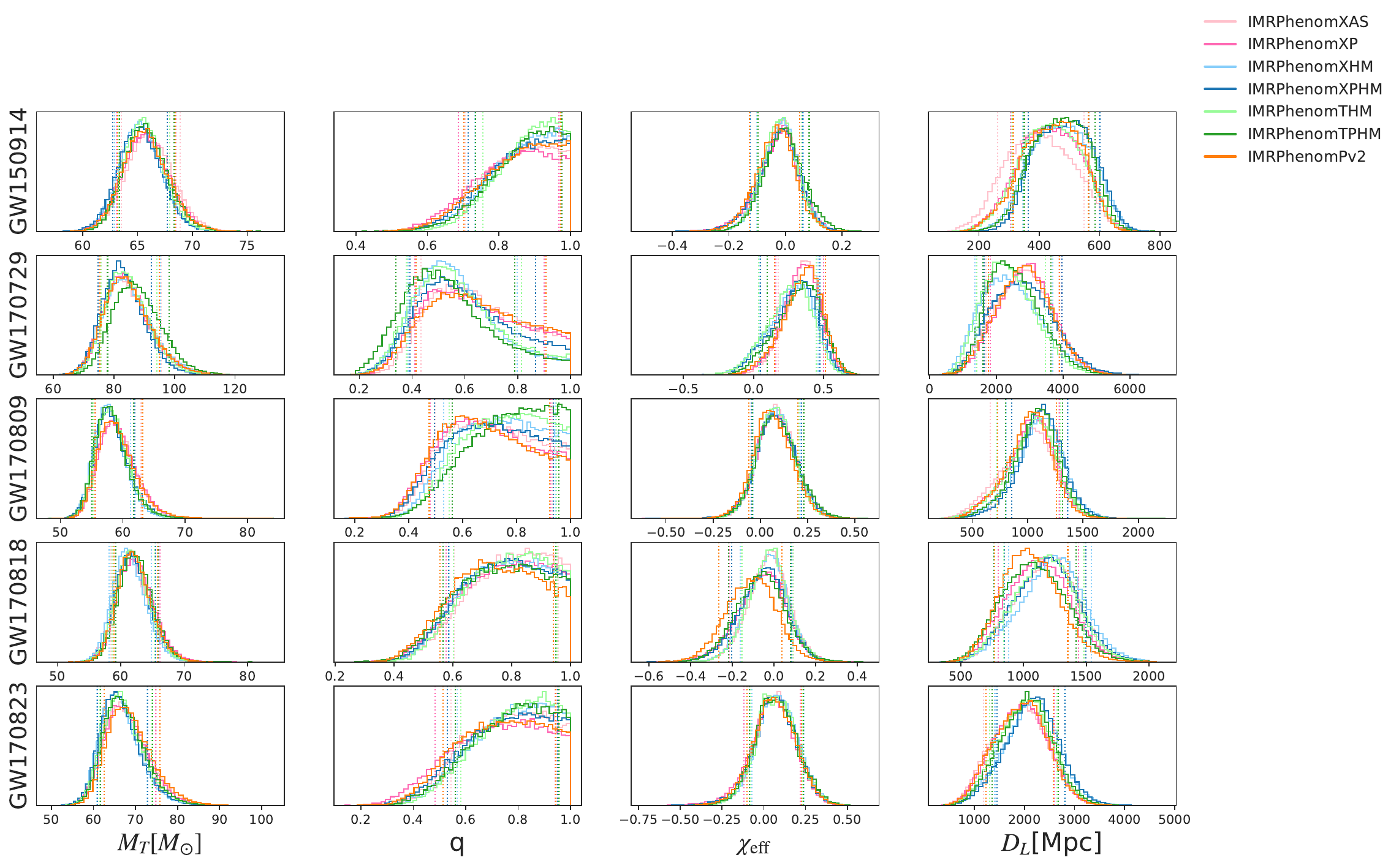}\\
 \caption{Posteriors for total mass, mass ratio, $\chieff$ and luminosity distance are shown for high-mass events and different waveform models.
  \label{fig:high_mass_collage}
 }
\end{figure*}

Fig.~\ref{fig:high_mass_collage} shows a comparison of posteriors for several key quantities. For high mass events, we also add the \phTPHM waveform model to our comparisons.

\subsubsection{GW150914}\label{sec:GW150914}

GW150914 was the first detection of a GW signal~\citep{Abbott:2016blz} and has the highest SNR of events consistent with BBH mergers to date (in the O1, O2 \citep{LIGOScientific:2018mvr} and O3a \citep{Abbott:2020niy} observing runs).

Bayes factors in Table \ref{tab:SNR_BF_table} are consistent between different models within error estimates, apart from 
\phTHM and \phTPHM, which are disfavored by factors of $0.57_{-0.11}^{+0.14}$ (THM) and $0.61_{-0.12}^{+0.15}$ (TPHM) with respect to the default \phXPHM version. Posterior results show only very marginal deviations, see Fig.~\ref{fig:high_mass_collage} and Tables \ref{tab:PEresults}, \ref{tab:SNR_BF_table}.
The same holds for all the eight versions of \phXPHM we have tested (NNLO and MSA Euler angles and four final spin versions for each angle description): the default version of \phXPHM has the largest Bayes factor, and the most disfavored version of \phXPHM (NNLO angles with final spin version FS2) has a Bayes factor of $0.69_{-0.13}^{+0.17}$ with respect to the default version, which is still broadly consistent within the error estimates. 
Posteriors between different \phXPHM versions are in general very similar and without noteworthy changes.

\subsubsection{GW170729}\label{sec:GW170729}

This is the most massive and most distant BBH merger detected during O1 and O2, with substantial support for unequal masses \citep{LIGOScientific:2018mvr}.
In consequence, a first set of parameter estimation runs with different waveform models with higher-mode content was performed in \citet{Chatziioannou:2019dsz}.
It is thus particularly interesting to compare our results with the \phenX and \phT families
against these earlier results, and also against those obtained by \citet{Payne:2019wmy} through approximate reweighting of \phD results to the \NRHybSur target waveform.
In addition, we test how posterior distributions for this event change with different mass and distance priors.
GW170729 is also interesting since its recovered effective spin is positive within error estimates.
Unfortunately the SNR is relatively low, at $\ensuremath{10.8_{-0.3}^{+0.3}}\xspace$ for \phXPHM and $\ensuremath{10.9_{-0.4}^{+0.3}}\xspace$ for \phTPHM.
Only GW151012 has lower SNR within our set of ten events. 
Consequently, posterior distributions for GW170729 are relatively broad.
We find that all eight versions of \phXPHM yield Bayes factors and posterior distributions that are consistent within error estimates.

The more massive component of the binary, with central values just above $50,\Msun$ from various posteriors, see Table \ref{tab:PEresults}, is situated near the lower end of the pair-instability supernova (PISN) mass gap.
That hypothetical gap is placed between approximately $52$ and $133$ solar masses by \citet{Woosley:2016hmi}, with the edges however uncertain --
see \citet{Woosley:2021xba} for a more recent overview.
\phTPHM, which recovers the largest Bayes factor,
also recovers the highest value for the larger mass, $m_1 = 57.3^{12.0}_{-10.9}\,M_\odot$, potentially situating it beyond the onset of the mass gap.

\emph{a. \hspace{5pt} Higher-mode content:} \hspace{5pt}  As demonstrated by the Bayes factors in Table \ref{tab:tabBF}, we find weak evidence of higher-order mode content in this event.

These Bayes factors in favour of higher modes are larger
(in the range 1.9--2.9 including uncertainties)
for aligned-spin \phenX and for precessing \phT models,
but below 2 for precessing \phenX and aligned-spin \phT models.
This different behaviour of the two model families is consistent
with \phenX having a more accurate aligned-spin sector
but \phT providing improved treatment of precession as described in Sec.~\ref{sec:waveforms}.
In either case, our Bayes factors are lower than the value of $5.1$ found by \citet{Chatziioannou:2019dsz} between the older aligned-spin models \phHM and \phD,
but the aligned-spin \phenX and precessing \phT values are slightly larger than the value of 1.4 obtained by \citet{LIGOScientific:2018mvr} using the RapidPE method \citep{Pankow:2015cra,Lange:2018pyp} with a catalog of NR and \NRSur \citep{Blackman:2017pcm} waveforms.
We thus find that when using improved waveform models 
the Bayes factors are more consistent with the RapidPE method than what was found by \citet{Chatziioannou:2019dsz}, and the Bayes factor in favor of higher-mode content is lower.

\begin{figure*} 
    \includegraphics[width=0.9\columnwidth]{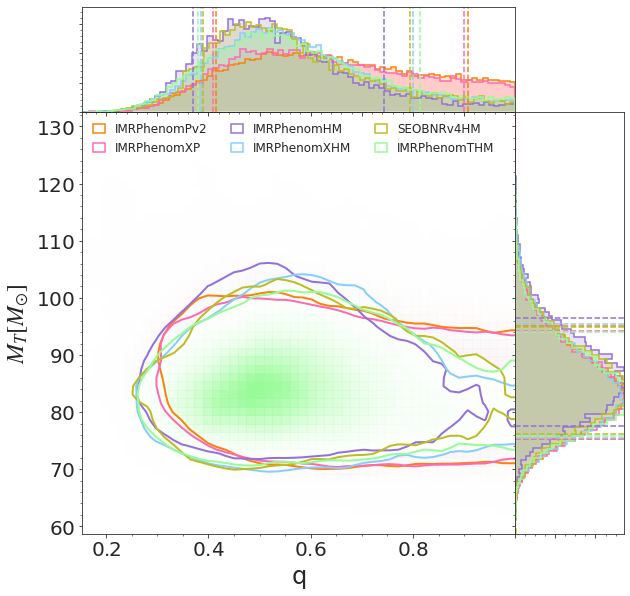}\includegraphics[width=0.925\columnwidth]{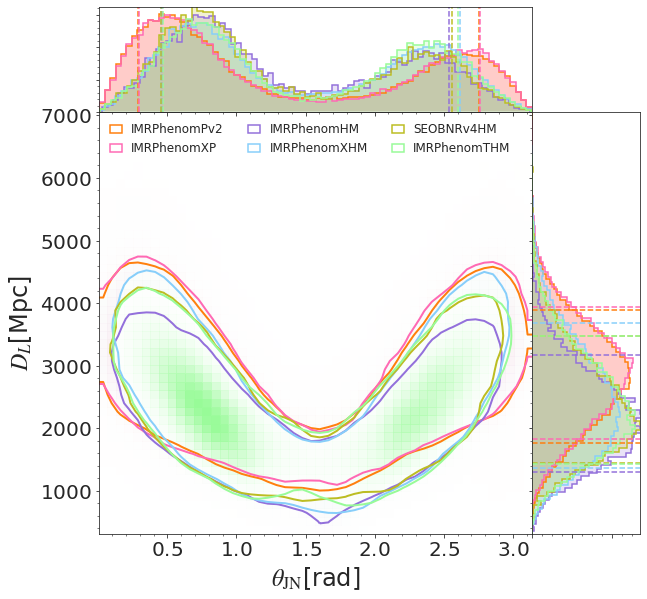}\\\includegraphics[width=0.9\columnwidth]{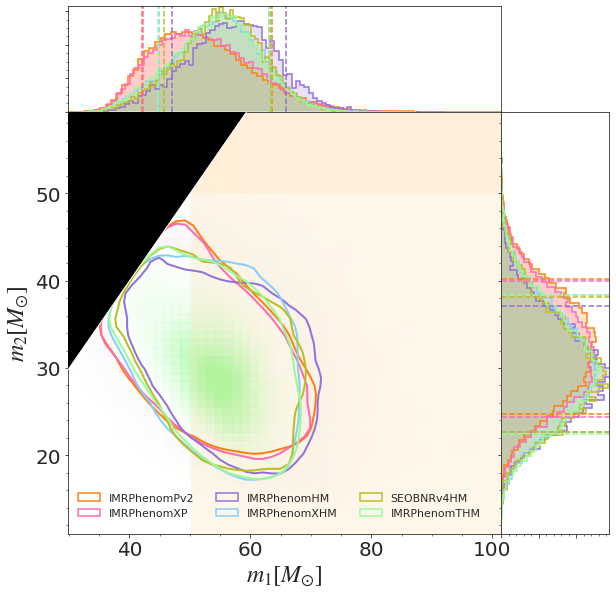}\includegraphics[width=0.9\columnwidth]{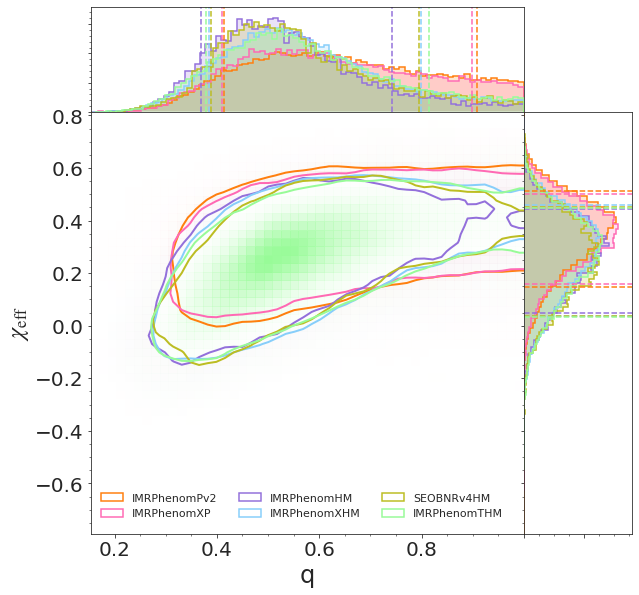}]
    \caption{Posterior distributions of GW170729 for the total mass, mass ratio, luminosity distance, inclination, component masses, mass ratio and effective spin, using the dominant-mode models \phPvtwo \citep[orange,][]{LIGOScientific:2018mvr} and \phXP (light pink), the subdominant-mode models \phHM (purple), \seobnrvforhm \citep[olive green,][]{Chatziioannou:2019dsz}, \phXHM (light blue) and  \phTHM (light green).
    All contours are at 90\% credible intervals.
    \label{fig:gw170729_old}
    }
\end{figure*}

\begin{figure*} 
    \includegraphics[width=0.9\columnwidth]{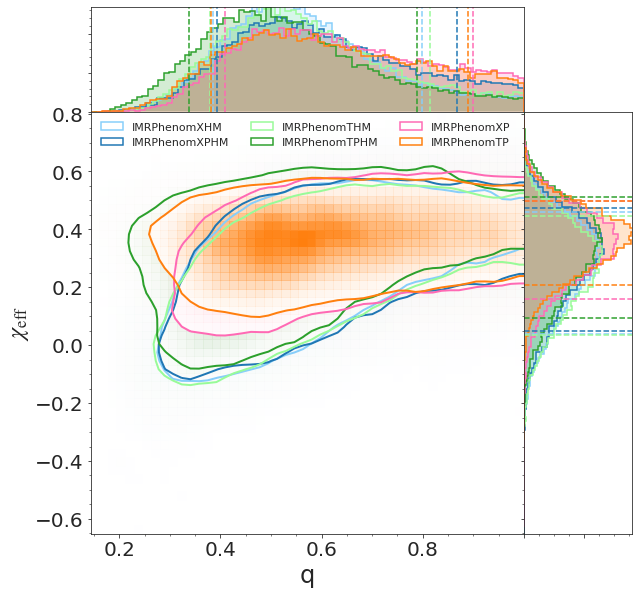}\includegraphics[width=0.9\columnwidth]{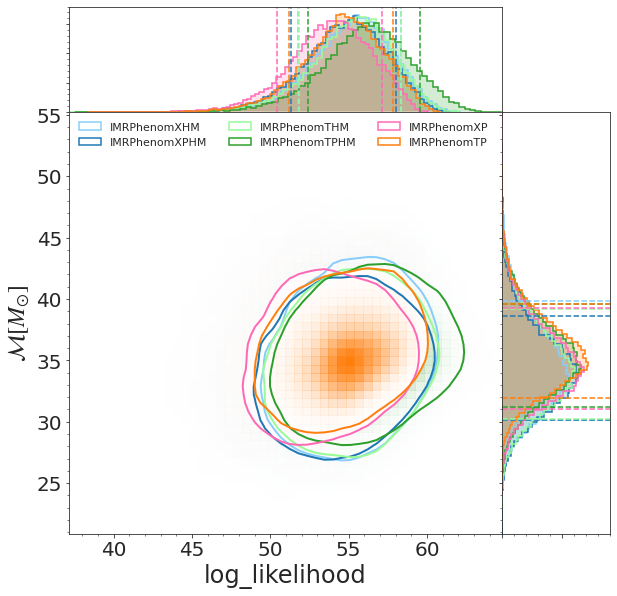}\\\includegraphics[width=0.9\columnwidth]{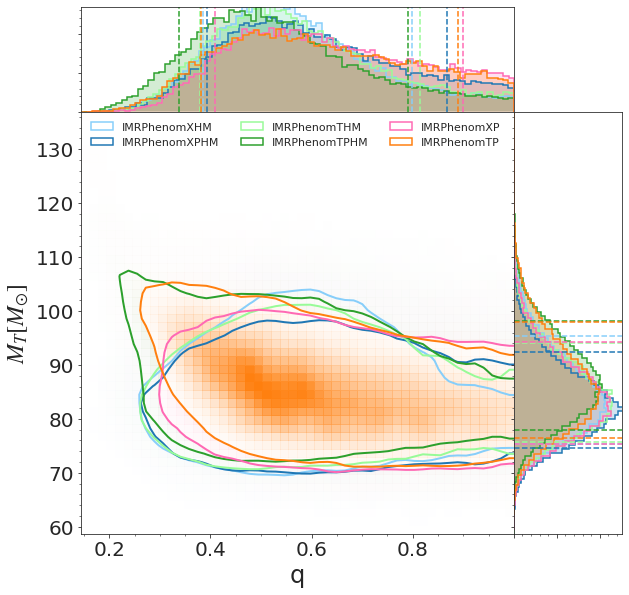}\includegraphics[width=0.9\columnwidth]{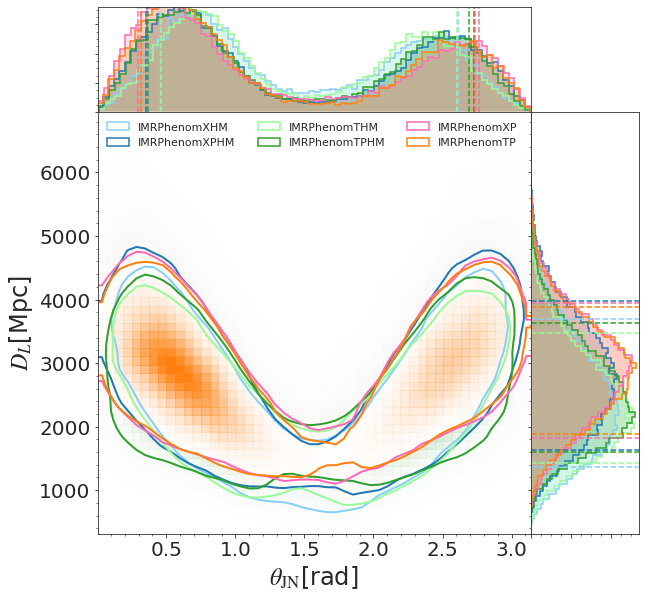}]
    \caption{Posterior distributions of GW170729 for the mass ratio, total mass, chirp mass, effective spin parameter, luminosity distance, inclination and $\log$ likelihood using the dominant-mode models \phXP and \phTP (light pink and orange respectively) and the subdominant-mode models \phXHM (light blue), \phXPHM (dark blue) and \phTHM (light green), \phTPHM (dark green).
    All contours are at 90\% credible intervals.
    \label{fig:gw170729_new}
    }
\end{figure*}

In Fig. \ref{fig:gw170729_old} we see that \ppvtwo and the corresponding newer precessing model for the dominant quadrupole, \phXP, give very similar results, while the non-precessing multi-mode waveform models recover lower mass ratio and higher luminosity distance, as was also observed by \citet{Chatziioannou:2019dsz} and \citet{Payne:2019wmy}.
The non-precessing likelihood-reweighted results of \citet{Payne:2019wmy} (not plotted)
are in very good agreement with our \phXHM and \phTHM posteriors for most parameters,
with the peak for $\chieff$ shifted down towards 0 but its posterior still well consistent
within 90\% credible intervals.
Adding not only higher modes but precession, we obtain more evidence for unequal masses for \phTPHM, as seen in Fig. \ref{fig:gw170729_new}.

In Figs. \ref{fig:gw170729_old} and \ref{fig:gw170729_new} we also directly compare the frequency and time domain families \phenX and \phT. We can see that the distribution of likelihood values across the posterior for \phTPHM is shifted toward larger values, and indeed we obtain a Bayes factor of $2.16_{-0.39}^{+0.47}$ in favor of \phTPHM when compared to \phXPHM, while the aligned-spin versions, \phTHM and \phXHM, give much more similar results.

\begin{table} 
%\begin{ruledtabular}
\begin{tabular}{cccc}
\hline\hline
Hypotheses & Model properties & PhenomX & PhenomT \\
\hline 
\multirow{2}{*}{HM vs $\ell=2=\vert m\vert $} & aligned  & \ensuremath{2.27_{-0.40}^{+0.48}}\xspace & \ensuremath{1.69_{-0.30}^{+0.36}}\xspace \\
 & precessing &  \ensuremath{1.35_{-0.24}^{+0.29}}\xspace & \ensuremath{2.31_{-0.41}^{+0.50}}\xspace \\
\hline\hline
\end{tabular}
%\end{ruledtabular}
\caption{
\label{tab:tabBF}
Comparison of Bayes factors between models of the \phenX and \phT families concerning the hypothesis that the signal of GW170729 contains higher modes, for the precessing and non-precessing waveform models.
The \phenX aligned check compares \phXHM against \phX,
the \phenX precessing check compares \phXPHM against \phXP,
the \phT aligned check compares \phTHM against \phT, 
and the \phT precessing check compares \phTPHM against \phTP.
}

\end{table}

\begin{figure*} 

    \includegraphics[width=0.9\columnwidth]{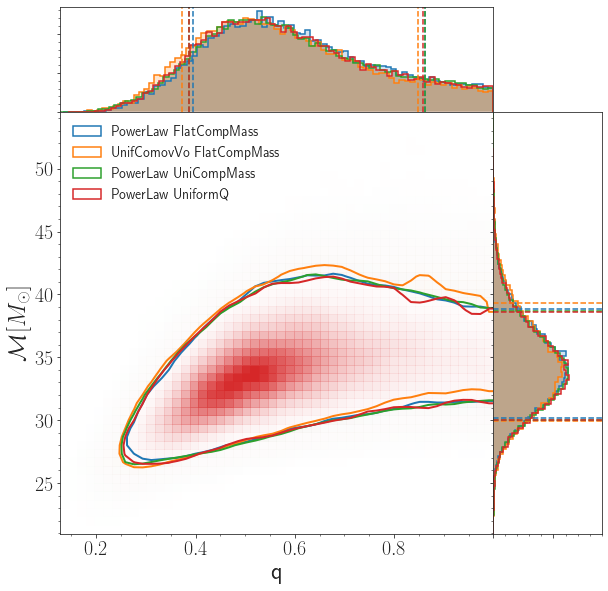}\includegraphics[width=0.925\columnwidth]{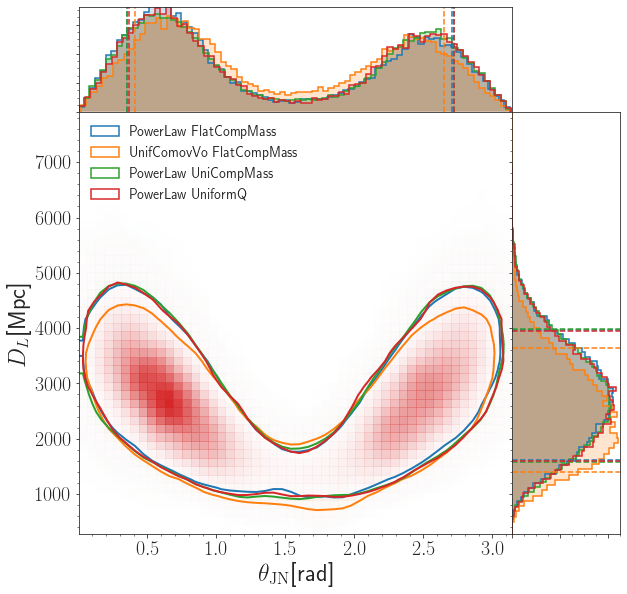}
    \caption{Comparison of posterior distributions for GW170729 estimated with \phXPHM using different distance and mass priors. Here the string {\tt PowerLaw} denotes our standard power-law distance prior. {\tt UnifComovVo} denotes a distance prior that is uniform in comoving volume. {\tt FlatCompMass} runs use a prior uniform in $q \leq 1$ and $\mathcal{M}_c$, with samples afterwards reweighted to a flat prior in component masses. {\tt UniCompMass} denotes the more recent method to directly ensure a flat prior in component masses while still sampling in $q$ and $\mathcal{M}_c$. Finally, {\tt UniformQ} denotes a flat prior in mass ratio $Q \geq 1$. All these priors are described in Sec.~\ref{sec:priors}. All contours are at 90\% credible intervals.}
    \label{fig:gw170729_priors}
\end{figure*}

\emph{b. \hspace{5pt} Comparison of different mass and distance priors:} \hspace{5pt} Given that GW170729 is the furthest event of the first catalog and has support for unequal masses, we compare the different distance and mass prior choices explained in Sec. \ref{sec:priors}, in order to see their influence on the posterior distributions. For this study we run \phXPHM using \mbox{{\tt nlive} $= 2048$} and \mbox{{\tt nact} $= 10$}. Fig.~\ref{fig:gw170729_priors} shows that there is a shift to lower distance and masses when we change from the power-law distance prior to the prior that is uniform in comoving source-frame volume.
As discussed in Sec.~\ref{sec:priors}, the standard power-law distance prior distributes the mergers uniformly throughout an Euclidean universe, which is only a good approximation at relatively close distances (small redshifts). However, at larger distances, it is not a good approximation, and including proper cosmological information in the prior is required, which the comoving volume prior achieves better~\citep{Abbott:2020niy,Romero-Shaw:2020owr}.
On the other hand, varying the mass priors does not affect the posteriors noticeably.
This is as expected because masses are more strongly constrained by the data,
hence the prior has a smaller influence on the posterior
-- at least for the GWTC-1 events.

In order to quantify how much the posteriors change, we compute the JS divergence between the results. In Table \ref{tab:tabPriorsJS} we show, for every possible combination of results, which parameter has the highest divergence. If we compare the results that have different mass priors, all divergences are below 0.003, meaning that the posterior results only vary very minimally. Comparing different distance priors we obtain higher divergences, where the luminosity distance is the parameter with the largest values, \mbox{$JS_{d_L} = 0.015$} and 0.012.

\begin{table*}
%\begin{ruledtabular}
\begin{tabular}{lcccc}
\hline\hline
  & {\tt PowerLaw} - {\tt FlatCompMass}  & {\tt UnifComovVo} - {\tt FlatCompMass} & {\tt PowerLaw} - {\tt UniCompMass} & {\tt PowerLaw} - {\tt UniformQ}\\
\hline
 {\tt PowerLaw} - {\tt FlatCompMass}& - & $JS_{D_L} = 0.0152$ &$JS_{ra} = 0.0029$ & $JS_{\mathcal{M}} = 0.0009$ \\
 {\tt UnifComovVo} - {\tt FlatCompMass} &  & - & $JS_{D_L} = 0.0128$ & $JS_{D_L} = 0.0119$ \\
 {\tt PowerLaw} - {\tt UniCompMass} & & & - & $JS_{ra} = 0.0030$ \\
 {\tt PowerLaw} - {\tt UniformQ} & & & & -\\
 \hline\hline
\end{tabular}
%\end{ruledtabular}
\caption{ Maximum Jensen-Shannon (JS) divergence values between the posterior distributions for GW170729 estimated with \phXPHM using different priors in distance and masses, as explained in Sec.~\ref{sec:priors}. {\tt PowerLaw} refers to a power-law distance prior. {\tt UnifComovVo} denotes a distance prior that is uniform in comoving volume. {\tt FlatCompMass} is used for a prior uniform in $q \leq 1$ and $\mathcal{M}_c$, with samples afterwards reweighted to a flat prior in component masses. {\tt UniCompMass} denotes directly using a flat prior in component masses while still sampling in $q$ and $\mathcal{M}_c$. Finally, {\tt UniformQ} is a prior flat in mass ratio $Q \geq 1$.}
\label{tab:tabPriorsJS}
\end{table*}

\subsubsection{GW170809}\label{sec:GW170809}

The posterior for this event is consistent with vanishing spins (both $\chieff$ and $\chi_\mathrm{p}$), and $\theta_{JN}$ is relatively well measured to correspond to a face-off orientation. The masses are slightly more unequal than for GW150914 and the total mass is slightly smaller. The event is however located at approximately twice the distance, corresponding to only about half the SNR of $\ensuremath{12.6_{-0.3}^{+0.2}}\xspace$ for \phXPHM. Similar to GW150914, the Bayes factors for the MSA-angle based versions of \phXPHM, \phXHM and \phX are also consistent within error estimates, while the NNLO-based versions of \phTHM and even more of \phTPHM are disfavored. MSA versions recover slightly higher $\chi_\mathrm{p}$ than the NNLO-based versions, otherwise we find no visible differences in posteriors between \phXPHM versions. 

In Fig. \ref{fig:high_mass_collage} one can see that the waveforms with subdominant modes have more support for equal masses. This effect is strongest for \phTPHM, which is however disfavored by the Bayes factor. In addition, the higher-mode models marginally shift the total mass to lower values, and the luminosity distance to higher values.

Using the RapidPE method with a catalog of higher-mode NR waveforms, \citet{LIGOScientific:2018mvr} found a revised $\chieff$ distribution which is symmetric about a median value of zero, while our higher-mode models show no such effect.

\subsubsection{GW170818}\label{sec:GW170818}

\begin{figure} 
    \centering
    \includegraphics[width=0.9\columnwidth]{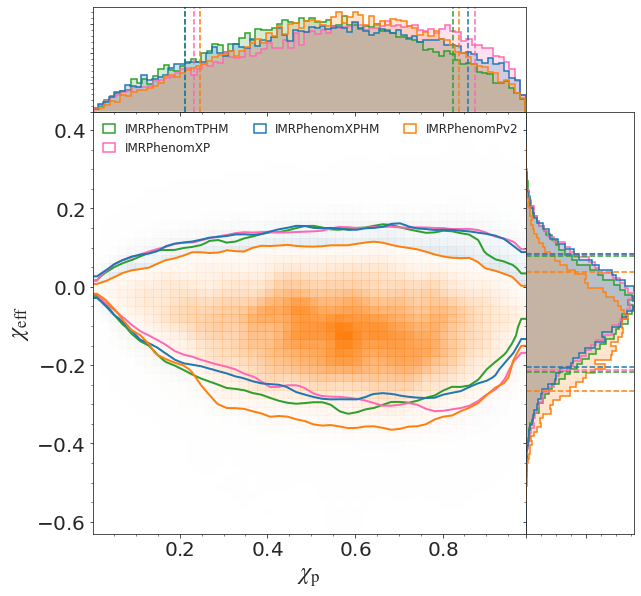}
    \caption{Comparison of posterior distributions for $\chi_\mathrm{p}$ and $\chieff$ of GW170818, estimated with \phTPHM (green), \phXP (pink), \phXPHM (blue) and \phPvtwo (orange, from \citet{LIGOScientific:2018mvr}).
    90\% credible intervals are indicated.
    \label{fig:gw170818_chip_chieff}
    }
\end{figure}

GW170718 is a heavy event most consistent with a negative effective spin, $\chieff=-0.11_{-0.16}^{+0.14}$ for \phPvtwo \citep{LIGOScientific:2018mvr}.
However, in our analysis, using the \phenX and \phT families, the effective spin parameter is shifted toward zero, as can be seen in Fig. \ref{fig:gw170818_chip_chieff}.
Unfortunately, this is also the third faintest event, which limits the information that can be extracted, compared to other events.
There is weak evidence for precession,
with a Bayes factor for XPHM against XHM of $2.54_{-0.48}^{+0.60}$, see Table 
\ref{tab:tabBF_prec}.

\begin{figure*} 
    \includegraphics[width=0.9\columnwidth]{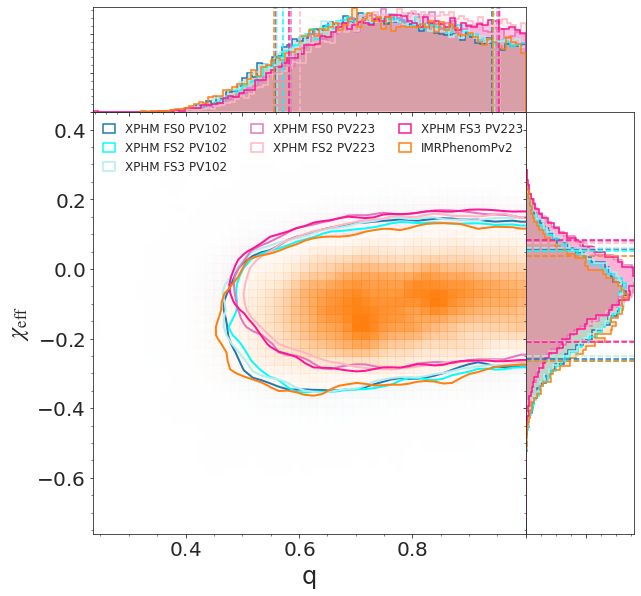}\includegraphics[width=0.9\columnwidth]{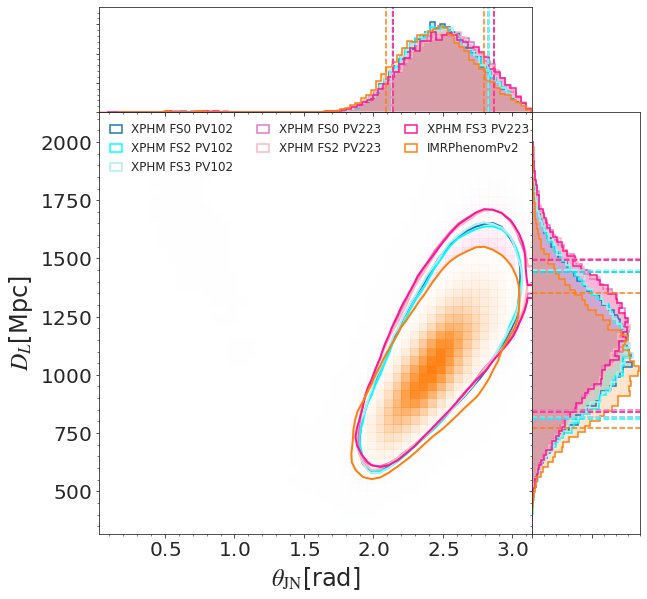}
    \caption{Comparison of posterior distributions of GW170818 for the mass ratio, the effective spin parameter, the luminosity distance and the inclination using the MSA (blues) and NNLO (pinks) \phXPHM versions and \phPvtwo (orange) results from \citet{LIGOScientific:2018mvr}.
    90\% credible intervals are indicated.
    \label{fig:gw170818_XPHM_versions}}
\end{figure*}

The \phXF and \phT families show good consistency in Figs. \ref{fig:high_mass_collage} and \ref{fig:gw170818_chip_chieff}, and a comparison of posteriors for different \phXPHM versions with the results for \phPvtwo is shown in 
Fig. \ref{fig:gw170818_XPHM_versions}. \phXPHM with NNLO Euler angles obtains a weakly higher Bayes factor than the default version, by a factor $1.64_{-0.31}^{+0.39}$ (see Tables \ref{tab:SNR_BF_table} and \ref{tab:tabBF_prec}), and the corresponding posteriors for $\chieff$ and the luminosity distance are more similar to the \phPvtwo results.

\subsubsection{GW170823}\label{sec:GW170823}

GW170823 is the second most massive and second most distant event. The SNR is relatively low at $\ensuremath{12.0_{-0.2}^{+0.1}}\xspace$ for \phXPHM, limiting the information that can be extracted. Indeed, in Fig. \ref{fig:high_mass_collage} we find very good agreement between the posteriors obtained with different waveform models.

\begin{figure} 
    \centering

\includegraphics[width=0.9\columnwidth]{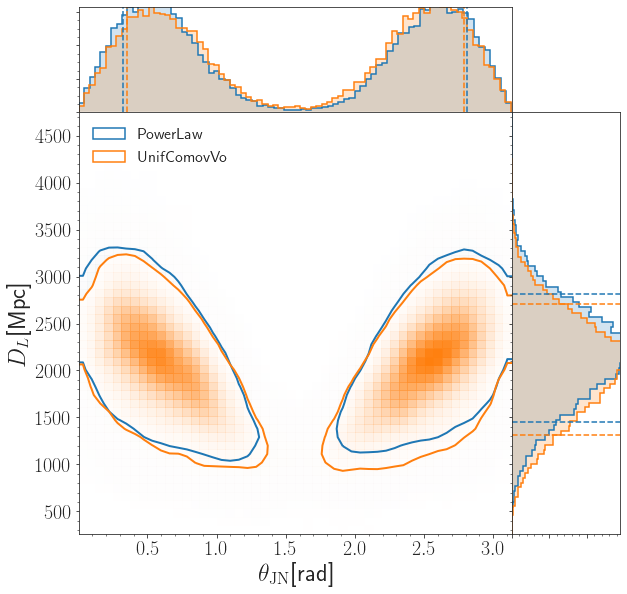}
    \caption{Comparison of the posterior distributions for GW170823 estimated with \phXPHM using different distance priors (proportional to the luminosity distance squared and uniform in the comoving volume and source frame time, labeled by {\tt PowerLaw} and {\tt UnifComovVo} respectively).
    \label{fig:gw170823_priors}
    }
\end{figure}

Due to the large distance we compare two distance priors, uniform in comoving volume and proportional to the luminosity distance squared, as we do in Sec. \ref{sec:GW170729} for GW170729. As expected, we find that the differences between both priors are smaller than for GW170729. The luminosity distance is the parameter with the highest JS divergence when comparing the two distance prior choices; however, all divergence values are lower than 0.007. The impact of the distance prior is very minor, as illustrated by Fig. \ref{fig:gw170823_priors}, where we compare the runs with different priors all using \phXPHM.

We find the Bayes factors between different \phXPHM versions to be consistent within error estimates, and find no notable differences in the respective posterior distributions for the source parameters.

\section{Conclusions}

\label{sec:conclusions}

In this paper we have obtained parameter estimation results for the BBH events in GWTC-1 with the new fourth generation of phenomenological waveform models.
This has served as a stress test before using these models for new events observed in the
second half of the third observing run (O3b), as presented in \citet{LIGOScientific:2021djp},
and in a reanalysis of O1, O2 and O3a events as part of \citet{LIGOScientific:2021usb}.
Our results overall confirm the expectations that subdominant harmonics and waveform systematics only have minor consequences for parameter estimation results for these ten events, and no major problems have been encountered using the \phXF and \phT families.

However, we have put particular emphasis on providing a more in-depth study of this set of events than previous works, considering different waveform models along with tests of convergence and comparisons of different sampling algorithms.
This has been particularly important for the event GW151226, where the initial arXiv version of \citet{Chia:2021mxq} had suggested that other results, including our analysis, may be inaccurate.
But 
various recent publications \citep{Nitz:2020mga,Vajpeyi:2022dvs}, as well as the final journal version of \citet{Chia:2021mxq}, now find broadly consistent results:
the event is compatible with a broader range of mass ratios than reported in the initial LVC studies,
and while results by different groups disagree in the relative posterior weight assigned to more or less unequal configurations, consensus appears to have been reached that these remaining differences are consistent with differences in priors.

For the other GWTC-1 event where results have been less stable in past studies, the heavy GW170729,
we find consistent results with previous, more limited higher-modes analyses by~\citet{Chatziioannou:2019dsz} and \citet{Payne:2019wmy}.
Results between different waveforms are consistent at 90\% credible level,
but the \phTPHM model tends to prefer higher primary masses, making it more likely to be a signal from a binary containing a black hole in the PISN mass gap.
Still, the uncertainties both on the GW parameter estimation and the astrophysical predictions are too large to make a decisive claim on this interpretation.

Given the high computational cost of Bayesian parameter estimation, not many other GW events have been analysed in similar detail. In two parallel papers \citep{Colleoni:2020tgc,GW190521}\footnote{
Note that we have updated our results for GW190412 to include the precessing time domain model \phTPHM in \citet{Estelles:TPHM}.

we have provided even more detailed analyses of two selected events from the third observing run O3: GW190412 \citep{Abbott:2020niy,LIGOScientific:2020stg,Abbott:2020tfl} and GW190521 \citep{Abbott:2020tfl,LIGOScientific:2020ufj}. The results found in these two papers complement our findings in the present paper:

GW190412 was the first event where subdominant spherical harmonic modes could be clearly identified.} Our re-analysis shows excellent agreement between the latest generations of non-precessing waveform models, where subdominant harmonics are calibrated to NR simulations, and broad consistency between different precessing models with subdominant harmonics. In particular we find good agreement between our \phXF and \phT families.
GW190521 is the shortest signal detected so far (only approximately 0.1 seconds long), and much more challenging: due to the lack of information in such a short signal, waveforms that correspond to very different sources fit the signal well, and the posterior distribution is multi-modal. We can however show good consistency between different sampling codes -- LALInference \citep{Veitch:2014wba} and Bilby \citep{Ashton:2018jfp,Smith:2019ucc} -- and different choices of mass priors. We also confirm our expectation that for events where only the merger and ringdown are observable, our description of precession in the time domain (\phTPHM) improves significantly over our frequency-domain model (\phXPHM). 

Meanwhile, the GWTC-1 events studied in this paper suggest that for lower-mass signals there is no simple preference for either the \phXF or \phT family:
The \phTHM and \phTPHM models are typically characterized by lower Bayes factors when compared to their \phXF counterparts. The reason is likely that
for lower-mass events the higher accuracy of the phase of the $(2,2)$ mode for \phXF makes the frequency-domain model more accurate than the current time-domain version. The exception is GW170814, which has a total mass of about 56 solar masses,
but does show some support for precession, see the Table \ref{tab:tabBF_prec} of Bayes factors. In this case  \phTPHM is also favored over \phXPHM.

Given that the \phenX models are both computationally cheaper and simpler to use (time-domain models require a careful choice of starting frequency to ensure consistent start times between different modes and a careful treatment of Fourier transforms when using matched filtering in the frequency domain), use of the \phXF
family is preferred for lower-mass events.
Still, the example of GW170814 suggests that a comparison with \phTPHM may provide further insight into such events when support for spin precession is found.

For higher masses, preference tilts toward \phTPHM,
as the accuracy of the phasing during inspiral is less relevant.
For  GW170729, which has a median value of the total mass in the detector frame above 120 solar masses ($122.7_{-15.7}^{+14.3}\,M_\odot$ for the \phXPHM default run), both \phXPHM  and \phTPHM provide consistent results, but support for \phTPHM is marginally larger than for \phXPHM.  
For GW170729 the better accuracy of the inspiral phasing for \phXF is much less relevant, while the treatment of precession in the time-domain model is superior. For GW170729 the advantage of \phTPHM is however still rather small, while for the even heavier GW190521 we find \citep{GW190521} that \phTPHM gives significantly better results.

Finally, the computational efficiency of the \phXF and \phT models has allowed us to also systematically test the influence of variations in priors and sampler settings.
We have verified that changes in the mass and distance priors only have very marginal influence on results. Perhaps most interestingly, we have checked that, choosing a prior that emphasizes large mass ratios, we have not found any multi-modal posteriors for any of the GWTC-1 events, as has been the case for GW190521~\citep{Nitz:2020mga,GW190521}.

In summary, we consider that detailed studies of waveform systematics and parameter estimation robustness,
as performed in this paper,
are essential for the robust interpretation of GW detections.
Along with providing publicly available data sets that allow to reproduce an analysis,
such investigations are an important step forward for the reliable astrophysical interpretation of GWs.

\section*{Acknowledgements}

This work was supported by European Union FEDER funds, the Ministry of Science, 
Innovation and Universities and the Spanish Agencia Estatal de Investigación grants PID2019-106416GB-I00/AEI/10.13039/501100011033,        % FPA2017-90687-REDC, FPA2017-90566-REDC. 
RED2018-102661-T,    % RENATA
RED2018-102573-E,    % REDES ESTRATÉGICAS: Participación Española en Estructuras Euro... 
FPA2017-90687-REDC,  % CPAN
Vicepresidència i Conselleria d’Innovació, Recerca i Turisme, Conselleria d’Educació, i Universitats del Govern de les Illes Balears i Fons Social Europeu, 
the Comunitat Autonoma de les Illes Balears through the Direcció General de Política Universitaria i Recerca with funds from the Tourist Stay Tax Law ITS 2017-006 (PRD2018/24 and PDR2020/11),
Generalitat Valenciana (PROMETEO/2019/071),  
EU COST Actions CA18108, CA17137, CA16214, and CA16104,
the Spanish Ministry of Education, Culture and Sport grants FPU15/03344 and FPU15/01319 and the Conselleria de Fons Europeus, Universitat i Cultura del Govern de les Illes Balears FPI/2215/2019.
M.C. acknowledges funding from the European Union's Horizon 2020 research and innovation programme, under the Marie Skłodowska-Curie grant agreement No. 751492 and from the Spanish Agencia Estatal de Investigación, grant IJC2019-041385.
D.K. is supported by the Spanish Ministerio de Ciencia, Innovaci{\'o}n y
Universidades (ref. BEAGAL 18/00148)
and cofinanced by the Universitat de les Illes Balears.
The authors thankfully acknowledge the computer resources at MareNostrum and the technical support provided by Barcelona Supercomputing Center (BSC) through Grants No. AECT-2019-2-0010, AECT-2019-1-0022,  from the Red Española de Supercomputación (RES).
Authors also acknowledge the computational resources at the cluster CIT provided by LIGO Laboratory and supported by National Science Foundation Grants PHY-0757058 and PHY-0823459.
This research has made use of data obtained from the Gravitational Wave Open Science Center~\citep{GWOSC}, a service of LIGO Laboratory, the LIGO Scientific Collaboration and the Virgo Collaboration. LIGO is funded by the U.S. National Science Foundation. Virgo is funded by the French Centre National de Recherche Scientifique (CNRS), the Italian Istituto Nazionale della Fisica Nucleare (INFN) and the Dutch Nikhef, with contributions by Polish and Hungarian institutes.

%%%%%%%%%%%%%%%%%%%%%%%%%%%%%%%%%%%%%%%%%%%%%%%%%%
\section*{Data Availability} %FIX ME

In order to complement this manuscript, we provide a data release \citep{Zenodo_release} which contains all the posterior samples and Bilby configuration files developed in this work. It also contains PSDs and calibration envelope files for the events where publicly released datasets from GWOSC \citep{GWOSC} are not sufficient.

%%%%%%%%%%%%%%%%%%%% REFERENCES %%%%%%%%%%%%%%%%%%

% The best way to enter references is to use BibTeX:

\bibliographystyle{mnras}
\bibliography{gwtc1} % if your bibtex file is called example.bib

% Alternatively you could enter them by hand, like this:
% This method is tedious and prone to error if you have lots of references
%\begin{thebibliography}{99}
%\bibitem[\protect\protect\citeauthoryear{Author}{2012}]{Author2012}
%Author A.~N., 2013, Journal of Improbable Astronomy, 1, 1
%\bibitem[\protect\protect\citeauthoryear{Others}{2013}]{Others2013}
%Others S., 2012, Journal of Interesting Stuff, 17, 198
%\end{thebibliography}

%%%%%%%%%%%%%%%%%%%%%%%%%%%%%%%%%%%%%%%%%%%%%%%%%%

%%%%%%%%%%%%%%%%% APPENDICES %%%%%%%%%%%%%%%%%%%%%

\appendix

\section{Convergence test and comparison of sampling methods}\label{subsec:convergence_test}

In order to set up appropriate sampler settings, we performed a convergence study for all the different events and waveform approximants. That convergence test consists in comparing several sampler settings from low to high resolution in order to get the most converged result but with an affordable computational cost. In our case we test two different numbers of live points for the Nested Sampling algorithm, \mbox{{\tt nlive} = 512} and 2048, and also the number of autocorrelation times: {\tt nact} = 10, 30, and for completeness we also test \mbox{{\tt nact} = 50} for \phXHM and \phTHM. 

\begin{figure*}
    \centering
    \includegraphics[width=0.9\columnwidth]{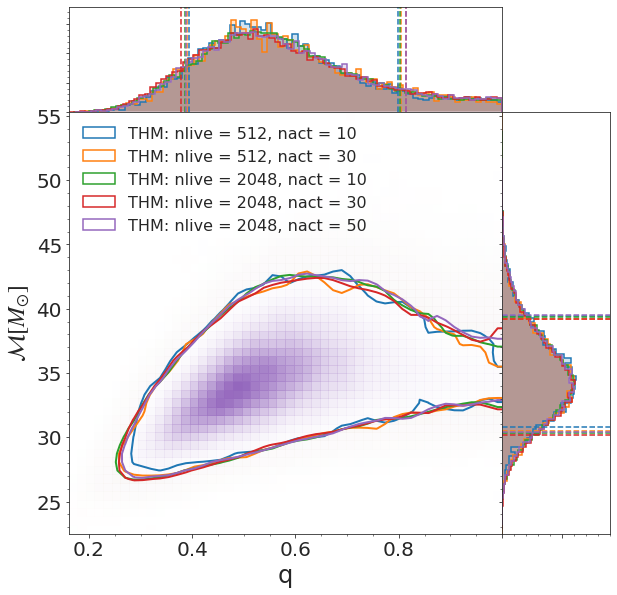}\includegraphics[width=0.9\columnwidth]{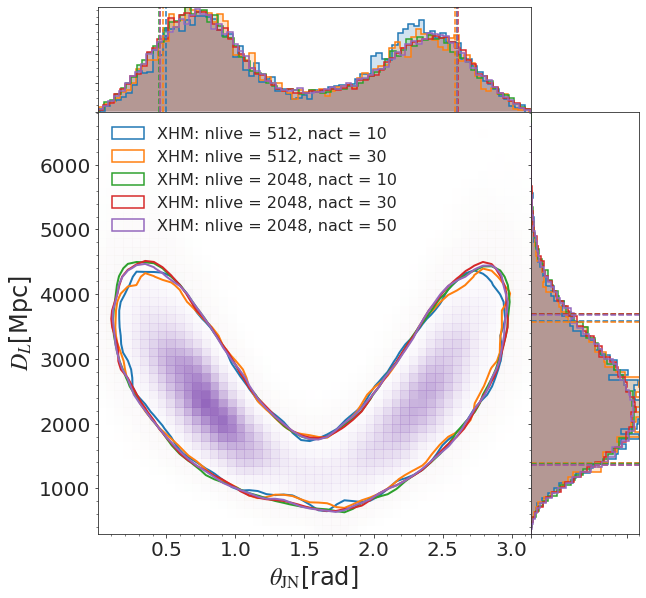}\\
    \includegraphics[width=0.9\columnwidth]{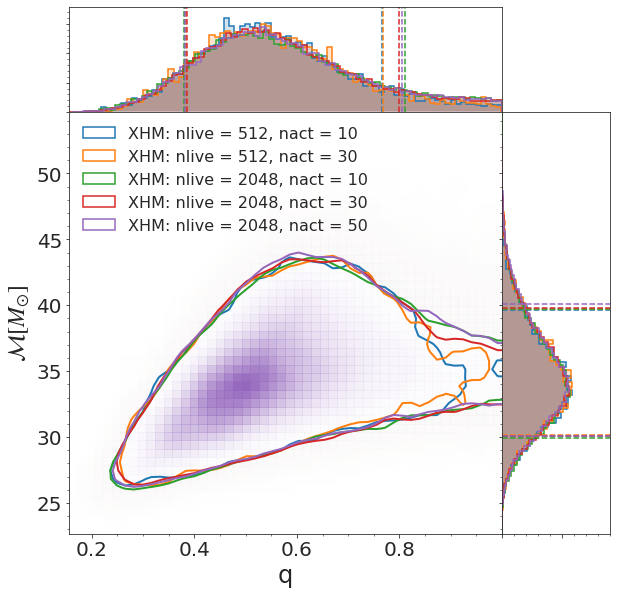}\includegraphics[width=0.9\columnwidth]{plots/convergence_GW170729/cornerPlot_theta_jn_luminosity_distance__conv_XHM_GW170729_v2.png}\\
    \includegraphics[width=0.9\columnwidth]{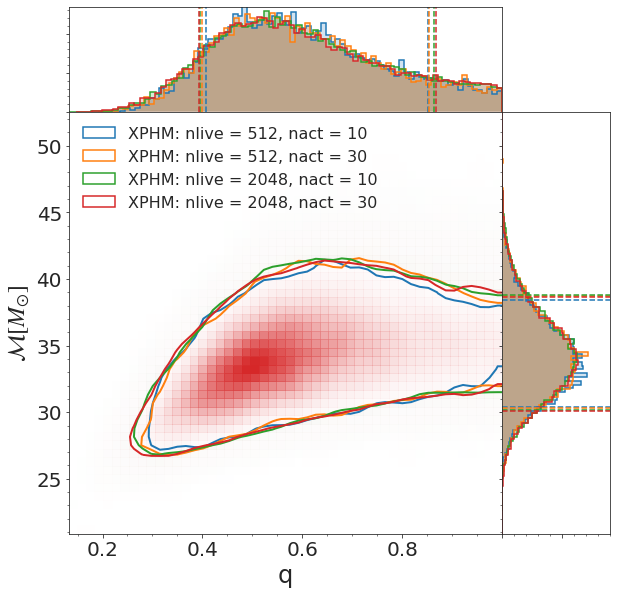}
    \includegraphics[width=0.9\columnwidth]{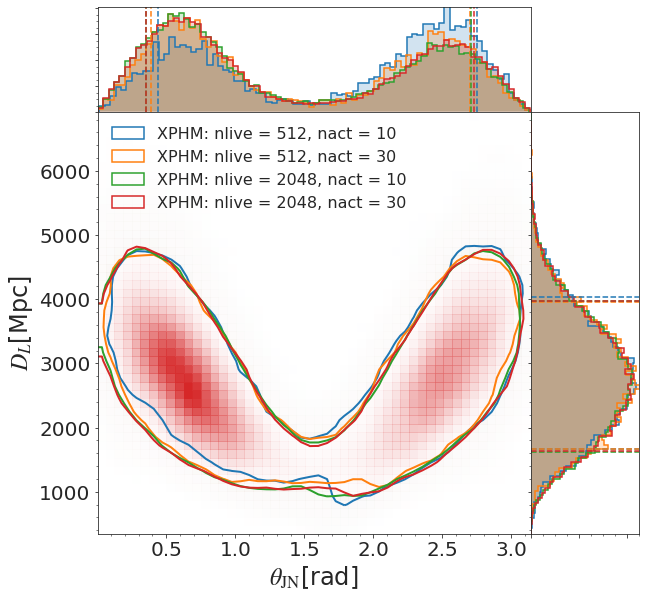}
    \caption{GW170729 convergence test comparing posterior distributions for the mass ratio, chirp mass, luminosity distance and inclination using \phTHM, \phXHM and \phXPHM
    (one per row) and with different sampler settings ({\tt nlive} = 512, 2048 and {\tt nact}=10,30,50) for each waveform, displaying 90\% credible contours.
    \label{fig:convergence}
    }
\end{figure*}

For all the BBH events where we performed the convergence test, we get the same conclusion: default settings of \mbox{{\tt nlive} = 2048} and \mbox{{\tt nact} = 30} result in sufficiently accurate posteriors at moderate computational cost. In Fig. \ref{fig:convergence} we show a comparison for GW170729 using different waveform models and sampler settings. Using \mbox{{\tt nlive} = 512} both the histograms and the 2D plots show fluctuations, indicating insufficient convergence. However, when we increase {\tt nlive} to 2048, the posterior distributions are much smoother. Comparing the number of autocorrelation times when we use a large number of {\tt nlive}, changing the {\tt nact} value does not make an important difference. 

We can get a more quantitative perspective from Table \ref{tab:tabConvergenceTHM}, where we show the maximum JS divergence values corresponding to the posteriors of Fig. \ref{fig:convergence}. As commented above, the highest differences appear when we change from low to high {\tt nlive}, but not when we change the {\tt nact} configuration. We decided to use \mbox{{\tt nlive} = 2048} and \mbox{{\tt nact} = 30} as the default setting because if we compare with the same run but using \mbox{{\tt nact} = 50}, we obtain an indistinguishable distribution: the divergences between both posterior distributions are smaller than 0.001 bits.

In addition we perform further tests for the three lowest-mass events,
motivated by the results in \citet{Nitz:2021uxj} and in particular the initial arXiv version of \citet{Chia:2021mxq} for GW151226, which are, or were, to some degree in tension with our results.

As mentioned in the introduction, a large part of this tension has been resolved with improvements in the sampling technique employed by \citet{Chia:2021mxq}, as they report in their final journal version.
A consensus seems to now have emerged that the event is consistent with a broader range of mass ratios than found by initial LVC studies,
but no clear preference for unequal masses persists,
and that remaining differences seem to be largely consistent with differences in prior choices.
In addition,
we note that the results of \citet{Nitz:2021uxj}
have been obtained at half our sampling rate
\citep[i.e. only 2048\,Hz, according to the data release of][]{Nitz:2021uxj},
and the discrepant results in the initial arXiv version of \citet{Chia:2021mxq} had been obtained at an even lower sampling rate of 1024\,Hz \citep[using the sampling code described in][]{Venumadhav:2019tad},
while for the final results \citep[using the updated method from][]{Roulet:2022kot}
they used 2048\,Hz.
See also the discussion of this point in the note at the end of the journal version of \citet{Chia:2021mxq}, phrased in terms of the Nyquist frequency (half the sampling rate).

A reduced sampling rate implies a lower cutoff frequency corresponding to the Nyquist frequency, and will reduce the higher-mode content.
While the effect on the recovered SNR is very small, we find a significant effect on the posterior distributions of some quantities, as we will discuss below.

We first compare the results for this event from our standard Bilby runs with alternative sampling choices: As previously shown in Fig.~\ref{fig:gw170729_priors} for GW170729, we again compare our standard method of reweighting the standard-prior runs to a prior that is flat in component masses, against directly sampling with the alternative flat-in-component-masses prior (using the two Bilby classes {\tt UniformInComponentsChirpMass} and {\tt UniformInComponentsMassRatio}, as discussed in Sec.~\ref{sec:priors}).
Furthermore we compare with results for the LALInference code \citep{Veitch:2014wba}, which is part of the LALSuite \citep{lalsuite} package for GW data analysis, using its implementation of parallel-tempered Markov Chain Monte Carlo (MCMC) sampling. LALInference samples in mass ratio and chirp mass, reweighting to a prior that is flat in component masses as described in \citet{Veitch:2014wba}. We use essentially standard LALInference settings with eight temperatures, but a large number of $60$ independent chains in order to have a broad distribution of chains covering the parameter space. These runs employ the same dataset, PSDs and calibration envelopes as the Bilby runs, discussed in Sec.~\ref{sec:lowmass}, but we do not employ the distance marginalization used for our Bilby runs.

\begin{figure*}
    \includegraphics[width=0.95\columnwidth]{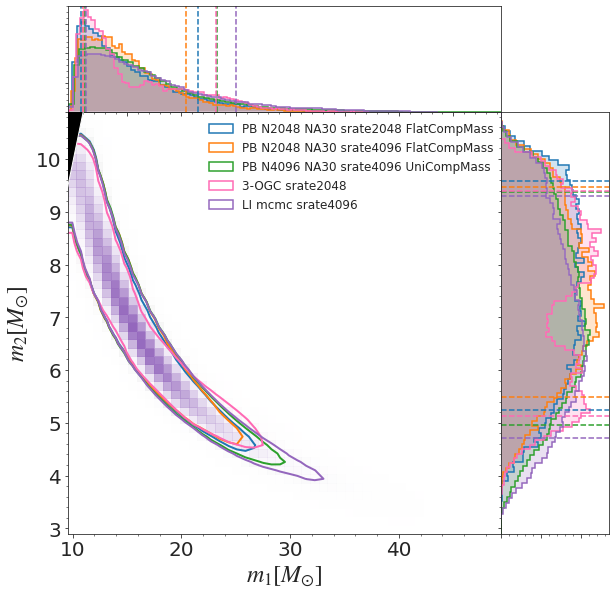}\includegraphics[width=0.95\columnwidth]{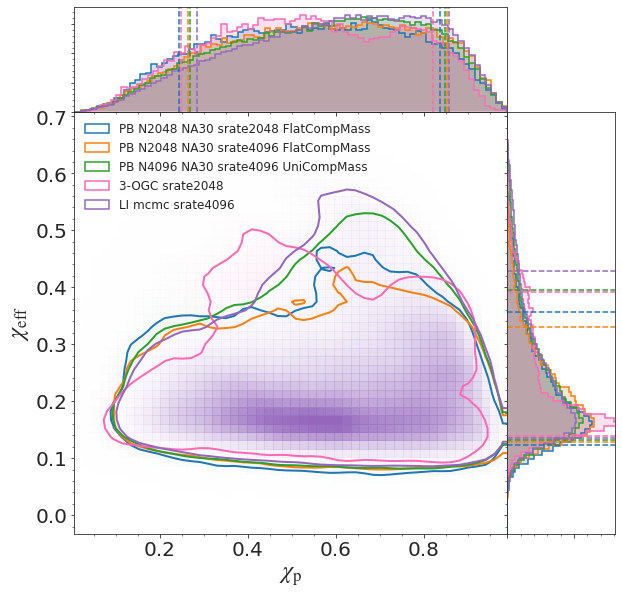}
    \caption{Comparison of the recovered posterior distributions for GW151226 for different values of the sampling rate ({\tt srate}) and mass priors in Parallel Bilby ({\tt PB}), a standard run in LALInference ({\tt LI mcmc}), and the results reported for this event in \citet{Nitz:2021uxj} (denoted as 3-OGC in the legend).
    For brevity, {\tt nlive} has been abbreviated as ``N'' and {\tt nact} as ``NA'' in the legend.
    {\tt FlatCompMass} refers to a run using a prior uniform in $q \leq 1$ and $\mathcal{M}_c$ and then reweighted to a flat prior in component masses.
    {\tt UniCompMass} refers to directly using a uniform prior in component masses.
    \label{fig:gw151226_convergence}}
\end{figure*}

In Fig.~\ref{fig:gw151226_convergence} we show a comparison between different runs for GW151226, employing
two different sampling rates (our default 4096\,Hz and a lower one at 2048\,Hz),
the two different mass priors,
and different sampling settings (\texttt{nlive} of 2048 and 4096 at fixed \mbox{\texttt{nact}=30}),
all with
Parallel Bilby,
as well as the standard LALInference run as described above,
and the results reported in \citet{Nitz:2021uxj}
(results data for \citet{Chia:2021mxq} have not been released).
We observe that while all our runs are consistent, employing the Bilby prior defined by the {\tt UniformInComponentsChirpMass} and {\tt UniformInComponentsMassRatio} classes (labelled ``UniCompMass" in the plot) reproduces LALInference results best, as expected,
meaning that our results are well converged.
For this event we also observe that differences with respect to the results reported in \citet{Nitz:2021uxj} are broadly consistent with the differences we observe between different sampling rates in our runs.

\begin{figure} 
    \includegraphics[width=0.9\columnwidth]{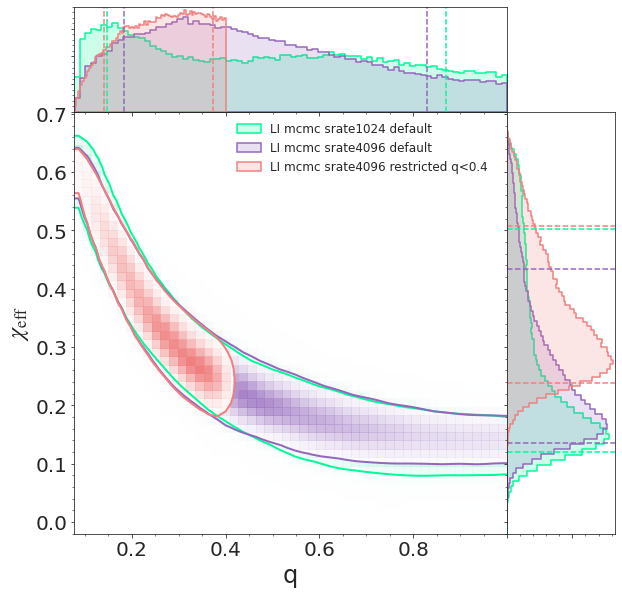}
    \caption{Comparison of the recovered posterior distributions for GW151226 in LALInference ({\tt LI mcmc}) for two different values of the sampling rate ({\tt srate} = 1024 and 4096\,Hz) and two different mass-ratio prior choices (the default one and one restricted to $q \leq 0.4$). The mass-ratio histograms have been normalized to equal height for ease of visibility.
    \label{fig:gw151226_LI}
    }
\end{figure}

We also perform runs that aim to reproduce the bimodal results of the initial arXiv version of \citet{Chia:2021mxq} with LALInference, using their initial sampling rate of 1024\,Hz.
In Fig.~\ref{fig:gw151226_LI} we can observe that indeed reducing our sampling rate by a factor of four
we find a bimodality in the mass ratio distribution, and increased support for smaller mass ratio $q$ (we can not show direct comparisons since the initial posterior data of \citet{Chia:2021mxq} have not been released). In this figure we also plot the results of a LALInference run with the standard 4096\,Hz sampling rate but with a prior restricted to $q \leq 0.4$, aimed at improving the sampling of more unequal mass ratios. We find that this restricted run shows very good agreement with that using the standard unrestricted prior,
enhancing the posterior region corresponding to higher spins, but remaining consistent with the low-q portion of the standard run and not producing an additional peak.
We thus conclude that the bimodality initially found by \citet{Chia:2021mxq} could at least partially be explained by the reduction of the higher-mode content when a low sampling rate is used that will cut off part of the merger--ringdown signal. 
However, while in the final journal version of \citet{Chia:2021mxq} they also increased their sampling rate (to 2048\,Hz, half ours), they report that the dominant effect in changing their results to the new, less discrepant posteriors was in improving their relative binning sampling algorithm.
For a full explanation of the results for GW151226 found by other studies it would also be interesting to compare results using the different PSD estimates employed, but the PSD used in \citet{Chia:2021mxq} is not publicly available. (The PSD used in \citet{Nitz:2021uxj} can in principle be reproduced from configuration files in their data release.)

To conclude the discussion of this event,
the recent work by \citet{Vajpeyi:2022dvs}
has directly compared the posterior weight between equal and unequal mass configurations,
contributing to the consensus picture hinted at above:
consistency with a range of mass ratios,
no strong preference for one end of the scale,
and some remaining dependence of results on prior choice.

\begin{table*}
%\begin{ruledtabular}
\begin{tabular}{llccccc}
\hline\hline
 & & N512 NA10 & N512 NA30 & N2048 NA10 & N2048 NA30 & N2048 NA50 \\
\hline 
\multirow{5}{*}{\phTHM} & N512 NA10 & - & $JS_{t_1} = 0.006$ &$JS_{t_2} = 0.020$ & $JS_{t_2} = 0.019$ & $JS_{t_2} = 0.019$\\
& N512 NA30 & & - & $JS_{t_2} = 0.023$ & $JS_{t_2} = 0.021$ &  $JS_{t_1} = 0.024$\\
& N2048 NA10 & & & - & $JS_{ra} = 0.001$ & $JS_{ra} = 0.001$ \\
& N2048 NA30 & & & & - & $JS_{t_1} = 0.001$\\
& N2048 NA50 & & & & & -\\
\hline \hline 
 & & N512 NA10 & N512 NA30 & N2048 NA10 & N2048 NA30 & N2048 NA50 \\
\hline 
\multirow{5}{*}{\phXHM} & N512 NA10 & - & $JS_{ra} = 0.007 $ &$JS_{t_2} = 0.019$ & $JS_{ra} = 0.024$ & $JS_{ra} = 0.025$\\
& N512 NA30 & & - & $JS_{t_2} = 0.023$ & $JS_{t_2} = 0.023$ &  $JS_{t_2} = 0.023$\\
& N2048 NA10 & & & - & $JS_{ra} = 0.005$ & $JS_{ra} = 0.004$ \\
& N2048 NA30 & & & & - & $JS_{ra} = 0.001$\\
& N2048 NA50 & & & & & -\\
\hline \hline 
 & & N512 NA10 & N512 NA30 & N2048 NA10 & N2048 NA30 &  \\
\hline 
\multirow{4}{*}{\phXPHM} &N512 NA10 & - & $JS_{ra} = 0.021 $ &$JS_{ra} = 0.039 $ & $JS_{ra} = 0.038$ & \\
& N512 NA30 & & - & $JS_{ra} = 0.012$ & $JS_{ra} = 0.009$ & \\
& N2048 NA10 & & & - & $JS_{ra} = 0.001$ & \\
& N2048 NA30 & & & & - & \\

\hline\hline
\end{tabular}
%\end{ruledtabular}
\caption{Maximum Jensen-Shannon (JS) divergence values between the posterior distributions for GW170729 estimated with \phTHM, \phXHM and \phXPHM using different sampler settings, where N is short for {\tt nlive} and NA is short for {\tt nact}.
\label{tab:tabConvergenceTHM}
}
\end{table*}

\begin{figure} 
    \includegraphics[width=\columnwidth]{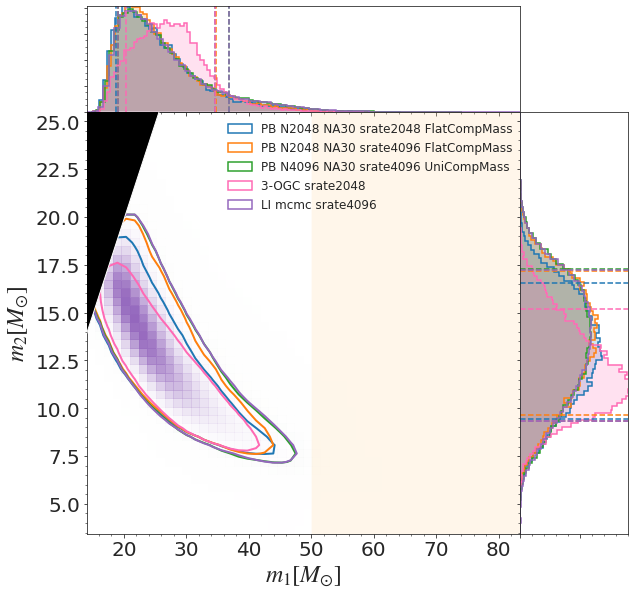}
    \includegraphics[width=\columnwidth]{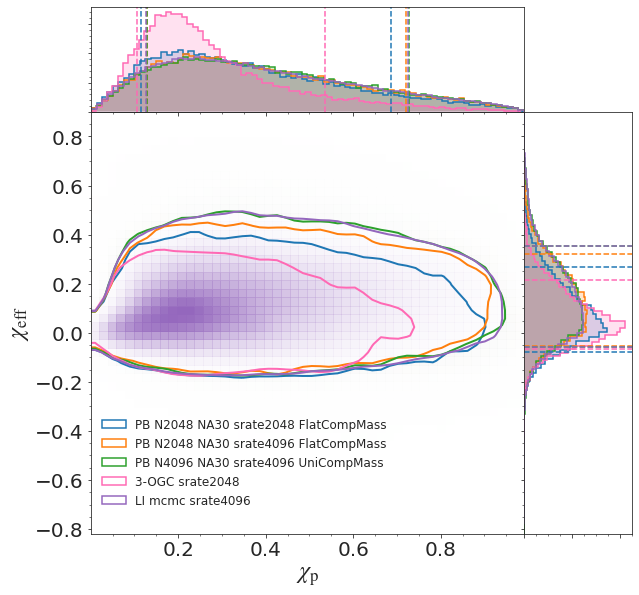}
    \caption{Comparison of the recovered posterior distributions for GW151012 for different values of the sampling rate ({\tt srate}) and mass priors in Parallel Bilby ({\tt PB}), a standard run in LALInference ({\tt LI mcmc}), and the results reported for this event in \citet{Nitz:2021uxj} (denoted as 3-OGC in the legend).
    For brevity, {\tt nlive} has been abbreviated as ``N'' and {\tt nact} as ``NA'' in the legend. {\tt FlatCompMass} refers to a run using a prior uniform in $q \leq 1$ and $\mathcal{M}_c$ and then reweighted to a flat prior in component masses. {\tt UniCompMass} refers to directly using a uniform prior in component masses.
    }
    \label{fig:gw151012_convergence}
\end{figure}

For the other two low-mass events, similar consistency is achieved between our different runs. In particular, for GW170608 our runs with a sampling rate of 4096\,Hz show excellent agreement, for both mass priors in Parallel Bilby and for LALInference, while for GW151012 small differences are visible between the standard prior Bilby run and the alternative prior and LALInference runs. However, differences between our different runs for this event are much smaller than differences with the results reported in \citet{Nitz:2021uxj}, as can be observed in Fig.~\ref{fig:gw151012_convergence}. Despite the small differences, the overall self-consistency of our results for GW151012 and GW170608 suggests that the results presented in \citet{Nitz:2021uxj} cannot be explained by different choices of the prior and sampling rate, and further examination of the convergence of those results and the influence of the choice of PSD estimation would be needed to understand the discrepancies.
%%%%%%

\section{Multibanding}\label{subsec:multibanding}

As mentioned in Sec.~\ref{sec:waveforms}, in order to accelerate the evaluation of the waveform model, \phXHM and \phXPHM implement the multibanding interpolation method described for non-precessing systems in \citet{Garcia-Quiros:2020qlt}, and extended to precession in \citet{Pratten:2020ceb}.
The method uses interpolation from an appropriately chosen unequally spaced coarse frequency grid to the equally spaced fine grid used  for data analysis. The grid spacing of the coarse grid is controlled by two threshold parameters, which are related to the local interpolation error for the phase and amplitude. The parameter controlling the accuracy of the non-precessing modes is called MB, and the parameter controlling the accuracy of the Euler angles evaluation used to construct the precessing waveform is PMB.

\begin{figure*} 
    \centering
    \includegraphics[width=\columnwidth]{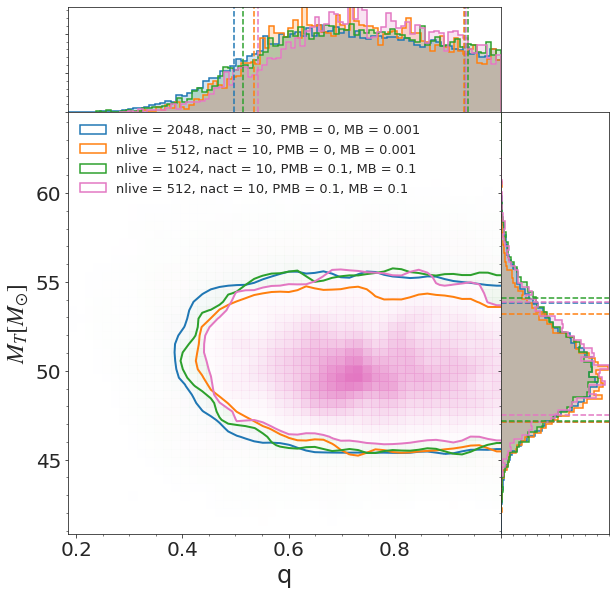}
    \includegraphics[width=\columnwidth]{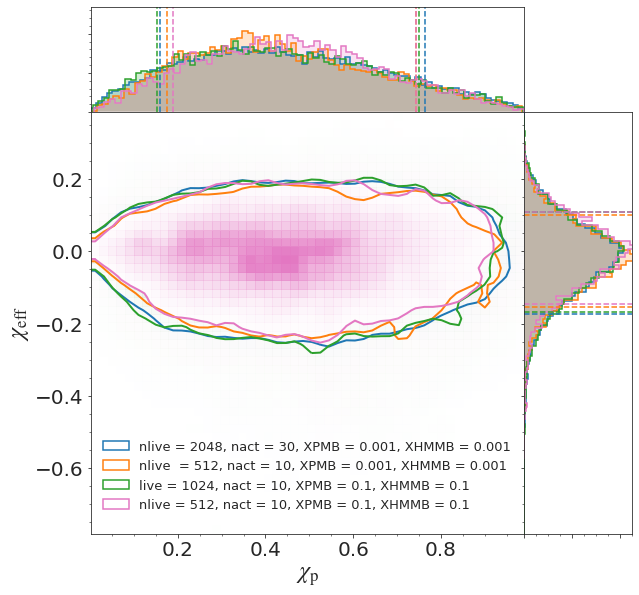}
    \caption{Comparison of the GW170104 posterior distributions for the total mass, mass ratio and effective spin obtained with \phXPHM using different sampler settings ({\tt nlive} and {\tt nact}) and 
     multibanding thresholds for the aligned-spin modes (XHMMB) and for the Euler angles (XPMB) where 0.001 is the multibanding default value.}
    \label{fig:MB}
    
\end{figure*}

\begin{table}
%\begin{ruledtabular}
\begin{tabular}{ccccccc}
\hline\hline
nlive & nact & MB & PMB & CPU h & L.eval. & Cost/L.eval.[ms]   \\
\hline
2048 & 30 & $10^{-3}$ & 0 & 1765 & $2.24 \times 10^8$ & 28.30 \\
512 & 10 & $10^{-3}$ & 0 & 154 & $1.67 \times 10^7$ & 33.11 \\
1024 & 10 & $10^{-1}$ & $10^{-1}$ & 157 & $3.46 \times 10^7$ & 16.34 \\
512 & 10 & $10^{-1}$ & $10^{-1}$ & 77 & $1.71 \times 10^7$ & 16.32 \\
\hline\hline
\end{tabular}
%\end{ruledtabular}
\caption{Computational cost comparison between runs on GW170104 using \phXPHM with different multibanding thresholds and sampler settings, where MB and PMB correspond to the multibanding thresholds of the aligned-spin modes and the Euler angles evaluation, respectively. The number of likelihood evaluations and the mean cost of each evaluation in ms are also shown.
\label{tab:tabMB}
}
\end{table}

We want to quantify the loss of accuracy and gain in computational speed for runs with a very aggressive multibanding threshold. To do that, we perform a comparison between the default version of \phXPHM (MB = $10^{-3}$ and PMB = $0$, i.e. no multibanding for the Euler angles) using {\tt nlive} = 512, 2048 and {\tt nact} = 10, 30 respectively, with two aggressive multibanding runs (MB = PMB = $0.1$) using {\tt nlive} = 512, 1024 and {\tt nact} = 10. In Fig. \ref{fig:MB} we show this comparison for
GW170104, which has the lowest total mass from the medium-mass events group. Results for the other events are contained in our data release \citep{Zenodo_release}. We find, not surprisingly, that for the low resolution runs ({\tt nlive} = 512 and {\tt nact} = 10) convergence is poor with and without the aggressive multibanding threshold. However, when we increase the number of live points to 1024 and 2048, the posterior distributions become smoother even using aggressive multibanding, and agree well with the distributions obtained with less aggressive multibanding. Comparing the computational resources of the runs in Table \ref{tab:tabMB}, we can observe that with an aggressive multibanding threshold the cost of each likelihood evaluation is reduced by approximately half. Although using a more aggressive multibanding threshold is not as accurate as runs without this acceleration, due to the approximations used, we see that in our case the change of the posterior distributions is very small. Thus, aggressive multibanding is particularly appropriate for quick exploratory runs, e.g. to tune prior bounds for masses or distance.

%%%%%%%%%%%%%%%%%%%%%%%%%%%%%%%%%%%%%%%%%%%%%%%%%%

% Don't change these lines
\bsp	% typesetting comment
\label{lastpage}
\end{document}